\documentclass[fleqn,usenatbib]{mnras}

\usepackage{newtxtext,newtxmath}

\usepackage{placeins}
\usepackage{tabularx}  
\usepackage{makecell}  
\usepackage{cleveref}
\crefrangeformat{figure}{#3#1#4--#5#2#6}

\DeclareRobustCommand{\VAN}[3]{#2}
\let\VANthebibliography\thebibliography
\def\thebibliography{\DeclareRobustCommand{\VAN}[3]{##3}\VANthebibliography}

\usepackage{float}       
\usepackage{subcaption}  
\usepackage{bm}          
\usepackage{array}       
\usepackage{booktabs}
\usepackage{array}
\newcolumntype{P}[1]{>{\raggedright\arraybackslash}p{#1}} 
\usepackage{url}
\usepackage{titlesec}
\titleformat{\section}
  {\normalfont\Large\bfseries} 
  {\thesection}{1em}{}
  
\usepackage{graphicx}	
\usepackage{amsmath}	
\usepackage{hyperref}  
\usepackage{xcolor}

\newcommand{\secondcolor}[1]{{\color{teal}#1}}

\usepackage{tikz}
\usetikzlibrary{shapes.symbols, shapes.geometric, arrows.meta}

\title[CC SN Rates in DES]{Core Collapse Supernova Rates in the Dark Energy Survey}

\author[K. Boyne et al.]{%
K. Boyne$^{1}$,\thanks{E-mail: kevin.boyne@port.ac.uk (UoP)}
O. Graur$^{1, 2}$,
M. Vincenzi$^{3}$,
P. Wiseman$^{4}$, 
M. Grayling$^{5}$, 
M. Smith$^{6}$, 
C. Frohmaier$^{1}$ and \newauthor
C. Lidman$^{7}$
\\
$^{1}$Institute of Cosmology and Gravitation, University of Portsmouth, Portsmouth, PO1 3FX, UK\\
$^{2}$Department of Astrophysics, American Museum of Natural History, Central Park West and 79th Street, New York, NY 10024-5192, USA\\
$^{3}$Department of Physics, University of Oxford, Parks Road, Oxford, OX1 3PU, UK\\
$^{4}$School of Physics and Astronomy, University of Southampton, Highfield Campus, Southampton, SO17 1BJ, UK\\
$^{5}$Kavli Institute for Cosmology, University of Cambridge, Madingley Road, Cambridge, CB3 0HA, UK\\
$^{6}$Physics Department, University of Lancaster, Physics Building, Lancaster, LA1 4YB, UK\\
$^{7}$Research School of Astronomy and Astrophysics and Centre for Gravitational Astrophysics, The Australian National University, Canberra, ACT, Australia
}

\date{Accepted XXX. Received YYY; in original form ZZZ}

\pubyear{\the\year{}}

\begin{document}
\label{firstpage}
\pagerange{\pageref{firstpage}--\pageref{lastpage}}
\maketitle

\begin{abstract}
Over its five-year campaign (2013–2017), the Dark Energy Survey (DES) supernova (SN) survey recorded 31,636 light curves. Using a custom-built, two-stage random forest classifier, we identified 2,445 likely core-collapse SNe (CC SNe) within the survey rate analysis footprint. With this sample, we measure the volumetric CC SN rate from redshift $z = 0$ to 1.3 in 11 redshift bins, representing the most detailed single-survey coverage to date across nine billion years of lookback time. Over this interval, the CC SN rates are well fit by a power law of the form $R_\text{CC}(z) = R_0(1 + z)^\alpha$, with $R_0 = 1.58 \pm 0.17 \times 10^{-4}\text{ yr}^{-1}\text{ Mpc}^{-3}$ and $\alpha = 2.85 \pm 0.25$. The rate turnover, expected at cosmic noon, is not yet apparent. Because CC SNe trace recent star formation, we compare our measurements with recent cosmic star-formation history (CSFH) models. Our measurements are $\sim 70$ per cent higher than CSFH models using a Salpeter initial mass function (IMF) to predict CC SN rates. Replacing the Salpeter IMF with a Baldry and Glazebrook IMF and a CC SN progenitor mass range of $8\text{--}40\,\text{M}_\odot$ bridges the gap. Alternatively, CSFHs may miss a hitherto unidentified population of low-luminosity galaxies. We further find that the local ratio of stripped-envelope (SE) SNe to all CC SNe is $31 \pm 5$ per cent, remaining constant from $z = 0$ to 1.3. Combined with an IMF analysis accounting for the "missing progenitor problem" of hydrogen-rich SNe, our results favour a mixture of single massive stars and binaries as SE SN progenitors, with the redshift-invariant ratio favouring binary channels.
\end{abstract}




\section{Intoduction}

A core-collapse supernova (CC SN) occurs when the core of a massive star exhausts its nuclear fuel and can no longer generate sufficient pressure to counteract gravity. The resulting explosion ejects several solar masses of material at velocities of $10^3$ to $10^4$ km s$^{-1}$, driving a powerful shock wave into the surrounding interstellar medium. What remains is a compact remnant — either a neutron star or, if the core mass exceeds the Tolman–Oppenheimer–Volkoff limit of $\sim$ 2.16~M\textsubscript{$\odot$}, a black hole \citep{OppenheimerVolkoff, Rezzolla2018}. Although all CC SNe share this basic mechanism, their explosion energies span a wide range, reflecting the diversity of their progenitor masses and explosion mechanisms \citep{Janka2012, Burrows2013}.  

CC SNe play several important roles in the Universe. They are major sources of heavy elements, driving stellar nucleosynthesis and producing the bulk of the Universe's $\alpha$-elements \citep{Burbidge1957, Nomoto2013}, as well as the building blocks of life, such as carbon, nitrogen, and oxygen.
Through their energetic feedback, CC SNe regulate star formation and chemical enrichment in galaxies \citep{Woosley2002, Hopkins2014}. They also produce neutrinos and accelerate cosmic rays \citep{Hirata1987, Blasi2013} and are potential sources of gravitational waves \citep{Kotake2013, Abbott2020CCSNTargeted}. Their compact remnants—neutron stars and stellar-mass black holes—are laboratories for extreme physics (e.g., \citealt{Psaltis2008, Lattimer2012, Ozel2016}). Mergers of neutron stars create kilonovae, which further enrich the Universe through r-process nucleosynthesis, producing heavier elements like gold and platinum (e.g., \citealt{Abbott2017, Kasen2017, Thielemann2017}).
\par
CC SNe are broadly classified into two families:
hydrogen-rich SNe (herewith H-rich SNe, comprising Types IIP, IIL, and IIn) and stripped envelope  SNe (SE SNe, comprising Types Ib, Ic, and IIb; see classification reviews by \citealt{Filippenko1997} and \citealt{GalYam2017}). Type~II SNe exhibit hydrogen in their spectra and include subtypes such as Type~IIP, which display a plateau phase in their light curves, and Type~IIL, which decline more linearly \citep{Arcavi2017}. Recent observations \citep[e.g.,][]{Anderson2014, Sanders2015, Valenti2016} suggest that IIP and IIL light-curve declines may form a continuum, driven mainly by how much of the hydrogen envelope the progenitor star retains before explosion \citep[e.g.,][]{Martinez2022, Martinez2022b, Martinez2022c}. Rarer subclasses include Type IIn, marked by narrow hydrogen emission lines from circumstellar interaction (e.g., \citealt{Kiewe2012, Nyholm2020, Moriya2024}), and Type II-87A–like events, which rise slowly to peak \citep{Arnett1989}.

SE~SNe show little or no hydrogen. Type Ib SNe are helium-rich, Type Ic are helium-poor, and Type IIb show transient hydrogen features that fade to reveal helium dominated
spectra \citep{PianMazzali2017}. Some Type Ic SNe exhibit unusually broad lines (Ic-BL), while Ca-rich SNe are fainter events with strong calcium features \citep{Perets2010, Kasliwal2012}. It should be noted that many Ca-rich events are now thought to be thermonuclear rather than core-collapse in origin.

The progenitors of CC~SNe are stars more massive than $8~\mathrm{M_{\odot}}$. Below this limit, stellar cores are supported by electron degeneracy pressure, halting collapse and leaving white dwarfs \citep{Karakas2017}. The upper mass limit is less well constrained; theoretical models suggest Type~II explosions up to $\sim 25~\mathrm{M_{\odot}}$ \citep{Heger2003, Eldridge2004}, but direct detections of Type~II progenitors imply a limit closer to $16.5\text{--}19~\mathrm{M_{\odot}}$ \citep{Smartt2009, davies2018}. This discrepancy, known as the ``red supergiant problem,'' remains an open question. Pre-explosion images have confirmed that Type IIP progenitors are red
supergiants, while Type IIL progenitors may
be partially stripped yellow or red supergiants \citep{Smartt2009, VanDyk2017}. The progenitor of
SN 1987A was directly observed to be a blue supergiant \citep{Walborn1987ApJL}.

Uncertainties remain regarding the progenitors of SE~SNe, for which two primary evolutionary channels are proposed. In the first scenario, massive single stars (typically $\gtrsim$25--30~M$_{\odot}$) lose their hydrogen-rich envelopes through strong stellar winds (e.g., \citealt{MaederLequeux1982, Heger2003, Georgy2012}) and episodic mass-loss events \citep{Smith2014}. Depending on the extent of this stripping, these progenitors may appear as Wolf--Rayet stars before exploding as Type~Ib or Ic~SNe \citep{Crowther2007, Gotberg2018}.

The second channel involves mass transfer to a binary companion via Roche lobe overflow, which strips the primary star's hydrogen (and potentially helium) layers (e.g., \citealt{Paczynski1971, Podsiadlowski1992, DeDonder1998, Vanbeveren1998NewAstronomy3}). For specific subclasses, such as SNe~Ic-BL associated with gamma-ray bursts, alternative mechanisms like chemically homogeneous evolution \citep{Yoon2005} or explosive common-envelope ejection \citep{Podsiadlowski2010} may be required.

\citet{Smith2011} concluded that both evolutionary channels are necessary to account for the observed fractions of SE~SNe. In this framework, mass transfer in binary systems is the dominant mechanism for lower-mass progenitors, while line-driven stellar winds from massive single stars stripped of their envelopes become significant at higher initial masses. \citet{Graur2017a} showed that measurements of the ratio between the mass-normalized rates of SE SNe and H-rich SNe were consistent with a binary progenitor channel for SE SNe but inconsistent with a scenario in which only single, massive stars ended as SE SNe. 

Because CC~SNe arise from short-lived massive stars, their rate of explosion closely traces ongoing star formation, providing an independent measure of the star-formation rate in galaxies. Measuring these explosion rates across cosmic time thus offers a unique, independent probe of the cosmic star formation history (CSFH) \citep{Madau1998, Baldry2003, Hopkins2006}. The CSFH has typically been derived using tracers such as (i) rest-frame UV emission, which reflects light from young, massive stars but is strongly affected by dust; (ii) total infrared luminosity, which captures dust-reprocessed starlight but can include contributions from older populations; and (iii) optical emission lines, which also suffer from extinction uncertainties. More recent examples of such CSFH models include those of \citet{Madau2014, Madau2017}, \citet{Harikane2022}, and \citet{Kim2024}. When corrected for host-galaxy extinction, CC~SN rates offer a near-instantaneous and complementary measure of star formation over cosmic time. Alhough optical CC~SN rates are themselves heavily impacted by dust obscuration---particularly within dense starburst environments where a significant fraction of explosions remain hidden---modelling this missing fraction \citep{Mattila2012} allows them to serve as an important independent test of the uncertainties that challenge traditional tracers.

The first systematic attempts to measure CC SN rates began in the 1990s, following Fritz Zwicky’s pioneering theoretical work in the 1930s \citep{Zwicky1938}; for a concise history of the early days of SN observations, see \citet{Graur2022}. Early studies such as \citet{Cappellaro1999} demonstrated that SN rates were strongly correlated with galaxy properties linked to star formation: bluer, late-type galaxies produced more events than red, early-type systems. Targeted programs such as the Lick Observatory Supernova Search (LOSS; \citealt{Leaman2010, Li2011a, Li2011b, Maoz2011, Graur2017a, Graur2017b, Shivvers2017}) implemented regular CCD monitoring and automated pipelines, enabling robust measurements of nearby SN rates normalized by galaxy luminosity and stellar mass. The advent of wide-field rolling surveys, including the Sloan Digital Sky Survey II Supernova Survey (SDSS-II SN Survey; \citealt{Frieman2008, Dilday2010}), the Palomar Transient Factory and intermediate Palomar Transient Factory (PTF/iPTF; \citealt{Law2009, Frohmaier2020}), and the Panoramic Survey Telescope and Rapid Response System (Pan-STARRS; \citealt{Kaiser2010, McCrum2015}), revolutionized the field by enabling direct measurements of volumetric SN rates.

In the early 2010s, an analysis by \citet{Horiuchi2011} highlighted a discrepancy between observed CC SN rates and those predicted from CSFH models of that time, such as \citet{Hopkins2006}, suggesting a missing fraction of highly obscured or faint SNe. Building on work by \citet{Mannucci2007MNRAS377}, \citet{Mattila2012} showed that 20-40 per~cent of CC SNe were being lost to high dust extinction, especially in highly star-forming galaxies. The corrections they estimated based on empirical calibrations of local starburst galaxies were then adopted by subsequent CC SN rate works, including \citet{Graur2011}, \citet{Mattila2012}, \citet{Melinder2012} and \citet{Dahlen2012}.

More recently, deep Hubble Space Telescope surveys such as the 
Cosmic Assembly Near-infrared Deep Extragalactic Legacy Survey 
\citep[CANDELS;][]{Grogin2011, Rodney2014} and the Cluster Lensing 
And Supernova Survey with Hubble \citep[CLASH;][]{Postman2012} 
extended measurements beyond a redshift of $1$ \citep{Graur2014, Strolger2015}. 
Building upon these space-based benchmarks, recent deep observations with the 
\textit{James Webb Space Telescope} (\textit{JWST}) as part of the \textit{JWST} Advanced Deep 
Extragalactic Survey \citep[JADES;][]{2026arXiv260712018D, Vassallo2026} have pushed 
transient and CC~SN rate estimations into unprecedented high-redshift regimes. 

However, while these space-based programs provide crucial high-redshift benchmarks, 
their small fields of view continue to limit statistical precision, leaving the high-$z$ 
regime susceptible to cosmic variance and small-number statistics. 
Meanwhile, modern all-sky transient surveys such as the All-Sky Automated 
Survey for SuperNovae \citep[ASAS-SN;][]{Shappee2014, Pessi2025}, the Asteroid 
Terrestrial-Impact Last Alert System \citep[ATLAS;][]{Tonry2018, Smith2020_ATLAS}, 
and the Zwicky Transient Facility \citep[ZTF;][]{Bellm2019, Perley2020, Dimitriadis2025} 
have achieved near-complete coverage of the nearby Universe, enabling precise 
measurements of local CC~SN rates. These recent surveys generally present rates which 
have not been corrected for dust obscuration in highly star-forming environments using 
the \citet{Mattila2012} methodology, and therefore predict lower rates than CSFH models 
and surveys in which rates have been corrected.

A key focus of the literature has been the tension between 
observed volumetric CC~SN rates and predictions derived from the cosmic star 
formation history \citep[CSFH; e.g.][]{Baldry2003, Hopkins2006}. Notably, 
\citet{Horiuchi2011} determined that the CC~SN rate predicted from models overpredicted empirical 
observations by approximately a factor of two. This tension has been somewhat alleviated by more recent CSFH models \citep[CSFH; e.g.][]{Madau2014}, which use the unbroken ($\alpha = 2.35$) \citet{Salpeter1955} IMF resulting in a lower normalization for the CSFH (by convention IMF high mass slope = $-\alpha)$. Intermediate redshifts 
($0.2 < z < 1.0$) represent a key epoch for comparison of CC SN rates with CSFH models, yet shallow local surveys cannot trace this evolution, and narrow deep space surveys lack the volume required to overcome cosmic variance. The Dark Energy Survey (DES) is therefore important as it uniquely bridges this intermediate redshift gap by combining high-sensitivity deep imaging with a large sky area, providing the statistical sample size necessary to robustly test the coupling between the CC~SN rate and the birth of massive stars.

The DES Supernova survey (DES-SN) conducted a deep, untargeted, five-season rolling search, assembling one of the largest high-redshift SN samples to date. \citet{Wiseman2021} analysed the SN Ia rate as a function of galaxy properties and inferred the SN Ia delay-time distribution; \citet{Toy2023} found subtle differences in the SN Ia rate between field and cluster galaxies at different masses. 
Recent work has modelled selection effects and contamination in the DES-SN sample, establishing the framework for robust cosmological analyses with SNe Ia \citep{Vincenzi2019, Vincenzi2021}, 
and detailed studies have characterised the luminosity functions and host galaxy properties of CC SNe discovered by DES \citep{Grayling2023}. These works culminated in the recent DES five-year SN~Ia cosmology results, which used a sample of $\sim$1635 SNe~Ia spanning $0.1 < z < 1.13$. These were classified spectroscopically or photometrically, and light curves were fit using the SALT empirical light-curve framework in order to standardize distances \citep{Guy2007}. This comprehensive sample placed tight constraints on cosmological parameters and probed the dark energy equation of state \citep{Vincenzi2024, DES2024, Popovic2026}.

In this paper, we present measurements of the volumetric CC SN rate, $R_{\mathrm{CC}}(z)$, using data from the Dark Energy Survey (DES; \citealt{Diehl2014_DES_strategy, Flaugher2015, DES2016_overview, Abbott2021_DESDR2, Sanchez2024_DES5yr}). While DES was primarily designed to probe cosmic acceleration via multiple cosmological probes---including weak lensing, large-scale structure, galaxy clusters (\citealt{DESoverview}), and Type~Ia SNe (SNe Ia; \citealt{Bernstein2012})
---its deep, multi-epoch imaging provides an ideal dataset for the systematic study of CC SNe. By utilizing 11 independent redshift bins to trace the rate evolution continuously from $0 < z < 1.3$, this work provides an unprecedentedly granular mapping of the cosmic CC SN rate across nine billion years of lookback time, offering a significantly higher sampling density than previous literature.

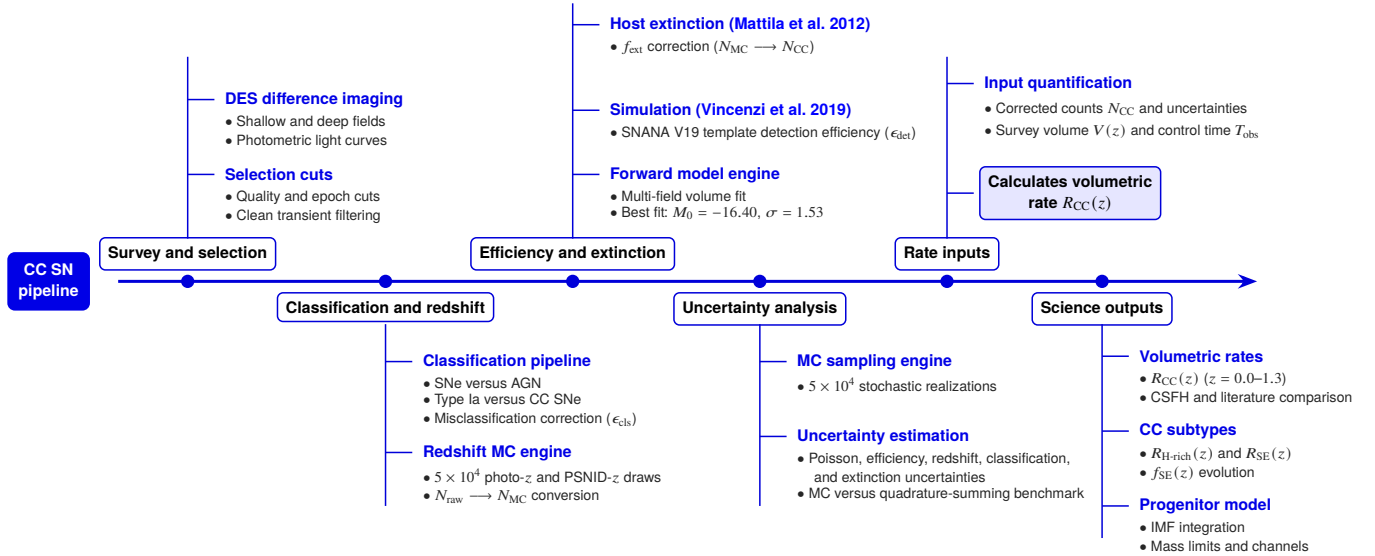
\begin{figure*}
\centering
\resizebox{\textwidth}{!}{%
\begin{tikzpicture}[
    font=\sffamily\small,
    >=Stealth,
    main axis/.style={draw=blue!85!black, ultra thick},
    axis point/.style={circle, fill=blue!85!black, inner sep=2.5pt},
    title box/.style={
        rectangle, fill=blue!90!black, text=white,
        font=\sffamily\bfseries\normalsize,
        rounded corners=4pt, inner sep=6pt, align=center
    },
    stage badge/.style={
        rectangle, draw=blue!85!black, fill=white, thick,
        font=\sffamily\bfseries\normalsize,
        rounded corners=3pt, inner sep=4.5pt, align=center
    },
    branch title/.style={
        font=\sffamily\bfseries\normalsize,
        text=blue!90!black, anchor=west
    },
    sub text/.style={
        font=\sffamily\small, text=black!85,
        align=left, anchor=north west
    },
    action box/.style={
        rectangle, draw=blue!85!black, fill=blue!10, thick,
        font=\sffamily\bfseries\normalsize,
        rounded corners=3pt, align=center, inner sep=4.5pt
    },
    stem line/.style={draw=blue!85!black, thick},
    tick line/.style={draw=blue!85!black, thick}
]

\draw[main axis, ->] (-8.8,0) -- (12.5,0);
\node[title box] at (-10.1,0) {CC SN\\pipeline};

\coordinate (x0) at (-7.5,0);
\node[axis point] at (x0) {};
\node[stage badge, anchor=south] at (-7.5,0.20)
    {Survey and selection};
\draw[stem line] (-7.5,0.80) -- (-7.5,4.2);

\draw[tick line] (-7.5,3.4) -- (-7.0,3.4);
\node[branch title] at (-6.9,3.4)
    {DES difference imaging};
\node[sub text] at (-6.9,3.2)
    {%
    \shortstack[l]{%
    $\bullet$ Shallow and deep fields\\
    $\bullet$ Photometric light curves%
    }%
    };

\draw[tick line] (-7.5,2.0) -- (-7.0,2.0);
\node[branch title] at (-6.9,2.0)
    {Selection cuts};
\node[sub text] at (-6.9,1.8)
    {%
    \shortstack[l]{%
    $\bullet$ Quality and epoch cuts\\
    $\bullet$ Clean transient filtering%
    }%
    };

\coordinate (x1) at (-3.8,0);
\node[axis point] at (x1) {};
\node[stage badge, anchor=north] at (-3.8,-0.20)
    {Classification and redshift};
\draw[stem line] (-3.8,-0.80) -- (-3.8,-4.2);

\draw[tick line] (-3.8,-1.5) -- (-3.3,-1.5);
\node[branch title] at (-3.2,-1.5)
    {Classification pipeline};
\node[sub text] at (-3.2,-1.7)
    {%
    \shortstack[l]{%
    $\bullet$ SNe versus AGN\\
    $\bullet$ Type Ia versus CC SNe\\
    $\bullet$ Misclassification correction
    ($\epsilon_{\mathrm{cls}}$)%
    }%
    };

\draw[tick line] (-3.8,-3.2) -- (-3.3,-3.2);
\node[branch title] at (-3.2,-3.2)
    {Redshift MC engine};
\node[sub text] at (-3.2,-3.4)
    {%
    \shortstack[l]{%
    $\bullet$ $5\times10^4$ photo-$z$ and PSNID-$z$ draws\\
    $\bullet$ $N_{\mathrm{raw}} \longrightarrow N_{\mathrm{MC}}$
    conversion%
    }%
    };

\coordinate (x2) at (-0.3,0);
\node[axis point] at (x2) {};
\node[stage badge, anchor=south] at (-0.3,0.20)
    {Efficiency and extinction};

\draw[stem line] (-0.3,0.80) -- (-0.3,5.2);

\draw[tick line] (-0.3,4.8) -- (0.2,4.8);
\node[branch title] at (0.3,4.8)
    {Host extinction \citep{Mattila2012}};
\node[sub text] at (0.3,4.6)
    {$\bullet$ $f_{\mathrm{ext}}$ correction
    ($N_{\mathrm{MC}} \longrightarrow N_{\mathrm{CC}}$)};

\draw[tick line] (-0.3,3.2) -- (0.2,3.2);
\node[branch title] at (0.3,3.2)
    {Simulation \citep{Vincenzi2019}};
\node[sub text] at (0.3,3.0)
    {$\bullet$ \textsf{SNANA} V19 template detection efficiency
    ($\epsilon_{\mathrm{det}}$)};

\draw[tick line] (-0.3,2.0) -- (0.2,2.0);
\node[branch title] at (0.3,2.0)
    {Forward model engine};
\node[sub text] at (0.3,1.8)
    {\shortstack[l]{%
    $\bullet$ Multi-field volume fit\\
    $\bullet$ Best fit: $M_0=-16.40$, $\sigma=1.53$%
    }};
\coordinate (x3) at (3.2,0);
\node[axis point] at (x3) {};
\node[stage badge, anchor=north] at (3.2,-0.20)
    {Uncertainty analysis};
\draw[stem line] (3.2,-0.80) -- (3.2,-4.2);

\draw[tick line] (3.2,-1.5) -- (3.7,-1.5);
\node[branch title] at (3.8,-1.5)
    {MC sampling engine};
\node[sub text] at (3.8,-1.7)
    {$\bullet$ $5\times10^4$ stochastic realizations};

\draw[tick line] (3.2,-2.9) -- (3.7,-2.9);
\node[branch title] at (3.8,-2.9)
    {Uncertainty estimation};
\node[sub text] at (3.8,-3.1)
    {%
    \shortstack[l]{%
    $\bullet$ Poisson, efficiency, redshift, classification,\\
    \hspace{1.2em}and extinction uncertainties\\
    $\bullet$ MC versus quadrature-summing benchmark%
    }%
    };


\coordinate (x4) at (6.7,0);
\node[axis point] at (x4) {};

\node[stage badge, anchor=south] at (6.7,0.20)
    {Rate inputs};

\draw[stem line] (6.7,0.80) -- (6.7,4.2);

\draw[tick line] (6.7,3.7) -- (7.2,3.7);
\node[branch title] at (7.3,3.7)
    {Input quantification};

\node[sub text] at (7.3,3.45)
    {%
    \shortstack[l]{%
    $\bullet$ Corrected counts $N_{\mathrm{CC}}$ and uncertainties\\[2pt]
    $\bullet$ Survey volume $V(z)$ and control time $T_{\mathrm{obs}}$%
    }%
    };

\draw[tick line] (6.7,1.65) -- (7.2,1.65);
\node[action box, anchor=west] at (7.3,1.65)
    {Calculates volumetric\\rate $R_{\mathrm{CC}}(z)$};

\coordinate (x5) at (9.6,0);
\node[axis point] at (x5) {};
\node[stage badge, anchor=north] at (9.6,-0.20)
    {Science outputs};
\draw[stem line] (9.6,-0.80) -- (9.6,-4.8);

\draw[tick line] (9.6,-1.4) -- (10.1,-1.4);
\node[branch title] at (10.2,-1.4)
    {Volumetric rates};
\node[sub text] at (10.2,-1.6)
    {%
    \shortstack[l]{%
    $\bullet$ $R_{\mathrm{CC}}(z)$ ($z=0.0$--$1.3$)\\
    $\bullet$ CSFH and literature comparison%
    }%
    };

\draw[tick line] (9.6,-2.8) -- (10.1,-2.8);
\node[branch title] at (10.2,-2.8)
    {CC subtypes};
\node[sub text] at (10.2,-3.0)
    {%
    \shortstack[l]{%
    $\bullet$ $R_{\mathrm{H\text{-}rich}}(z)$ and $R_{\mathrm{SE}}(z)$\\
    $\bullet$ $f_{\mathrm{SE}}(z)$ evolution%
    }%
    };

\draw[tick line] (9.6,-4.2) -- (10.1,-4.2);
\node[branch title] at (10.2,-4.2)
    {Progenitor model};
\node[sub text] at (10.2,-4.4)
    {%
    \shortstack[l]{%
    $\bullet$ IMF integration\\
    $\bullet$ Mass limits and channels%
    }%
    };

\end{tikzpicture}%
}
\vspace{0.2cm}
\caption{Timeline and mind-map schematic of the analysis
pipeline. The analysis proceeds from left to right along the main
timeline, with each stage branching vertically. Both upper and lower
branches follow a top-to-bottom reading order. The branches summarize
data selection, classification and redshift estimation, efficiency and
extinction corrections, uncertainty analysis, rate inputs, and final
science outputs.}
\label{fig:mnras_pipeline_flowchart_horizontal}
\end{figure*}

In Section~\ref{sec:observations} we describe the DES imaging and follow-up spectroscopy used in this work. Section~\ref{sec:sandc} outlines our SN selection and classification procedures, including candidate selection (Section~\ref{sec:selection}), classification (Section~\ref{sec:classification}), and subtype classification (Section~\ref{sec:subtypes}). In Section~\ref{sec:zandext} we correct the observed CC~SN counts in each redshift bin for redshift uncertainties (Section~\ref{sec:redshift_uncertainties}) and incompleteness due to host-galaxy dust extinction (Section~\ref{sec:extinction}). In Section~\ref{sec:rdeu} we describe the rate calculation framework, beginning with our overall methodology (Section~\ref{sec:rates}) and the determination of survey detection efficiency via pipeline simulations (Section~\ref{sec:efficiency}). We then model the underlying survey selection function and detection thresholds using a physical forward model (Section~\ref{sec:survey_selection}) before outlining the calculation of the final CC~SN rates (Section~\ref{sec:uncertainties}). A comprehensive treatment of systematic errors via a  Monte Carlo (MC) propagation framework is detailed in Section~\ref{sec:mc_error_propagation}. We then present H-rich and SE~SN subtype rates, finding that SE~SNe represent $31\pm5$ per cent of all CC~SNe, with no apparent evolution of this fraction out to $z = 1.3$ (Section~\ref{sec:subtype_rates}). In Section~\ref{sec:discussion} we compare our extinction-corrected CC~SN rates with previous rate measurements and find that they trace a consistent increasing trend with redshift. We find that our CC~SN rates are $\sim$70~per~cent higher than predicted by recent CSFH models, and discuss the origin of this offset (Section~\ref{sec:ccsn_csfh_tracers}). We then use our CC~SN and subtype rates to constrain progenitor models for SE~SNe, finding that they are consistent with a mixed progenitor channel model (Section~\ref{sec:progenitors}). Finally, we summarise our conclusions in Section~\ref{sec:conclusions}. A comprehensive schematic summarizing the complete operational workflow executed throughout this analysis is shown in Figure~\ref{fig:mnras_pipeline_flowchart_horizontal}.

Throughout this work, we assume a $\Lambda$-cold-dark-matter ($\Lambda$CDM) cosmological model
with parameters $\Omega_{\Lambda} = 0.7$, $\Omega_{\rm m} = 0.3$, and
$H_0 = 70~\mathrm{km~s^{-1}~Mpc^{-1}}$. Unless noted otherwise, all magnitudes
are on the AB system.

\section{Observations}
\label{sec:observations}

DES-SN operated from 2013 to 2018, monitoring ten fields — eight shallow and two deep — covering a total area of $\sim 27~\mathrm{deg}^2$, which reduces to $22.912~\mathrm{deg}^2$ after taking account of dead CCDs, borders, masked pixels and field overlaps \citep{Lasker2020}. The shallow fields had a single-epoch 50~per~cent completeness depth of $\sim 23.5$\,mag, while the deep fields extended this to $\sim 24.5$\,mag \citep{Kessler2015, Vincenzi2021}. Further details of the observing strategy, cadence, seeing, and overall depth distributions are provided by \citet{Smith2020}. 

Transients were detected in real time using difference imaging \citep{Kessler2015}, and artifacts were removed with the \texttt{autoScan} machine-learning classifier \citep{Goldstein2015}. An extensive spectroscopic follow-up campaign, mainly using the AAOmega spectrograph on the Anglo-Australian Telescope as part of the OzDES survey (e.g., \citealt{Yuan2015, Childress2017, Lidman2020}), provided redshifts and host-galaxy spectroscopy for thousands of transients.

Photometric calibration relied on a catalogue of around 17 million tertiary standard stars within the DES footprint, built using the Forward Global Calibration Method \citep{Stubbs2006}, implemented for DES by \citet{Burke2018}. This method achieved residuals below 3\,mmag \citep{Sevilla-Noarbe2021, Rykoff2023}, ensuring a uniform flux scale across all survey fields and epochs.

Initial flux measurements were obtained using point-spread-function (PSF) fitting photometry on difference images with the \texttt{DIFFIMG} pipeline \citep{Kessler2015}. Deep templates were made by co-adding optimal exposures with low sky noise and small PSF. Each nightly image was aligned to the template and convolved to match its PSF and photometry. The resulting difference image contained only pixels with significant flux changes, from either genuine transients or residual artifacts. Sources detected at $>3.5\sigma$ above the background were evaluated by \texttt{autoScan}, which assigned probabilities of being real astrophysical events.

High-precision photometry was obtained with the Scene Modelling Photometry (SMP) pipeline \citep{Brout2019, Sanchez2024_DES5yr}, applied to all 31,636 \texttt{DIFFIMG} candidates. SMP fits all single-epoch images simultaneously, accounting for spatial and temporal variations in the source within a consistent noise model. Not all fits converged successfully, mainly due to multi-seasonality; ultimately, 19,706 high-quality SMP light curves were produced and released in the DES-SN five-year dataset. While these SMP data are optimized for high-purity SN~Ia cosmology, the non-convergence of the remaining $\sim$12,000 candidates represents an important population difference for a volumetric rate analysis. This approach preferentially impacts long-lived or unusual profiles that deviate from typical SN~Ia light-curve behaviors. To ensure maximum sample completeness across all types of core-collapse explosions, this work utilizes the initial, flux-complete tracking from the real-time \texttt{DIFFIMG} pipeline as our base catalog.

Transients were linked to host galaxies using the directional light radius (DLR) method \citep{Sullivan2006, Gupta2016}. Here, the DLR is the host's projected radius in the direction of the transient, as defined by its elliptical Kron profile. The normalized distance, $d_{\mathrm{DLR}}$, is the angular separation between the transient and host center divided by this radius, and is therefore a dimensionless ratio. Host galaxy candidates were initially identified using the SVA1-COADD GOLD catalogue \citep{Rykoff2016}, requiring a directional light radius of 
$d_{\mathrm{DLR}} < 7$ for a valid association. However, this catalogue was only utilized for preliminary targeting; for the final data releases and definitive $d_{\mathrm{DLR}}$ measurements, all host galaxy properties were 
derived from the deep, stacked DES SN field imaging as detailed by \citet{Wiseman2020}.

Spectroscopic redshifts were obtained primarily through the OzDES program, which utilized the 2dF fibre positioner and AAOmega spectrograph on the 3.9-m Anglo-Australian Telescope (AAT; \citealt{Yuan2015, Childress2017, Lidman2020}). OzDES targeted transient host galaxies for spectroscopic follow-up based on an automated selection framework combining directional light radius cuts, transient detection significance, and light-curve duration constraints. These exact targeting and data quality criteria are described comprehensively in \citet{Yuan2015}, \citet{Childress2017}, and \citet{Vincenzi2021}.

This strategy ensured the efficient use of spectroscopic resources. Additional redshifts were collected from public catalogues overlapping the DES fields, with final classifications and redshifts derived from both SN features and host-galaxy emission and absorption lines \citep{Smith2020}. Further details regarding the host-galaxy redshift sample are provided in Section~\ref{sec:zandext}.

\section{Selection \& Classification}
\label{sec:sandc}

In this section, we describe how we identified our CC SN sample out of the 19,792 passing the \texttt{transient\_status} criteria through the DES-SN DIFFIMG pipeline (Section~\ref{sec:selection}), the classification process we developed to distinguish CC SNe from AGNs and SNe Ia (Section~\ref{sec:classification}), and the further subclassification of our CC SN sample into H-rich and SE SNe (Section~\ref{sec:subtypes}). 

\subsection{Candidate Selection}
\label{sec:selection}

The starting point for this analysis is the full set of 31,636 transient light curves identified by the DES difference-imaging pipeline (\texttt{DIFFIMG}) over the five-year DES-SN survey \citep{Smith2020, Vincenzi2021}. This sample was progressively reduced using a sequence of quality and consistency cuts designed to isolate genuine astrophysical transients suitable for classification.

Candidate identification and initial transient selection in DES follow the difference-imaging and machine-learning protocols described by \citet{Smith2020}, utilizing the \texttt{transient\_status} flag framework. This initial selection process---including spatial coincidence thresholds, \texttt{autoScan} quality cuts, and multi-season bitmask logic---is detailed comprehensively in \citet{Smith2020}, while the parallel SMP data workflows are described by \citet{Sanchez2024_DES5yr}. Rather than restricting our input baseline to the subset of transients that achieved successful convergence within the cosmology-optimized SMP framework, we begin with the broader, unclipped \texttt{DIFFIMG} sample. This approach ensures that our downstream selection function remains independent of external pipeline convergence constraints, which was considered important for mapping the full, diverse core-collapse population. Applying the \texttt{transient\_status} criteria to this complete base catalog reduces our sample to 19,792 transients, tracking closely with the SMP selection, while allowing for the diversity of physical core-collapse behaviours—ranging from short-lived stripped-envelope events to extended interacting profiles—for robust, independent classification.

While starting from the complete \texttt{DIFFIMG} baseline preserves sample diversity, it was recognized that targeted selection cuts were necessary to filter out non-SN contaminants and unphysical artifacts from this broad dataset. Further selection cuts were applied using the \texttt{numepoch\_ml} flag, which tracks the number of distinct nights with robust detections. Candidates were required to have detections on more than one night and fewer than 40 detection epochs. Because the Dark Energy Survey observing cadence samples fields roughly every seven days over a five-month season, an upper limit of 40 epochs corresponds to more than a full year of active monitoring. This threshold effectively removes extremely long-lived variables, ongoing active galactic nuclei (AGNs), or repeatedly variable instrumentation artifacts that persist across multiple seasons. Finally, transients lacking redshift information and those fainter than $M = -14$\,mag across all bands were excluded. These ($\sim 25$) faint events are localized at $z < 0.05$, an M-z regime where non-SN stellar outbursts—such as luminous blue variables (LBVs), luminous red novae (LRNs) or intermediate-luminosity red transients (ILRTs; \citealt{Karambelkar2023}),—significantly outnumber true core-collapse events. Following these sequential cuts, a clean sample of 19,512 transients remained for downstream classification.

The redshift information used in this work is derived from three distinct sources: high-precision spectroscopic redshifts ($\text{spec-}z$) primarily sourced from host-galaxies via OzDES \citep{Lidman2020}; deep host-galaxy photometric redshifts ($\text{photo-}z$) derived via template-fitting from broad-band \textit{griz} catalogs \citep{Hartley2022}; and real-time Photometric SuperNova IDentification redshifts ($\text{PSNID-}z$) calculated directly from template-matching light curve properties during the survey \citep{Sako2011, Sako2018}. A rigorous comparative evaluation of these three redshift frameworks, along with the MC error propagation modeling used to handle their respective uncertainties, is described fully in Section~\ref{sec:redshift_uncertainties}.

\subsection{Classification}
\label{sec:classification}

The purpose of the classification process is twofold: (i) to remove AGN contaminants and (ii) to separate SNe~Ia from CC~SNe. While several DES-developed photometric classifiers are optimised for high-purity 
\text{SN~Ia} samples for cosmological analyses \citep[e.g.][]{Moller2020, Qu2021, Vincenzi2021, DES2024}, these approaches were not considered ideally suited to the present study, as they were designed primarily to maximise \text{SN~Ia} purity rather than to isolate a representative CC~SN sample.  Our analysis instead requires a framework 
dedicated to the robust identification and classification of CC~SNe across the full DES redshift range, while maintaining rigorous rejection of non-SN contaminants such as AGNs. 

We therefore evaluated a range of machine-learning techniques, 
including artificial neural networks, gradient boosting, support vector machines, k-nearest neighbours, and random forests. Random forests were adopted owing to their strong performance, robustness to noisy and correlated features, and interpretability for 
heterogeneous datasets \citep{breiman2001, richards2012}. 
Artificial neural networks were found to be computationally 
expensive for iterative feature testing, while other methods 
performed less well than random forests.

The \texttt{RandomForestClassifier} used in this work is 
implemented as part of the \texttt{sklearn.ensemble} module in the widely used Scikit-learn machine learning library for Python 
\citep{pedregosa2011scikit}. This classifier is a meta-estimator designed for classification tasks. The classifier 
functions by fitting a large number of decision tree classifiers on 
various sub-samples of the dataset and using averaging to improve 
predictive accuracy and control over-fitting \citep{breiman2001}.

The algorithm introduces randomness in two key ways to create a 
diverse ensemble of trees, as described in \citet{breiman2001} by:
\begin{enumerate}
    \item Bootstrap sampling: Each tree is trained on a 
    different random sample drawn with replacement (a bootstrap 
    sample) from the original training set.
    \item Random feature selection: When splitting a node 
    during the construction of a tree, the best split is chosen 
    from a random subset of the features, rather than considering 
    all features.
\end{enumerate}

This ensemble approach, which combines the predictions of many 
randomized trees, generally results in a model with lower variance 
and better generalizability compared to a single decision tree 
\citep{hastie2009elements}.

All data loading, tabular parsing, and array processing were 
handled via the \texttt{pandas} and \texttt{numpy} libraries 
\citep{harris2020numpy}, while core algorithms, feature scaling 
strategies, and validation partitions were constructed using the 
Scikit-learn framework \citep{pedregosa2011scikit}.

Initial tests using a single multi-class classifier to distinguish AGNs, SNe~Ia, and CC~SNe were found to perform poorly in separating CC SNe from SNe Ia. This limitation arose largely from biases in the available training data. The spectroscopically confirmed CC~SN sample is volume-limited to $z<0.25$ \citep{Grayling2023}, while the DES SN~Ia sample has undergone additional cosmology-driven quality cuts. In addition, redshift dependent effects such as Malmquist bias, cosmological time dilation, flux dimming, and rest-frame wavelength shifting across the DES filter set introduce systematic differences between low- and high-redshift populations. 

These considerations motivated a two-stage classification process utilizing the redshift sources described in Section~\ref{sec:redshift_uncertainties}. In the first stage, AGN contaminants are removed using a binary SN--AGN classifier trained on real samples, many of which are spectroscopically confirmed. In the second stage, the remaining SN candidates are classified as SNe~Ia or CC~SNe using redshift bin-specific classifiers trained on simulated DES light curves spanning the full survey redshift range.

Classifier performance in all redshift bins is summarised using precision, recall, F1-score, and overall accuracy, derived from the confusion matrix. Precision quantifies the purity of a class (the fraction of predicted objects that are truly members), while recall measures completeness (the fraction of true objects recovered). The F1-score is the harmonic mean of precision and recall, summarising both purity and completeness. Support denotes the number of true objects per class. Macro-averaged metrics treat each class equally, whereas weighted averages account for class imbalance via the class support. For rate determination, adopting a balanced classification threshold is important to minimize systematic bias by equalizing the trade-off between completeness (recall) and purity (precision) \citep{Chen2004, He2009}. 

While a standard Bayesian classifier might default to a threshold 
of \( P > 0.5 \), the optimal threshold for a Random Forest often 
deviates from this value due to the asymmetric features of SN light 
curves and the relative densities of the training classes. To 
address this, we optimized the classifier configuration with an 
exhaustive hyperparameter search via \texttt{GridSearchCV}, 
evaluated using 5-fold stratified cross-validation 
(\texttt{StratifiedKFold}). We implemented an iterative selection 
process for the probability thresholds (\( P_{\mathrm{SN}} \) for 
stage 1 and \( P_{\mathrm{CC\,SN}} \) for stage 2). Specifically, 
we calculated the absolute difference between precision and recall 
across the full range of probability outputs using the automated 
\texttt{precision\_recall\_curve} API, selecting the threshold that 
minimized this difference (precision \(\approx\) recall). The final 
classification was then executed using these optimized boundaries, 
ensuring that the resulting rates are not biased by an 
over-representation of contaminants or an under-recovery of true 
transients.

\begin{figure*}
    \centering
    \includegraphics[width=\textwidth]{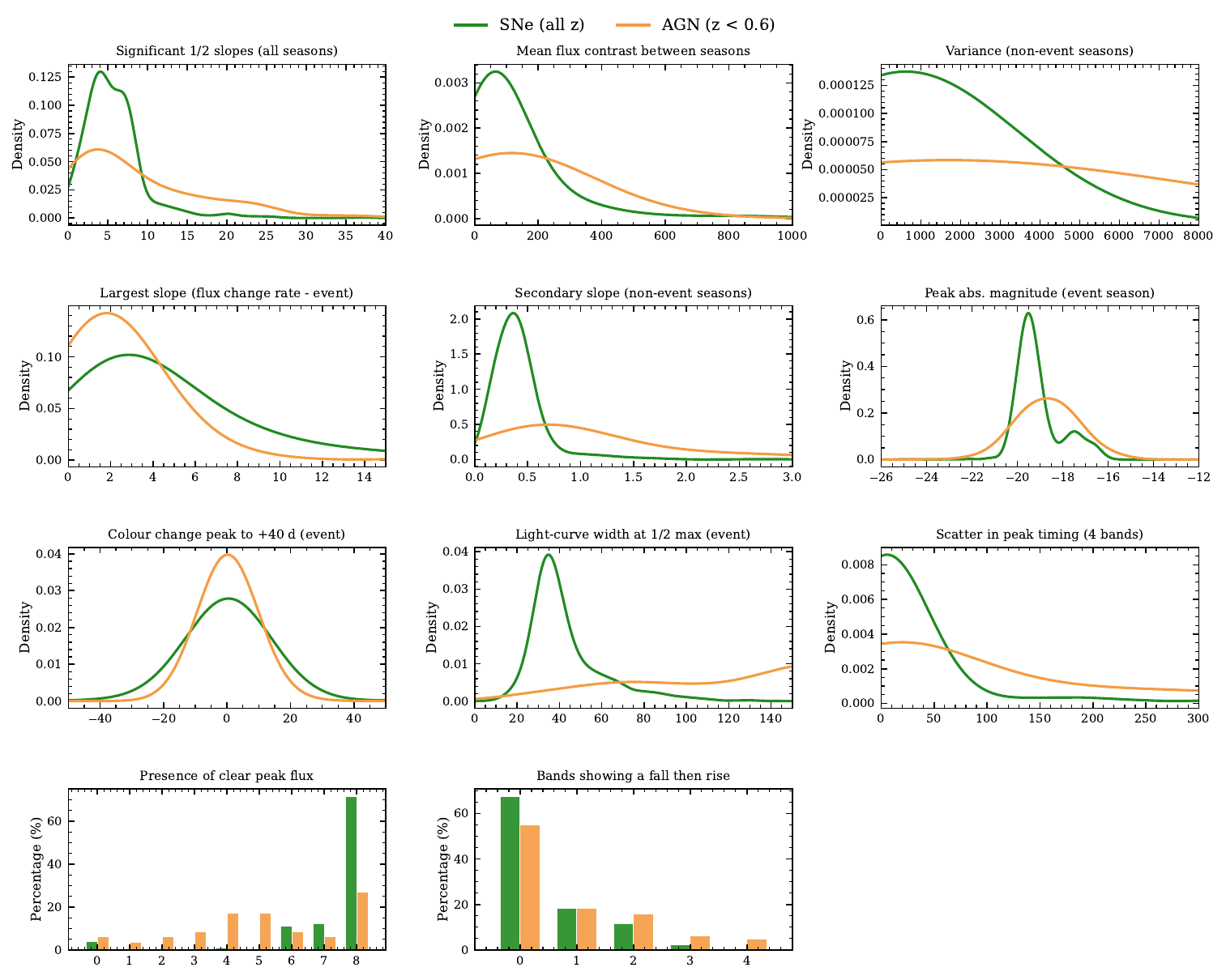}
    \caption{Feature distributions for the SN--AGN classifier in the $z = 0.7$--$0.8$ to $1.0$--$1.1$ transient classification interval. SN distributions are shown in green and AGN distributions in orange.}
    \label{fig:features1}
\end{figure*}

\begin{figure*}
    \centering
    \includegraphics[width=\textwidth]{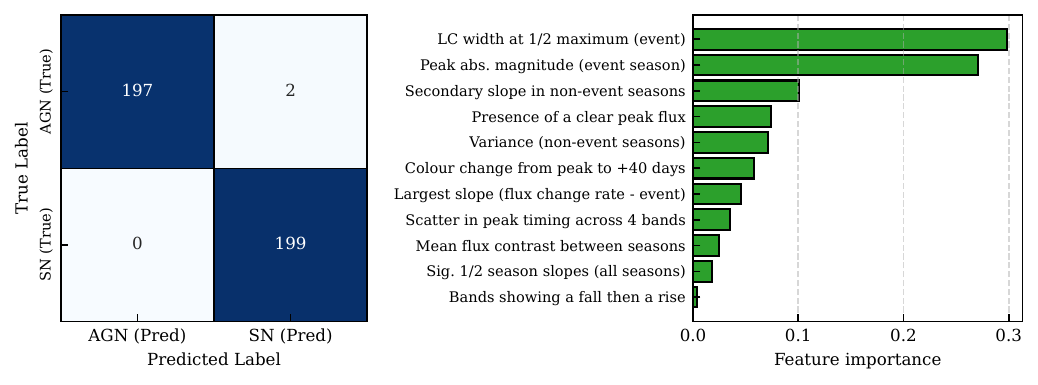}
    \caption{Confusion matrix (left) and Random Forest feature importance rankings (right) for the SN--AGN classifier across the $z = 0.7$--$0.8$ to $1.0$--$1.1$ transient classification interval.}
    \label{fig:cm_fi1}
\end{figure*}

\subsubsection{Stage~1: SN--AGN Classification}

The first stage of the classification pipeline removes AGN contaminants. Existing DES AGN rejection methods based on light-curve properties \citep{Pursiainen2018, Pursiainen2020, Wiseman2020} were found to exclude a non-negligible fraction of confirmed CC~SNe and were therefore not adopted. Instead, we developed a new probabilistic SN--AGN classifier based on a random forest algorithm.

The classifier uses 11 light-curve features designed to capture differences between transient SN behaviour and longer-term AGN variability. these features quantify:

\begin{itemize}
    \item[(i)] light-curve shape, including:
    \vspace{-0.22cm}
    \begin{itemize}
        \item[a)] the width at half peak magnitude,
        \item[b)] the largest slope (flux change rate), and
        \item[c)] whether bands showed a decline then a rise in the light curve;
        \item[d)] the average (of 4 bands) peak absolute magnitude;
    \end{itemize}

\item[(ii)] $g-r$ colour evolution from peak to peak + 40 d;
    \vspace{0.15cm}
    \item[(iii)] season-to-season variability, including:
    \vspace{-0.20cm}
        \begin{itemize}
            \item[a)] non-event season variance,
            \item[b)] flux contrast between seasons,
            \item[c)] number of non-null half slopes,
            \item[d)] steepness of secondary (non-event) slopes, and
            \item[e)] scatter in peak timing; and
        \end{itemize}
    
    \item[(iv)] the presence of distinct peaks or monotonic trends.
    \end{itemize}
collectively assessing whether the observed variability is consistent with a single explosive event or ongoing nuclear activity.

Training was performed using a hybrid sample consisting of a downsized subset of 542 SNe~Ia (drawn from the baseline pool of 1,635 transients; \citealt{DES2024, Vincenzi2024}), 119 CC~SNe \citep{Grayling2023}, and 548 AGNs from OzDES \citep{Lidman2020}. Downsizing the SN~Ia sample was required to achieve a more equitable class balance across the three training classes. While all selected objects possess high-confidence, spectroscopically confirmed host-galaxy redshifts, the transient classifications themselves rely on a mix of spectroscopic and photometric data. Specifically, the downsized SN~Ia sample contains 71 spectroscopically confirmed events and 471 photometric events (30 of which were identified only by Spectral Adaptive Light-Curve Template (SALT) which was run on all candidate SNe Ia). Similarly, the CC~SN sample comprises 42 spectroscopically confirmed and 77 photometrically classified events. All 548 AGN are drawn from the OzDES spectroscopic program. To ensure that this AGN selection does not introduce significant selection bias, its representativeness was verified by demonstrating consistent coverage with the Sloan Digital Sky Survey (SDSS) DR17 AGN luminosity function \citep{Chanchaiworawit2024}. Finally, the remaining class imbalance within the consolidated training matrices were mitigated by applying a combined Synthetic Minority Over-sampling Technique and Edited Nearest Neighbors (\texttt{SMOTE--ENN}) resampling pipeline \citep{Chawla2002, batista2004study} from the \texttt{imbalanced-learn} toolbox \citep{imbalanced-learn}, supplemented with basic class-imbalance bootstrap techniques via the \texttt{resample} utility inside \texttt{scikit-learn}.

\begin{table}
    \centering
    \caption{Redshift partitioning bounds for the AGN training sample and corresponding step-wise transient classification intervals.}
    \label{tab:redshift_bins}
    \small 
    \begin{tabular}{>{\centering\arraybackslash}p{2.8cm} >{\centering\arraybackslash}p{4.2cm}}
    \toprule
    \textbf{AGN Training Sets ($\Delta z = 0.6$)} & \textbf{Transient Classification Intervals ($\Delta z = 0.1$)} \\
    \midrule
    $0.0 \le z \le 0.6$ & 0.0--0.1 to 0.4--0.5 \\
    \addlinespace
    $0.3 \le z \le 0.9$ & 0.5--0.6 to 0.6--0.7 \\
    \addlinespace
    $0.6 \le z \le 1.2$ & 0.7--0.8 to 1.0--1.1 \\
    \addlinespace
    $0.9 \le z \le 1.5$ & 1.1--1.2 to 1.2--1.3 \\
    \bottomrule
    \end{tabular}
\end{table} 

To account for redshift-dependent selection effects and cosmic evolution, a stratified classification approach was implemented. Due to the absence of intermediate and high-redshift CC~SNe, and the sparsity and risk of selection bias for high-redshift SNe~Ia training data, only the AGN sample was partitioned into broad, overlapping redshift bins of width $\Delta z = 0.6$ (Table~\ref{tab:redshift_bins}). This partitioning ensures adequate sample sizes and accurately captures background AGN behavior across the redshift range.

Although adjacent bins share overlapping AGN baselines, classification was executed sequentially in non-overlapping steps of $\Delta z = 0.1$, evaluating the DES transients within each step as a localized group. Because all classification steps within a given transient classification interval (such as the $z = 0.7$--$0.8$ to $1.0$--$1.1$ range mapped in Table~\ref{tab:redshift_bins}) utilize the exact same training set configuration, their underlying feature distributions, feature importances, confusion matrices, and classification performances are identical. Consequently, the $z = 0.7$--$0.8$ to $1.0$--$1.1$ transient classification interval feature distribution is shown in Figure~\ref{fig:features1}, with the corresponding performance metrics detailed in Table~\ref{tab:rf_classification_report} summarizes the performance metrics of the first-stage classifier. The associated confusion matrix and random forest feature importance rankings are provided in Figure~\ref{fig:cm_fi1}. Application of this first-stage classifier reduced the sample to 9,352 likely SNe. 

A detailed breakdown of the Stage~1 classification performance across individual redshift bins is provided in Appendix~\ref{sec:appendix_A}. All four confusion matrices, one for each of the AGN redshift bins, have been included in Figure~\ref{fig:stage1_cms}

\begin{table}
    \centering
    
    \caption{Classification report for the SN--AGN classifier across the $z = 0.7$--$0.8$ to $1.0$--$1.1$ transient classification interval. Metrics shown include precision, recall, $F_1$-score, and support, evaluated at the balanced threshold where precision $\approx$ recall for SNe.}
    \label{tab:rf_classification_report}
    \begin{tabular}{lcccc}
    \hline
     & Precision & Recall & F1-score & Support \\
    \hline
    SN & 0.99 & 0.98 & 0.99 & 199 \\
    AGN & 0.99 & 0.99 & 0.99 & 199 \\
    Macro Avg & 0.99 & 0.99 & 0.99 & 398 \\
    Weighted Avg & 0.99 & 0.99 & 0.99 & 398 \\
    Overall Accuracy & & 0.99 & & 398 \\
    \hline
    \end{tabular}
\end{table}

\subsubsection{Stage~2: CC~SN - SN~Ia Classification}
\label{sssec:cc_sn_ia_classification}

In the second stage, SN candidates were classified as CC~SNe or SNe~Ia using 13 redshift-bin-specific random forest classifiers trained on simulated DES light curves \citep{Vincenzi2019, Vincenzi2021}, which accounted for rest-frame to observer-frame shifts and time dilation. The simulations included a range of SN~Ia (Ia-Normal, Iax, Ia-91bg) and CC~SN (Ib, Ic, Ic-BL, IIb, IIP, IIL, IIn) subtypes. Training data (simulated SNe light curves) from each individual redshift bin were used to train each redshift-specific classifier individually.  

Classification was based on six event-season focused features describing:
\begin{itemize}
    \item[(i)] Light-curve shape, including:
    \vspace{-0.22cm}
    \begin{itemize}
        \item[a)] the largest slope (flux change rate),
        \item[b)] the width at half peak magnitude, and
        \item[c)] whether bands showed a fall then a rise;
    \end{itemize}
    \vspace{-0.1in}
    \item[(ii)] The average (of 4 bands) peak absolute magnitude;
    \item[(iii)] The $g-r$ colour evolution; and
    \item[(iv)] The presence of distinct peaks or monotonic trends.
\end{itemize}

An increase in the average peak absolute magnitude of recovered CC~SNe is observed at higher redshifts. This trend is driven by three independent selection and measurement effects: (i) classic Malmquist bias, where the survey becomes increasingly incomplete to low-luminosity CC~SNe at $z > 0.1$; (ii) at low redshift in particular ($z < 0.1$), the DES pipeline is unsuited for capturing the brightest events due to CCD saturation and subsequent peak clipping; and (iii) at higher redshifts, the intrinsic rest-frame UV deficit—caused by metal line blanketing in the ejecta—shifts into the observer $g$ and $r$ bands. This drastically suppresses the observable peak flux, and higher random noise fluctuations are misinterpreted by the classifier as the true peak flux.  This introduces an  upward bias in the average observed peak brightness, which increases with redshift up to $\sim 1.5\,\mathrm{mag}$ at high $z$.

Despite these effects, the random forest classifier remains robust. As the peak average absolute magnitude (4-band average) consistently ranks as the most important feature across most redshift bins, Anderson-Darling tests were performed to evaluate how reliably this feature differentiates simulated CC~SNe from SNe~Ia against our observed CC~SN sample (see Appendix~\ref{sec:appendix_ad_tests} for full details and figures). Furthermore, peak average absolute magnitude is only one of six features evaluated in Stage~2 (alongside eleven in Stage~1); while other shape, colour, and trend metrics are also subject to redshift-dependent effects, they are affected to a lesser degree. While redshift-dependent effects like cosmological time dilation and spectral energy shifting impact these features, they affect both SNe~Ia and CC~SNe symmetrically, preserving the relative discrimination between the two populations.

Potential contamination from superluminous SNe (SLSNe) is also minimal. Because SLSNe represent an extreme subclass of core-collapse events, their inclusion does not fundamentally corrupt the CC~SN selection. Moreover, their quantitative impact is negligible: very few spectroscopically confirmed SLSNe are classified as CC~SNe by our pipeline (see below), and their intrinsic volumetric rate is roughly three orders of magnitude lower than the global CC~SN rate \citep[e.g.,][]{Quimby2013, Frohmaier2020}.

We implemented the Stage 2 random forest classifiers using the \texttt{scikit-learn} package, applying balanced class weights and our optimized probability thresholds to classify the transient candidates in each redshift bin.

\begin{figure*}
    \centering
    \includegraphics[width=\textwidth]{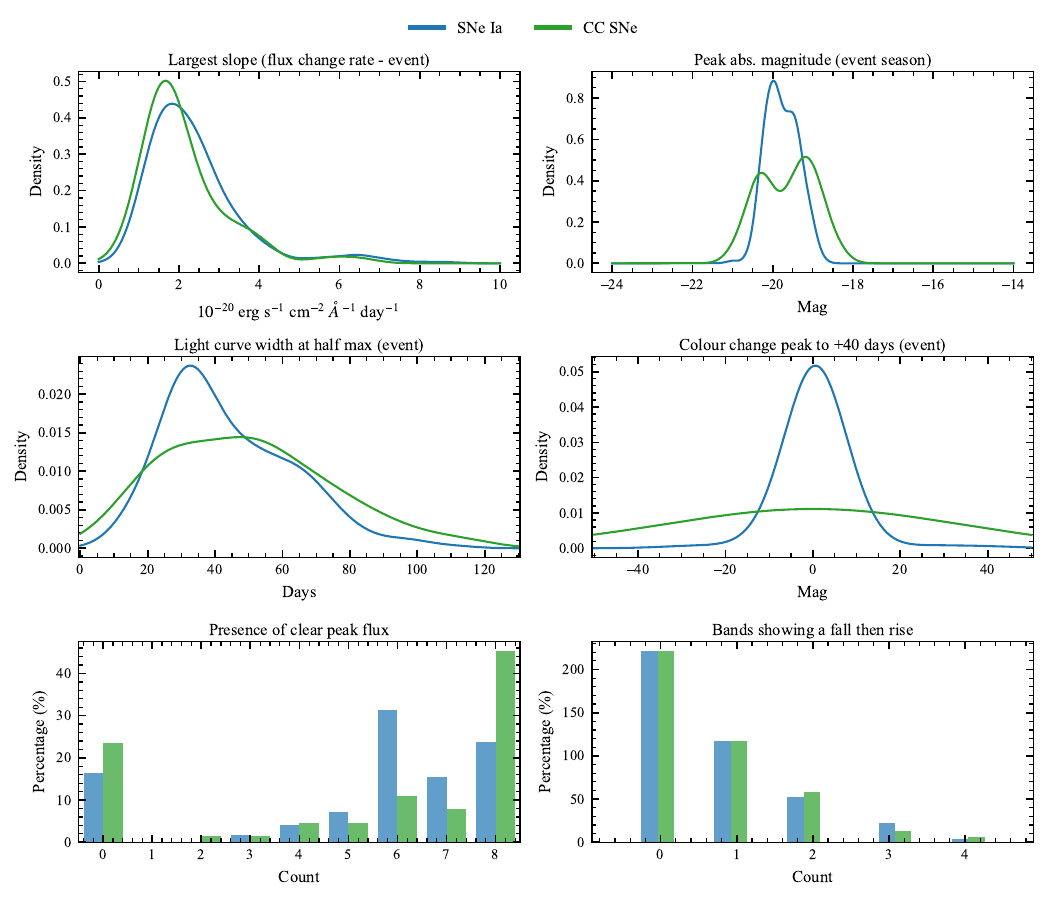}
    \caption{Second stage random forest classifier: feature distributions for SNe classified as SNe~Ia and CC~SNe for the $z = 0.7$--$0.8$ to $1.0$--$1.1$ transient classification interval. SNe Ia distributions are shown in blue, CC~SNe distributions in green.}
    \label{fig:features}
\end{figure*}

\begin{figure*}
    \centering
    \includegraphics[width=\textwidth]{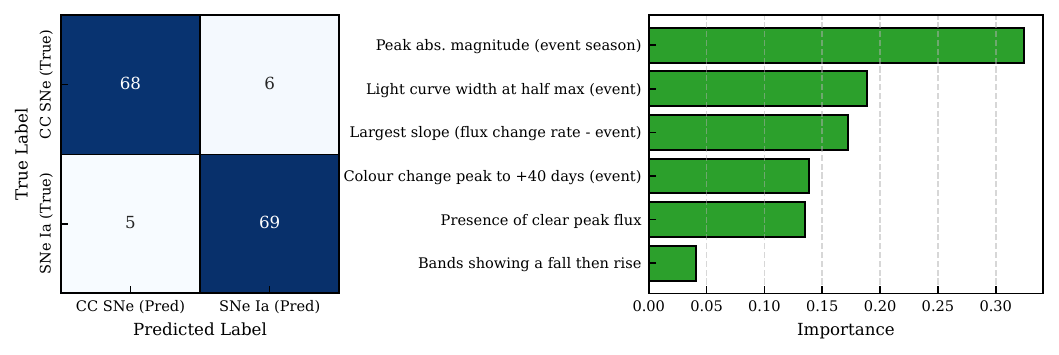}
    \caption{Confusion matrix (left) and feature importance (right) for the CC~SN -- SN~Ia random forest classifier in the $z=0.8$--$0.9$ bin.}
    \label{fig:rf_performance}
\end{figure*}

As an example, Figure~\ref{fig:features} illustrates the feature distributions for the $z=0.8$--$0.9$ bin, with classifier performance summarised in Table~\ref{tab:cc_classification_report}. The corresponding confusion matrix and feature importance rankings are shown in Figure~\ref{fig:rf_performance}. Across all redshift bins, the mean weighted accuracy is $87.5$ per~cent, with no significant redshift dependence observed across the primary sample distribution. This stage yielded 3,169 transients classified as likely CC~SNe.

\begin{table}
\centering
\caption{Classification report for the CC~SN--SN Ia classifier in the $z=0.8$--$0.9$ bin. Metrics shown include precision, recall, F1-score, and support, evaluated at the balanced threshold where precision $\approx$ recall for CC~SNe.}
\label{tab:cc_classification_report}
\begin{tabular}{lcccc}
\hline
{} & Precision & Recall & F1-score & Support \\
\hline
CC~SN  & 0.93 & 0.92 & 0.93 & 74 \\
SN~Ia  & 0.92 & 0.93 & 0.93 & 74 \\
Macro Avg & 0.93 & 0.93 & 0.93 & 148 \\
Weighted Avg & 0.93 & 0.93 & 0.93 & 148 \\
Overall Accuracy & & 0.93 & & 148 \\
\hline
\end{tabular}
\end{table}

\begin{figure*}
    \centering
    
    \begin{subfigure}[b]{1.0\linewidth}
        \centering
        \includegraphics[width=\textwidth, trim=0in 0in 0in 0.1in, clip]{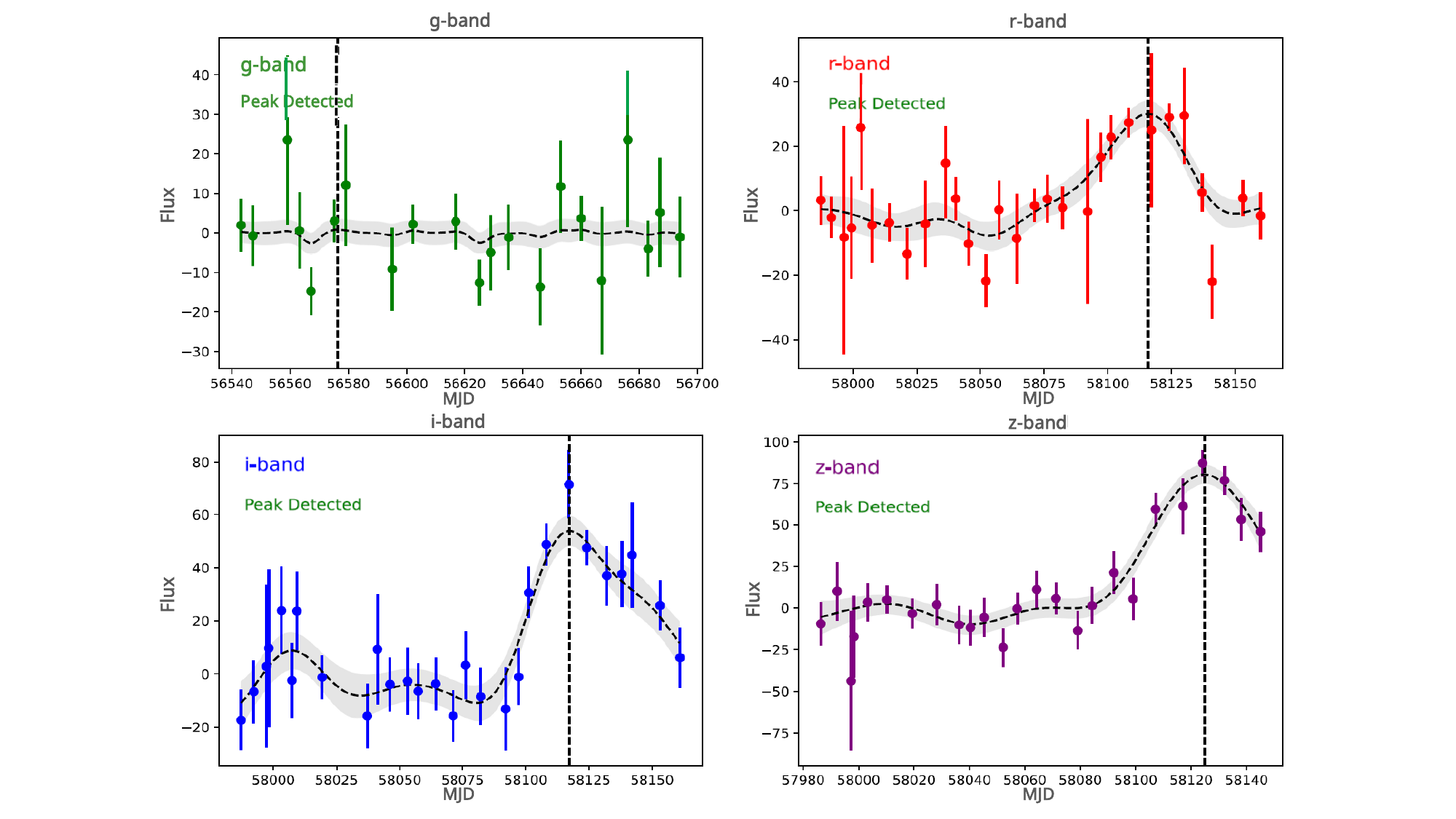}
        
        \vspace{-0.0cm} 
        \caption{Light curve of transient SNID 1935109 classified as an H-rich CC SN (z = 1.179).}
        \label{fig:lc_hrich}
    \end{subfigure}
    
    \vspace{0.5cm} 
    
    \begin{subfigure}[b]{1.0\linewidth}
        \centering
        \includegraphics[width=\textwidth, trim=0in 0in 0in 0.1in, clip]{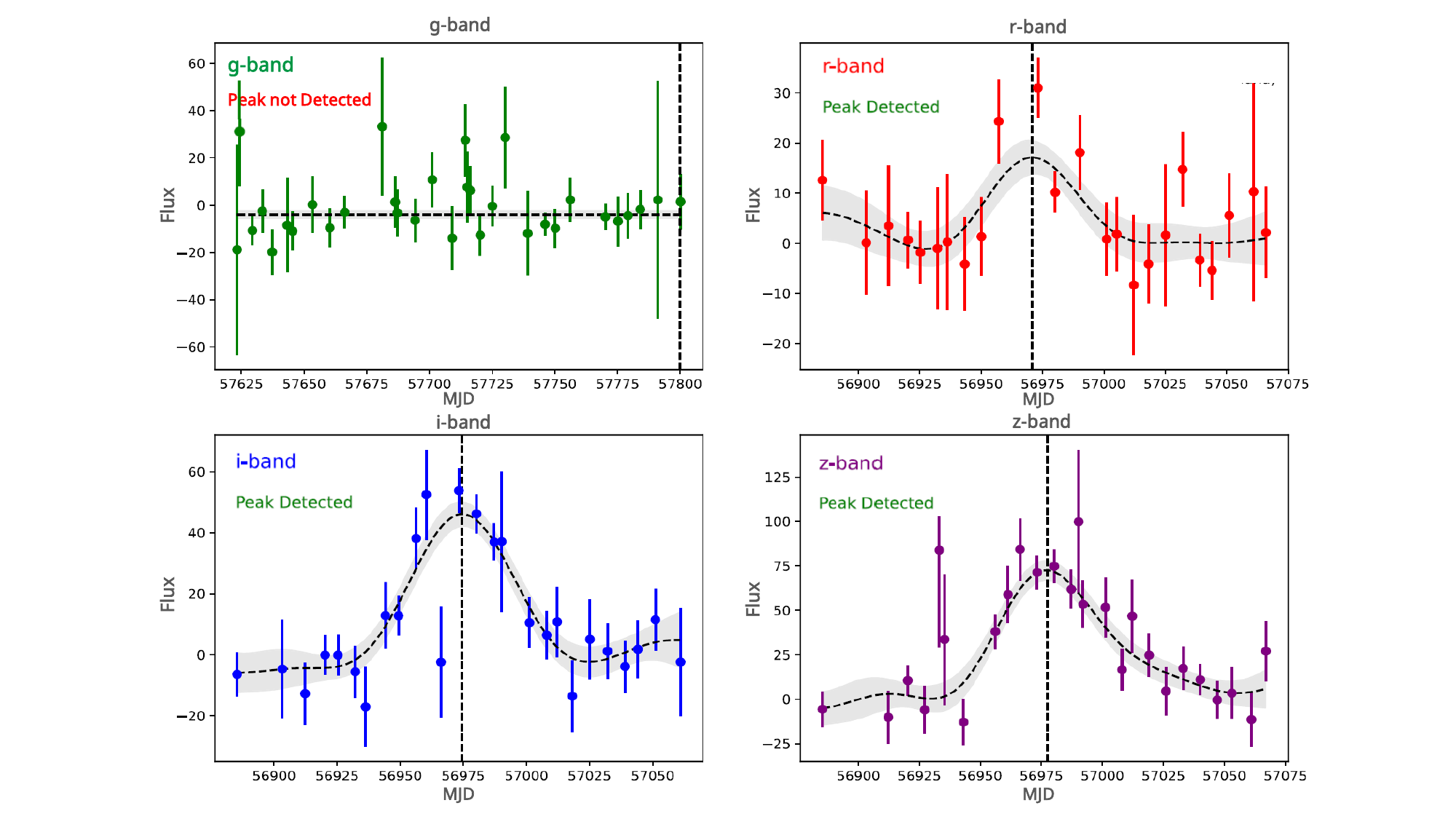}
        
        \vspace{0.0cm} 
        \caption{Light curve of transient SNID 1309198 classified as an SE SN (z = 1.272).}
        \label{fig:lc_sesne}
    \end{subfigure}

    \vspace{0.3cm} 
    
\caption{Multi-band ($g, r, i, z$) \texttt{DIFFIMG} light curves for a representative transient classified as an H-rich CC SN (a) and an SE SN (b). Data points are shown in wavelength-space flux density units ($10^{-20}\text{ erg cm}^{-2}\text{ s}^{-1}\text{\AA}^{-1}$), scaled directly from the raw pipeline frequency flux measurements ($1\,\mu\text{Jy} \approx 3.39 \times 10^{-20}\text{ erg cm}^{-2}\text{ s}^{-1}\text{\AA}^{-1}$ at the DECam pivot wavelength). Observed data points are overlaid with automated GP regression models (dashed black curves, shaded $1\sigma$ envelopes) and filter peak epochs (vertical dashed lines).}
    \label{fig:example_lcs}
\end{figure*}

A detailed breakdown of the Stage~2 classification performance across individual redshift bins is provided in Appendix~\ref{sec:appendix_A}. All twelve confusion matrices, one for each of the $\Delta z = 0.1$ redshift bins from $z = 0 - 1.3$ ($z = 1.1 - 1.3$ is combined), have been included in Figure~\ref{fig:figA5}.

To assess potential SN~Ia contamination within our automated CC~SN sample, we compare the performance of our classifier against the machine-learning classification probabilities from the DES 5-Yr (DES-SN5YR) data release \citep{Sanchez2024_DES5yr}. The \texttt{DES\_classification.csv} file within this release provides SN~Ia probabilities for all DES transients with an SMP light curve, utilizing both the \texttt{SuperNNova} (\textsc{SNN}) recurrent neural network classifier \citep{Moller2020} and the \textsc{SCONE} convolutional neural network framework \citep{Qu2021}. Given the overall precision and recall statistics of our classifier, a baseline contamination level of $5\text{--}15$ per~cent is anticipated. Cross-matching our sample with this catalog, we find that roughly $10$ per~cent of our selected CC~SNe exhibit an SN~Ia probability of $P_{\text{Ia}} > 0.50$, fully in line with our expected classification performance indicators.

The DES-SN5YR data release also contains files detailing spectroscopically confirmed transients, providing an additional independent check on the classifier performance and contamination rates. Within this subset, only 12 out of 412 spectroscopically confirmed SNe~Ia were classified as CC~SNe by our pipeline. Similarly, only 3 out of 15 spectroscopically confirmed SLSNe were classified as CC~SNe. Conversely, because the DES spectroscopic program was optimized to target SN~Ia-like candidates, the spectroscopically confirmed CC~SN sample is naturally unrepresentative of the broader population, consisting primarily of atypical, SN~Ia-like events. Our pipeline nevertheless successfully identified 56 out of 82 of these highly challenging objects as CC~SNe.

The impact of contamination on our volumetric rate estimations has been mitigated by employing a balanced threshold classification technique designed to maintain $\text{recall} \approx \text{precision}$ during our initial selection phase, and accounting for any residual contamination effects with our post-classification correction factor. The 3,169 photometrically classified CC~SNe were then checked against the official survey footprint, removing 145 candidates located outside the strict $22.912~\mathrm{deg}^2$ area \citep{Lasker2020} (which excludes dead CCDs and field borders) to leave a final sample of 3,024 CC~SNe.These comparisons and mitigating measures give confidence that our final CC~SN efficiencies and rates are well protected against contamination from both SNe~Ia and SLSNe.

These remaining CC~SN candidates were visually inspected to ensure consistency with expected CC~SN light-curve behaviour. Visual inspection focused on identifying coherent peaks or monotonic trends within a single observing season, consistency across bands, and the absence of the long-term variability characteristics of AGNs.

The visual inspection of these candidates rejected 579 objects, primarily comprising low-luminosity AGNs missed by the Stage~1 classifier, detector artifacts, or unusual non-SN transients. Potential rare non-SN transient contaminants included narrow-line AGNs (the primary contaminant, at a few per cent), luminous blue variables (LBVs), luminous red novae (LRNs), intermediate-luminosity red transients (ILRTs), tidal disruption events (TDEs), and SLSNe. The number of these rare contaminants remains tightly constrained by their low volumetric rates and distinct photometric signatures. For transients with $M < -14$, LRNs, ILRTs, and LBVs occur at rates $\lesssim 10^{-5}\,\mathrm{yr^{-1}\,Mpc^{-3}}$, while TDEs and SLSNe are rarer still ($\sim 10^{-7}$--$10^{-6}\,\mathrm{yr^{-1}\,Mpc^{-3}}$), implying overall contamination fractions $\lesssim 1$ per~cent (e.g., \citealt{Humphreys1994, Smartt2009, Angus2019_DES, Weis2020, Karambelkar2023, Callow2024, Callow2025, Clark2026}).

Following both survey footprint and visual inspection cuts, a final baseline sample of 2,445 CC~SNe was retained for the rate analysis. Representative examples of multi-band \texttt{DIFFIMG} light curves classified as CC~SNe, illustrating an H-rich CC~SN stacked directly above a stripped-envelope (SE) CC~SN—are presented with their automated Gaussian Process (GP) peak-flux fits in Figure~\ref{fig:example_lcs}. At these higher redshifts, g-band detections are usually absent and r-band fluxes are suppressed because cosmological redshift shifts the rest-frame far-UV—where flux is intrinsically suppressed by heavy iron-group line blanketing—into bluer optical bands, driving the signal-to-noise (S/N) ratio below the detection threshold. (as illustrated in Figure~\ref{fig:figC4_schematic}).

\subsection{CC~SN Subtypes}
\label{sec:subtypes}

The final CC SN sample was subdivided into SE SNe and H-rich SNe using a binary classifier trained on the simulated light curves of \citet{Vincenzi2019, Vincenzi2021}. The same six features and number of classification intervals (\(13 \times \Delta z = 0.1\)) as used in the second-stage binary classifier described in 
Section \ref{sssec:cc_sn_ia_classification} were employed. The subtype classification was 
carried out using a \texttt{RandomForestClassifier}, implemented via the Scikit-learn library \citep{pedregosa2011scikit}, and cross-validated under a \texttt{StratifiedKFold} architecture.

Despite various attempts to improve the classifier in this more 
difficult classification environment — including boundary-engineering 
techniques such as \texttt{SMOTE} \citep{Chawla2002} and 
\texttt{SMOTE--ENN} \citep{Batista2004} implemented via the \texttt{imbalanced-learn} library \citep{lemaitre2017} to resolve overlapping feature spaces, and exhaustive grid searches using 
\texttt{GridSearchCV} to optimize the ensemble's structural 
hyperparameters — these exploratory optimization variations yielded 
no statistically significant performance changes for the final 
subtype validation boundaries. Consequently, comparison of the sub-classification with the low-redshift ($z < 0.25$) DES CC~SN reference sample from \citet{Grayling2023} revealed a systematic bias. \citet{Grayling2023} selected a volume-limited sample of 119 CC SNe discovered by DES at $z<0.25$ and confirmed either spectroscopically (42 SNe) or photometrically (77 SNe), with quality checks applied in both cases. This reference sample constitutes $\sim 20$ per cent of the total population of $z<0.25$ CC SNe classified in this work. Given this substantial overlap, a comparison between our sample and that from \citet{Grayling2023} provides a statistically meaningful assessment of our classifier's performance. We find that our classification procedure underestimates the number of SE SNe out of the total sample of CC SNe at $z<0.25$ by $\sim 25$ per cent.

To correct this bias, we apply an SE-specific calibration. The calibrated SE counts ($N_{\text{SE, cal}}$) are obtained by multiplying the uncalibrated counts by an under-classification correction factor of $1/0.75 \approx 1.33$. To preserve the total number of observed CC~SNe per redshift bin ($N_{\text{CC}}$), the H-rich counts are defined as the residual:
\begin{equation}
    N_{\text{H-rich, cal}} = N_{\text{CC}} - N_{\text{SE, cal}}.
\end{equation}

For the higher-redshift sample ($z > 0.25$), where no comparable reference sample is available, we adopt the same constant SE calibration factor of 1.33. Although this extrapolation is not directly constrained by high-$z$ data, redshift-dependent biases are expected to affect H-rich--SE classification less strongly than CC~SN--SN~Ia classification. The former involves subclasses within the same core-collapse population, with broadly similar luminosities and photometric properties, and thus likely comparable selection effects. In contrast, CC~SN--SN~Ia classification is more susceptible to redshift-dependent bias due to the substantial luminosity difference between those populations. Ignoring this systematic bias—measured directly from a 20~per~cent empirical census of the survey's low-redshift population—would result in a known underestimation of the SE~SN rate. Applying this constant correction across the full redshift range is therefore preferable to leaving the counts uncorrected, anchoring our results to the observed performance of the classifier on real data.

\begin{table}
\centering
\caption{Data processing steps and transient classification for the DES SN sample.}
\small
\setlength{\tabcolsep}{4pt}
\begin{tabular}{llcc}
\hline
Step & Description & Number & Comments \\
\hline
1 & DIFFIMG file list / metadata & 31\,636 & DIFFIMG files \\
2 & $\texttt{transient\_status}$ & 19\,792 & \citet{Smith2020} \\
3 & $\texttt{numepochs\_ml}<40$ & 19\,623 & $<40$ epochs \\
4 & $M<-14$ with redshift & 19\,512 & Remove faint / no-$z$ \\
\hline
\textbf{Prelim.} & AGN & {10\,160} & SN--AGN classifier \\
 & SNe Ia & 6\,183 & Ia/Iax/91bg \\
 & CC~SNe & 3\,169 & Binary CC classifier \\
\hline
\textbf{Final} & CC~SNe & \textbf{2\,445} & Visual inspection/area \\
\cline{1-4}
\textbf{Subtypes} & H-rich~SNe & 1\,686 & After calibration with \\
 & SE~SNe     & 759    & \citet{Grayling2023}\\
\hline
\end{tabular}
\label{tab:classification_steps}
\end{table}

A recap of the data processing steps, selection, and transient classification for the DES SN sample is shown in Table~\ref{tab:classification_steps}. The number of transients remaining after each step is shown, with brief comments on the basis for each step.

\section{Redshifts \& Extinction Corrections}
\label{sec:zandext}

In this section, we correct the observed number of CC SNe in each redshift bin for two key effects: uncertainties in the assigned redshifts (Section~\ref{sec:redshift_uncertainties}) and incompleteness caused by high dust extinction, mainly in highly star-forming host galaxies (Section~\ref{sec:extinction}).

\subsection{Redshift sources and uncertainties}
\label{sec:redshift_uncertainties}

As introduced during the initial sample selection (Section~\ref{sec:selection}), redshift estimates for the DES transients were obtained from three primary sources:

\begin{enumerate}
    \item Spectroscopic redshifts (spec-z) were derived either from SN spectra or from narrow emission or absorption lines in host-galaxy spectra. These were mainly obtained by OzDES as previously described in Section~\ref{sec:observations}. 

    \item Photometric redshifts (photo-$z$) were primarily obtained from the deep host-galaxy catalogues of \citet{Hartley2022}. These estimates utilize deep \textit{griz} photometry supplemented with available $u$-band and near-infrared ($JHK_s$) photometry, with redshifts derived via template fitting using the \textsc{eazy} code \citep{Brammer2008}. Several studies have validated the performance of host-based photometric redshifts for DES SN samples \citep[e.g.][]{Brout2019, Wiseman2021}.

    \item PSNID redshifts (PSNID-z) were estimated as part of the real-time preliminary classification process carried out during the observational phase, using the PSNID code \citep{Sako2011, Sako2018}. PSNID compares observed light curves to simulated SN templates, treating redshift as a free parameter and marginalizing over a grid of redshift values when computing classification probabilities. These measurements were performed using one of three redshift prior modes: a flat prior (\texttt{ZPRIOR=0}), a spectroscopic redshift prior (\texttt{ZPRIOR=1}), or a photometric host-galaxy redshift prior (\texttt{ZPRIOR=2}).  CC~SNe that rely on PSNID-$z$ as their redshift source necessarily adopt the mean of their \texttt{ZPRIOR=0} measurements as no other priors are available. To prevent external priors from biasing the PSNID-$z$ uncertainty evaluation, the spectroscopic sample used must be explicitly filtered to retain strictly unconstrained \texttt{ZPRIOR=0} measurements (excluding spec-$z$ and photo-$z$ priors). Because our primary redshift hierarchy ($\text{spec-}z > \text{photo-}z > \text{PSNID-}z$) automatically prioritizes higher-quality constraints whenever present, this explicit filtering is only required for the uncertainty evaluation subset.
\end{enumerate}

\begin{figure}
    \centering
    \includegraphics[width=\linewidth]{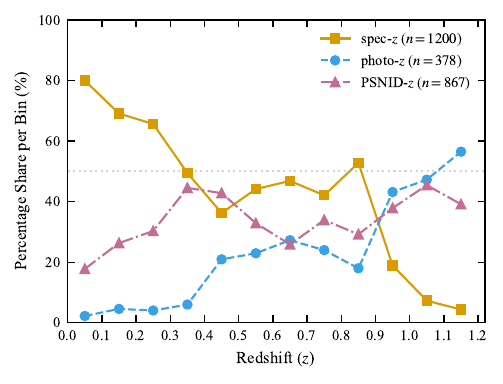}
    \caption{The fractional breakdown of the CC~SN sample as a function of redshift ($z$), divided into bins of $\Delta z = 0.1$. The lines track the percentage share of spectroscopic redshifts (spec-$z$; orange squares), host-galaxy photometric redshifts (photo-$z$; blue circles), and template-marginalized PSNID redshifts (PSNID-$z$; purple triangles). Total sample sizes ($n$) for each subset across the entire redshift range are listed in the legend.}
    \label{fig:redshift_source_share}
\end{figure}

The relative contribution of each chosen redshift source evolves significantly as a function of redshift, as illustrated in Fig.~\ref{fig:redshift_source_share}. When multiple estimates coexist for a single transient, we adopt a strict priority hierarchy: spec-$z$ is retained as the primary source, followed by photo-$z$ as secondary, and PSNID-$z$ as a tertiary fallback. 

Across the entire redshift range, the fallback PSNID-$z$ contribution remains quite stable, hovering between 30 and 50~per~cent without a strong monotonic trend. In contrast, the other two redshift sources show strong evolutionary trends. At low redshifts ($z \lesssim 0.3$), the sample is heavily dominated ($\sim 70$~per~cent) by high-confidence spec-$z$ measurements, reflecting the high completeness of targeted spectroscopic host campaigns for nearby, bright transients. 

In the intermediate regime ($0.4 \lesssim z \lesssim 0.8$), all three methods contribute significantly: spec-$z$ remains the leading contributor at $35$--$55$~per~cent, PSNID-$z$ provides a steady secondary share of $25$--$45$~per~cent, and host-galaxy photo-$z$ accounts for $20$--$30$~per~cent of the sample. At higher redshifts ($z \gtrsim 0.9$), the spec-$z$ fraction drops rapidly from $\sim 55$~per~cent at $z \approx 0.85$ to under $20$~per~cent at $z = 0.95$ and less than $10$~per~cent beyond $z > 1.0$, driven by the magnitude limits of spectroscopic target selection. In this deep regime ($z > 0.9$), photometric redshifts increase sharply to become the dominant source, accounting for nearly $60$~per~cent of the utilized sample data in the highest redshift bin ($z \approx 1.15$). This shift occurs because deep host-galaxy photometric catalogues successfully provide robust template fits at these depths, which, by virtue of our selection hierarchy, take precedence over tertiary PSNID-$z$ estimates (which supply most of the remaining $\sim 40$~per~cent share at $z > 1.0$).

However, because photo-$z$ becomes the dominant source at $z > 0.9$, our ability to directly validate its performance is limited. The number of overlapping data points with both spec-$z$ and photo-$z$ measurements becomes small at $z > 1.0$. While we find no evidence of a systematic shift where these sub-samples do overlap, there is inherently less empirical data available to anchor the validation at these high redshifts. We mitigate this limitation by propagating significantly larger, conservative uncertainties for these high-redshift bins (minimum error floors). Although unidentified systematic effects may still persist due to the sparse validation data, we consider that our comprehensive uncertainty framework and subsequent MC simulations account for these issues as robustly as possible given the observational constraints of the data set.

\begin{figure}
    \centering
    \includegraphics[width=\linewidth]{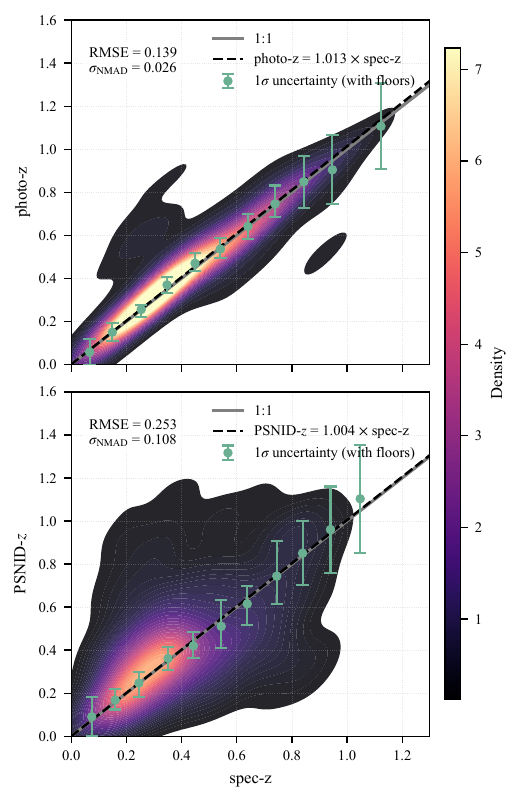}
    \caption{Contour density maps comparing spectroscopic redshifts ($\mathrm{spec}\text{-}z$) with alternative redshift estimates for CC~SNe. The upper panel shows $\mathrm{photo}\text{-}z$ versus $\mathrm{spec}\text{-}z$, while the lower panel shows $\mathrm{PSNID}\text{-}z$ versus $\mathrm{spec}\text{-}z$. Contours represent the relative density of events, and green markers indicate the mean and $1\sigma$ uncertainty range within each redshift bin. Minimum error floors have been applied to the lowest ($z \le 0.1$) and highest ($z \ge 0.8$) bins to mitigate the impact of sparse sampling in the paired $\mathrm{spec}\text{-}z$ reference sample, preventing unphysically narrow or non-monotonic uncertainty bounds during bin-level error estimation (cf. Fig.~\ref{fig:figC5} for the unfloored implementation).}
    \label{fig:contour final}
\end{figure}

For the purposes of this analysis, spec-$z$ measurements are assumed to have negligible uncertainty. For CC~SNe with multiple redshift estimates from different sources, we quantify the performance of photo-$z$ and PSNID-$z$ relative to spec-$z$ using contour density maps constructed in thirteen redshift bins spanning $0 \le z \le 1.3$. These maps illustrate the scatter and asymmetry of the redshift residuals and are shown in Fig.~\ref{fig:contour final}.

For CC~SNe with both spectroscopic and alternative redshift estimates, photo-$z$ shows a strong correlation with spec-$z$: 337 out of 461 events (73.1~per~cent) lie within $\Delta z < 0.05$, and 407 out of 461 (88.3~per~cent) within $\Delta z < 0.1$. This corresponds to a Root Mean Square Error (RMSE, representing the standard deviation of the residuals) of $0.139$ and a Normalised Median Absolute Deviation ($\sigma_{\rm NMAD}$) of $0.026$. The latter is defined as $1.48 \times \text{median}(|\Delta z| / (1+z_{\rm spec}))$, and is a metric specifically chosen for its resilience to catastrophic outliers \citep[e.g.][]{Ilbert2006, Hildebrandt2010}.

PSNID-$z$ exhibits a broader core scatter across the unselected sample, with 337/1199 events (31.2 per~cent) within $\Delta z < 0.05$ and 613/1199 (51.1 per~cent) within $\Delta z < 0.1$, yielding an RMSE of $0.254$ and $\sigma_{\mathrm{NMAD}} = 0.108$. This increased scatter reflects lower light-curve S/N, limited phase coverage, and intrinsic template degeneracies. Nevertheless, the overall linear scaling remains excellent at $\text{PSNID-}z = 1.004 \times z_{\mathrm{spec}}$, confirming that the redshift recovery is virtually unbiased across the sample, and well-suited for rate modeling when forward-propagated via our MC framework (we explicitly confirm the reliability and unbiased nature of these PSNID-$z$ estimates in Appendix~\ref{app:PSNID_independence}).

To propagate redshift uncertainties into the CC~SN rate calculation, we perform an MC simulation. Each transient is assigned a redshift $z$ and asymmetric uncertainties $\sigma_z^-$ and $\sigma_z^+$ based on its redshift source (spec-$z$, photo-$z$, or PSNID-$z$), informed by the empirical distributions shown in Fig.~\ref{fig:contour final}.

For transients with non-zero uncertainty (photo-$z$ or PSNID-$z$), redshift realizations are drawn from a skew-normal distribution:
\begin{equation}
z^{(i)} \sim \mathrm{SkewNormal}(\xi = z, \omega = \bar{\sigma}_z, \alpha),
\end{equation}
where $\bar{\sigma}_z = (\sigma_z^- + \sigma_z^+)/2$ is the mean uncertainty and the skewness parameter is defined as $\alpha = (\sigma_z^+ - \sigma_z^-)/\bar{\sigma}_z$. Spectroscopic redshifts are treated as fixed values with no uncertainties.

Due to the relatively small number of CC~SNe having both spec-$z$ and either photo-$z$ or PSNID-$z$ measurements in the lowest ($z < 0.1$) and highest ($z > 0.8$) redshift bins, sample variance makes raw error estimates vulnerable to underestimation. To prevent underrepresenting the uncertainty in these sparse regimes---and to ensure realistic bin migration without artificial clustering at the survey boundaries during MC resampling---we apply empirical error floors to reflect broader scatter, template degeneracies, and survey completeness limits \citep{Sako2018}. Redshift MC analyses without error floors yield broadly consistent overall results. However, there is a significant transfer of CC~SNe from the central bins to the lowest and highest redshift bins, driven by unfloored scatter and edge-truncation effects in the boundary regimes during resampling. The unfloored results are presented in Appendix~\ref{app:no_ef_comparison}.

We generate 10,000 realizations of the full redshift sample. For each realization, events are binned in redshift to produce a distribution of counts. The final bin statistics are given by
\begin{equation}
N_{\mathrm{MC}} (z) = \frac{1}{N_{\mathrm{sim}}} \sum_{i=1}^{N_{\mathrm{sim}}} N_z^{(i)}, \quad
\sigma_{N_{\mathrm{MC}}}^{-} = P_{50} - P_{16}, \quad
\sigma_{N_{\mathrm{MC}}}^{+} = P_{84} - P_{50},
\end{equation}
where $N_z^{(i)}$ is the number of events in the $i\text{-th}$ realization, and $P_{16}$, $P_{50}$, and $P_{84}$ denote the 16th, 50th, and 84th percentiles of the generated distributions. These $N_{\mathrm{MC}}$ values are presented in Table~\ref{tab:CC_SN_rates_redshift}.

\subsection{High-extinction CC SNe}
\label{sec:extinction}

Observed CC~SN rates are known to underestimate the true event rate due to dust extinction  in highly star-forming galaxies (e.g., \citealt{Mannucci2007MNRAS377, Horiuchi2011, Botticella2012AA543A132, Mattila2012, Xiao2015}). This effect is particularly severe in luminous and ultraluminous infrared galaxies (LIRGs and ULIRGs), where intense star formation is accompanied by large dust columns.

Near-infrared surveys and radiative transfer modelling indicate that extinction in these environments can reach $A_V \sim 10$--$50$~mag, rendering most CC~SNe invisible to optical surveys. \citet{Mattila2012} estimated that 19~per~cent of CC~SNe are missed in normal star-forming galaxies, rising to 83~per~cent in starburst LIRGs/ULIRGs. Recent work by \citet{Mantynen2025} finds a total undetectable fraction in starburst galaxies of $89.7^{+2.6}_{-4.4}$~per~cent, consistent with these results.

In normal star-forming galaxies, the missing fraction is driven primarily by viewing geometry, as in highly inclined systems, the line of sight intersects a much larger dust column. MC modelling shows that most normal galaxy extinctions occur in edge-on or near edge-on galaxies, with mean values reaching $A_B \simeq 19.8$~mag \citep{Hatano1998} if fully edge-on.

The DES CC~SN simulations use the V19 templates \citep{Vincenzi2019, Vincenzi2021}, which are based on the luminosity function of \citet{Li2011a} and do not include host-galaxy extinction corrections. As a result, the simulated detection efficiency reflects the observed DES sample. Highly extinguished CC~SNe are therefore accounted for separately using the missing fractions reported by \citet{Mattila2012}.

Milky Way extinction is minimal in the DES-SN fields, with median $E(B-V)_{\mathrm{MW}} < 0.02$ \citep{Vincenzi2024}, corresponding to an average extinction of $\sim0.04$~mag across the DES filters. The resulting additional incompleteness is small.

For the $i$th redshift bin, the extinction-corrected number of CC~SNe is given by
\begin{equation}
N_{\mathrm{CC}}(z_i) =
\frac{N_{\mathrm{MC}}(z_i)}
{1 - \mathrm{MF}_{\mathrm{HG}}(z_i) - \mathrm{MF}_{\mathrm{MW}}(z_i)},
\end{equation}
where $N_{\mathrm{MC}}(z_i)$ is the MC--corrected number of CC~SNe, while $\mathrm{MF}_{\mathrm{HG}}(z_i)$ and $\mathrm{MF}_{\mathrm{MW}}(z_i)$ represent the missing fractions due to host-galaxy and Milky Way extinction, respectively. The values for $\mathrm{MF}_{\mathrm{HG}}(z_i)$ are adopted from \citet[Table~10]{Mattila2012}. The final extinction-corrected CC~SN counts, $N_{\mathrm{CC}}(z_i)$, are summarized in Table~\ref{tab:CC_SN_rates_redshift}.

\section{Rates, Detection Efficiency \& Uncertainties}
\label{sec:rdeu}

In this section, we derive the DES CC~SN volumetric rates. We outline the core mathematical framework used to calculate the rates (Section~\ref{sec:rates}) and describe the determination of the survey detection efficiency via pipeline simulations (Section~\ref{sec:efficiency}). We then model the underlying survey selection function and completeness boundaries using an analytical forward model to provide a physical sanity check on our simulation limits (Section~\ref{sec:survey_selection}). Section~\ref{sec:uncertainties} details the final CC~SN rates evaluated across 11 redshift bins, while Section~\ref{sec:mc_error_propagation} describes our hybrid MC framework used to robustly propagate joint statistical and asymmetric systematic uncertainties. Finally, we present the volumetric rates for the H-rich and SE CC~SN subtypes and examine whether the SE~SN fraction of the global core-collapse population exhibits any evolutionary trend out to a redshift of $z = 1.3$ (Section~\ref{sec:subtype_rates}).

\subsection{Volumetric CC~SN rate  calculation}
\label{sec:rates}

After propagating redshift uncertainties via MC simulations and correcting the fraction of CC SNe missed due to dust extinction, the final corrected number of events in each redshift bin, $N_\mathrm{CC}(z_i)$, is used to compute the volumetric CC~SN rate.

The volumetric CC~SN rate in redshift bin $i$ is
\begin{equation}
R_\mathrm{CC}(\langle z \rangle_i) =
\frac{N_\mathrm{CC}(z_i)}{
\epsilon_\mathrm{class}(z_i)\,
\epsilon_\mathrm{detect}(z_i)\,
V(z_i)\,
T_\mathrm{rest}^{\mathrm{control}}(z_i)
}.
\label{eq:Rcc_integral}
\end{equation}

where  $\epsilon_\mathrm{class}(z_i)$ is the classification correction factor, $\epsilon_\mathrm{detect}(z_i)$ is the detection efficiency, $V(z_i)$ is the comoving volume in the bin, $T_\mathrm{rest}^{\mathrm{control}}(z_i)$ is the rest-frame control time.

\begin{figure}

    \centering

    \includegraphics[width=\linewidth]{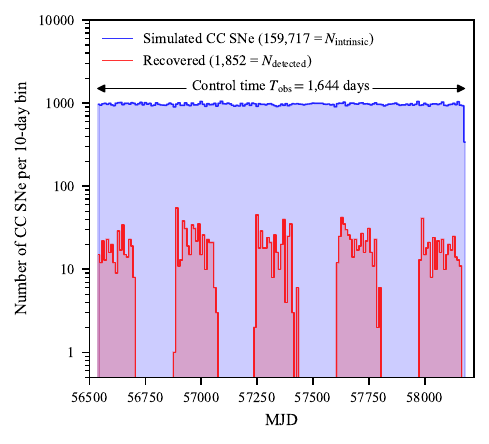}

    \vspace{-10pt}

    \caption{Observer-frame control time $T_{\mathrm{obs}}$, derived from the simulated explosion period (which mirrors the actual overall DES-SN survey period, including non-observing periods/seasons).}

    \label{fig:control}

\end{figure}

The rest-frame control time is related to the observer-frame control time, $T_\mathrm{obs}$, through
\[
T_\mathrm{rest}^{\mathrm{control}}(z_i) = \frac{T_\mathrm{obs}}{1+\langle z \rangle_i},
\]
accounting for cosmological time dilation. The observer-frame control time corresponds to the time span over which simulated explosions were generated, mirroring the overall DES-SN survey period, including non-observing periods. As shown in Figure~\ref{fig:control}, this totals 1,644 days.

The comoving volume of each redshift bin is computed as
\begin{equation}
V(z_i) = \int_{z_{\mathrm{low},i}}^{z_{\mathrm{high},i}} 
\frac{\mathrm{d}V}{\mathrm{d}z} \, \mathrm{d}z ,
\end{equation}
where $\mathrm{d}V/\mathrm{d}z$ includes the DES\secondcolor{-SN} survey solid angle of
$22.912~\mathrm{deg}^2$ \citep{Lasker2020}.

Each bin is assigned an effective redshift, $\langle z \rangle_i$, defined as the volume-weighted mean redshift, which represents the average redshift of the survey volume within the bin:
\begin{equation}
\langle z \rangle_i =
\frac{\int_{z_{\mathrm{low},i}}^{z_{\mathrm{high},i}} z \, \mathrm{d}V}
{\int_{z_{\mathrm{low},i}}^{z_{\mathrm{high},i}} \mathrm{d}V}.
\end{equation}
A more general definition of the effective redshift would include a weighting term $(1+z) \eta(z)$ to account for cosmological time dilation and redshift-dependent SN rate evolution \citep{Graur2014}. In our analysis, however, time dilation is explicitly corrected through the rest-frame control time, and the narrow redshift bins ($\Delta z = 0.1$) limit the impact of rate evolution within each bin. Under these conditions, adopting a simple volume-weighted mean provides an adequate representation of $\langle z \rangle_i$.

\subsection{Detection efficiency and classification correction factor}
\label{sec:efficiency}

The detection efficiency, $\epsilon_{\rm detect}$, quantifies the fraction of intrinsic SNe successfully detected by the survey. We estimate this efficiency by conducting a simulation that follows the methodology described in \citet[Section 4]{Vincenzi2021}, as implemented to evaluate CC~SN contamination for the 5-yr DES cosmology analysis \citep{DES2024}. This process generates large samples of both fake SNe Ia and fake CC SNe through the SNANA/PIPPIN pipeline, mirroring how real transients are selected in the DES-SN survey. Observed fluxes are modeled using DES-specific noise—including the anomalous noise introduced by the DIFFIMG pipeline—and events are recovered only if they meet the operational trigger requirement of detections on at least two separate nights. This simulation allows us to calculate $\epsilon_{\rm detect}$ by comparing the number of recovered events to the total injected population. 

SN surveys typically use one of two types of detection efficiency: a magnitude-dependent efficiency, $\epsilon_\mathrm{detect}(m)$, or a redshift-dependent efficiency, $\epsilon_\mathrm{detect}(z)$. A magnitude-based efficiency cannot be used in our case, because many of the simulated events were not assigned an observer-frame magnitude. This could have occurred for several possible reasons, including trigger failures, seasonal gaps,  incomplete K-corrections, or high host-galaxy surface brightness. This means that, for a given apparent magnitude bin, the intrinsic number of simulated CC~SN explosions,
$N_{\mathrm{intrinsic}}(m)$, is incomplete, and $\epsilon_{\mathrm{detect}}(m)$ calculated using the subsample of simulated CC SNe assigned a magnitude would not be physically meaningful.

\begin{figure}
    \centering

    \includegraphics[width=\columnwidth]{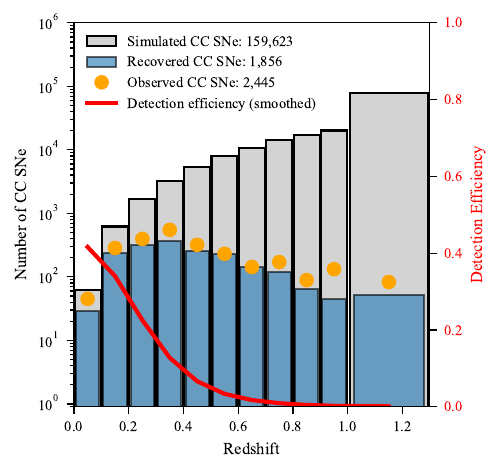}
    \caption{Detection efficiency of DES CC~SNe as a function of redshift $z$, showing the fraction of simulated CC SNe detected as a fraction of all intrinsic simulated explosions.}
    \label{fig:detection_combined}
\end{figure}

Since all simulated SNe were assigned a redshift, we adopt a redshift-based detection efficiency:

\begin{equation}
\epsilon_\mathrm{detect}(z) =
\frac{N_\mathrm{detected}(z)}{N_\mathrm{intrinsic}(z)},
\end{equation}

where $N_\mathrm{detected}$ and $N_\mathrm{intrinsic}$ are the numbers of detected and simulated CC~SNe, respectively ( Figure~\ref{fig:detection_combined}). 

This method has two main advantages: (i) each simulated event is assigned a redshift, so the detection efficiency is computed by directly counting recovered and injected events in each redshift bin, without relying on magnitude–redshift conversions or intermediate assumptions, and (ii) the method directly measures the fraction of the comoving volume in which SNe are detectable given the survey's cadence, depth and selection criteria. The primary limitation is that the efficiency is averaged over all magnitudes within each redshift bin. However, this effect is reduced by adopting narrow bins ($\Delta z = 0.1$).

Following detection, the classification algorithm is run on the combined sample of simulated SNe Ia and CC SNe to determine the classification correction factor, $\epsilon_{\rm class}$. This factor represents the ratio of the total number of objects classified as CC SNe to the total number of simulated CC SNe:

\begin{equation}
\epsilon_{\rm class} = \frac{TP + FP}{TP + FN} = \frac{\text{Classified CC SNe}}{\text{Simulated CC SNe}}.
\label{eq:eps_class}
\end{equation}

In this framework, $TP$ (True Positives) are correctly identified CC~SNe, $FP$ (False Positives) are SNe~Ia misclassified as CC~SNe, and $FN$ (False Negatives) are CC SNe misclassified as SNe~Ia. These values are determined by comparing classifier predictions against ground-truth labels in our simulated samples. This factor effectively scales the observed counts by the ratio of sample completeness (recall) to purity (precision) \citep{Moller2020}.  

The combination of $\epsilon_{\mathrm{detect}}$ and $\epsilon_{\mathrm{class}}$ accounts for both survey selection and algorithmic classification biases. While $\epsilon_{\mathrm{detect}}$ corrects for events lost due to survey cadence, depth and selection criteria $\epsilon_{\mathrm{class}}$ accounts for CC SNe missed by the classifier and overcounting from SNe Ia contamination. The detection efficiencies and classification correction factors for each redshift bin for both CC SNe and H-rich and SE SNe  subtype samples are shown in Table \ref {tab:CC_SN_rates_redshift}. Applying these terms in Equation~\ref{eq:Rcc_integral} allows us to recover the volumetric rate from our observed photometric sample.

\subsection{Survey Selection Function and Completeness}
\label{sec:survey_selection}

To evaluate the physical plausibility of the completeness corrections applied to our measured CC~SN rates, we model the survey selection function following the parameterized formalism introduced by \citet{Strolger2004} and adapted by \citet{Melinder2012}. The probability of detecting a SN of apparent magnitude $m$ is given by the logistic function:
\begin{equation}
    \epsilon(m) = \frac{\epsilon_0}{1 + \exp\left(\frac{m-m_c}{s}\right)},
\end{equation}
where $\epsilon_0$ is the efficiency ceiling, $m_c$ is the effective survey depth at which completeness drops to $0.5\,\epsilon_0$, and $s$ defines the selection smoothness.

In our forward modeling, $m_c$ is held fixed at $23.96$\,mag (a weighted average across shallow and deep fields; \citealt{Kessler2015, Vincenzi2021}) to prevent optimization instability caused by degeneracy with the characteristic absolute magnitude. The remaining selection parameters are fixed based on known pipeline characteristics:
\begin{itemize}
    \item Efficiency ceiling ($\epsilon_0 = 0.5$): Reflects the low-redshift detection asymptote ($z = 0.0\text{--}0.2$), accounting for non-magnitude constraints such as seasonal gaps and observing cadence.
    \item Selection smoothness ($s = 0.5$\,mag): Captures the sharp, probabilistic detection threshold characteristic of automated DES difference-imaging pipelines.
\end{itemize}

The intrinsic CC~SN population is approximated by a single-Gaussian luminosity function in rest-frame absolute magnitude:
\begin{equation}
    \Phi(M) = \frac{1}{\sigma\sqrt{2\pi}} \exp\left[ -\frac{(M-M_0)^2}{2\sigma^2} \right],
\end{equation}
where $M_0$ is the characteristic mean absolute magnitude and $\sigma$ is the intrinsic population dispersion. Convolving $\Phi(M)$ with the selection function yields the overall redshift-dependent detection efficiency, $\bar{\epsilon}(z)$:
\begin{equation}
    \bar{\epsilon}(z) = \int_{-\infty}^{+\infty} \Phi(M)\,\epsilon(m[M,z])\,\mathrm{d}M,
\end{equation}
where $m(M,z) = M + \mu(z) + K(z) + A$, where $\mu(z)$ is the cosmological distance modulus, $K(z)$ is the K-correction, and $A$ represents host and Galactic extinction. Here $A$ is set to zero to remain consistent with our template library assumptions \citep{Li2011a}. 

At high redshift ($z \gtrsim 0.6$), the survey encounters a classic ``tip-of-the-iceberg'' selection effect: flux limits restrict detections exclusively to the extreme, intrinsically bright tail of the supernova luminosity function. As redshift increases, detection efficiency declines rapidly—from $\bar{\epsilon} = 0.0133$ ($z = 0.6\text{--}0.7$) down to $\bar{\epsilon} = 0.0022$ ($z = 0.9\text{--}1.0$), ultimately dropping to $\bar{\epsilon} = 0.0006$ in our highest bin ($z = 1.0\text{--}1.3$). In this high-$z$ regime, recovering the true intrinsic rate requires large volumetric selection corrections in the order of $> 1,000\times$ ($1/\bar{\epsilon}$). Here, the slope of the volumetric correction factor $\log_{10}(1/\bar{\epsilon})$ reaches a stable, quasi-linear equilibrium resulting from the competing evolution of expanding population penetration into the Gaussian tail and the combined rates of cosmological dimming ($\mathrm{d}\mu/\mathrm{d}z$) and chromatic shifting ($\mathrm{d}K/\mathrm{d}z \approx 1.2$). The complete theoretical derivation of this analytical boundary is detailed in Appendix~\ref{sec:appendix_b5}.

\begin{figure}
    \centering
    \includegraphics[width=\columnwidth]{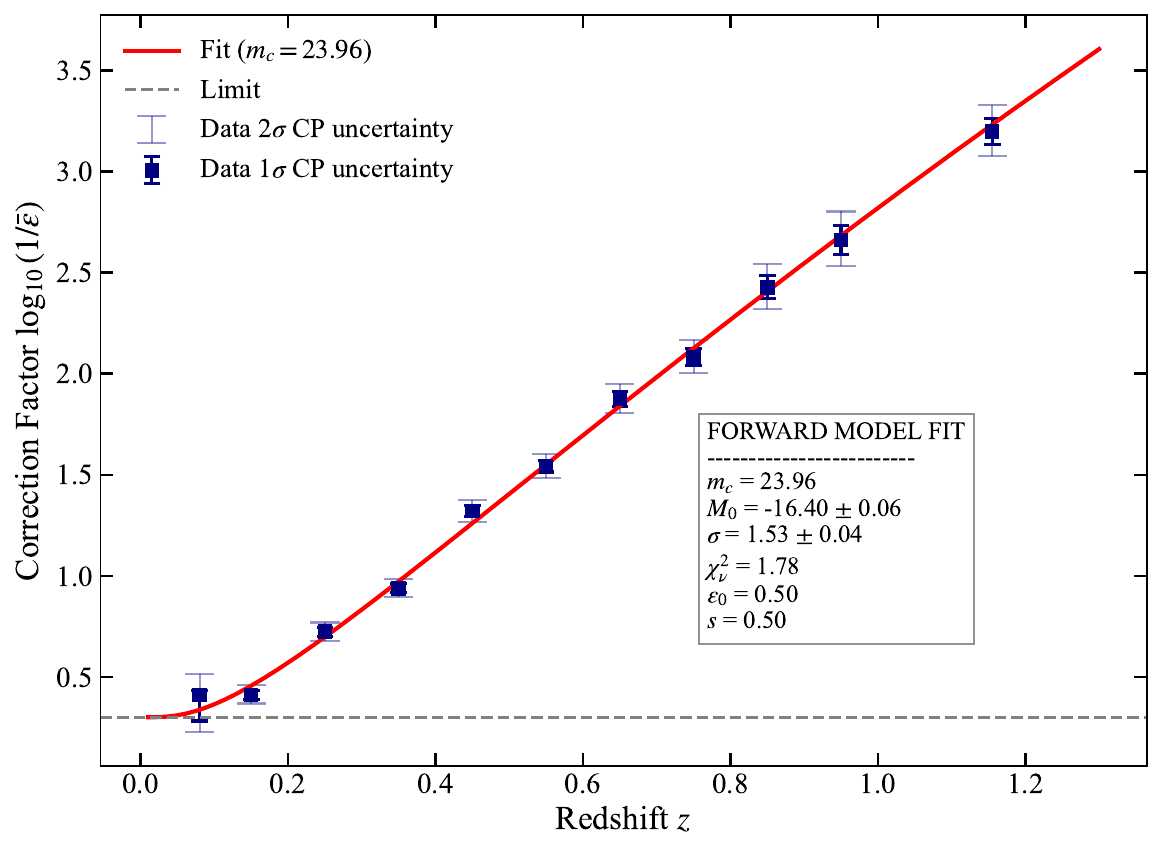}
    \caption{Physical forward model fit to the binned CC~SN selection correction factors ($\log_{10}(1/\bar{\epsilon})$) as a function of redshift $z$. Navy square markers represent empirical correction factors derived from \textsc{snana} simulations, with inner and outer error bars depicting $1\sigma$ and $2\sigma$ Clopper--Pearson uncertainties, respectively. The solid red curve displays the best-fit continuous forward model ($\epsilon_0 = 0.5$, $m_c = 23.96$\,mag) optimized via $\chi^2$ minimization.}
    \label{fig:forward_model_fit}
\end{figure}

We fit the physical parameters ($M_0$, $\sigma$) via an \textsc{iminuit} $\chi^2$ minimization against the recovery fractions derived from our full \textsc{snana} \citep{Kessler2009SNANA,  Vincenzi2019, Vincenzi2021} survey simulations. As shown in Figure~\ref{fig:forward_model_fit}, our baseline All Fields configuration converges to $M_0 = -16.40 \pm 0.06$\,mag and $\sigma = 1.52 \pm 0.04$\,mag, yielding a reduced $\chi^2_\nu = 1.78$. This confirms that while our simplified parameterized form leaves minor local scatter, it provides an excellent global description of survey completeness across all fields and redshifts.

Importantly, the reliability of the All Fields forward model in the highest-redshift regime is anchored by a combined empirical bin spanning $z = 1.0\text{--}1.3$. This high-$z$ bin captures 50 valid detections out of approximately 80,000 simulated events, establishing an empirical detection efficiency baseline of $\bar{\epsilon} \approx 0.0006$. By pooling detections across this broad high-$z$ range, the optimization mitigates severe Poisson noise that would otherwise destabilize this regime where the $> 1,000\times$ correction factor applies.

Extensive robustness checks—including field-depth variations (deep vs. shallow fields), efficiency ceiling sensitivity ($\epsilon_0 = 0.55$), high-$z$ Poisson bin pooling, and analytic sensitivity to luminosity function evolution—are detailed in Appendix~\ref{sec:app_fitting_details} and Appendix~\ref{sec:app_lf_sensitivity}.

Ultimately, this forward model acts as an analytical reality check on our pipeline selection corrections. By fitting the entire survey volume simultaneously, the model anchors the underlying SN population using complete, low-redshift data. High-redshift completeness is then deterministically predicted by flux-limit thresholding and cosmic dimming, mathematically validating the large "tip-of-the-iceberg" selection corrections derived from our simulations at high redshift.

Our forward model and empirical efficiency corrections show broad consistency, with model predictions remaining within the $2\sigma$ binomial uncertainties across all redshifts (Figure~\ref{fig:forward_model_fit}). This agreement demonstrates that our pipeline selection function introduces no major systematic bias into the derived volumetric rates. Although the model falls slightly outside $1\sigma$ bounds at a few redshift points, it accurately reproduces the overarching trend across the survey volume, confirming the robustness of our simulation-based efficiency corrections. Importantly, no luminosity function evolution was assumed in the \textsc{snana} simulation, based on the V19 templates, that underpins these efficiency corrections. Nevertheless, sensitivity tests incorporating a progressive luminosity function brightening of $0.1\,\mathrm{mag}$ per unit redshift (Appendix~\ref{sec:app_lf_sensitivity}) confirm that our primary finding of strong volumetric rate growth with redshift remains valid. In the absence of evidence for luminosity function evolution, high-redshift rate uncertainties stem primarily from finite simulation sampling noise (Table~\ref{tab:rate_uncertainties}, Efficiency column), the scale of which is validated by the forward model fit remaining within $2\sigma$ across all redshifts.

\subsection{Rate uncertainties and CC SN rates}
\label{sec:uncertainties}

Uncertainties are evaluated independently within each redshift bin. The total uncertainty on $R_\mathrm{CC}$ combines five contributions: Poisson counting statistics, classification corrections, detection efficiency, MC redshift propagation, and extinction corrections.

\begin{table*}
    \centering
    \small
    \caption{Full uncertainty breakdown by redshift bin on corrected CC~SN rates.}
    \label{tab:rate_uncertainties}
    \vspace{2pt} 
    \begin{tabular}{l c c c c c c c c}
        \hline
        \hline
        Redshift Bin & 
        $R_{\mathrm{CC}}$ (Det) & 
        Poisson $\pm$ & 
        Classification $\pm$ & 
        Efficiency $\pm$ & 
        MC $\pm$ & 
        Extinction $\pm$ & 
        Total $\pm$ & 
        Total absolute \\
        & ($10^{-4}~\mathrm{yr}^{-1}~\mathrm{Mpc}^{-3}$) & (\%) & (\%) & (\%) & (\%) & (\%) & (\%) & ($+$/$-$) \\
        \hline
        0.0--0.1 & 2.3  & 13.2/15.1 & 12.0/17.1 & 16.9/17.4 & 6.6/7.3 & 11.3/29.3 & 30.6/39.6 & 0.9/0.7 \\
        0.1--0.2 & 2.6  & 5.8/6.2   & 3.8/4.3   & 5.2/5.3   & 2.3/2.3 & 11.5/28.3 & 14.8/29.8 & 0.8/0.4 \\
        0.2--0.3 & 3.0  & 4.8/5.1   & 3.1/3.4   & 5.1/5.3   & 2.2/2.2 & 11.1/27.0 & 13.8/28.2 & 0.8/0.4 \\
        0.3--0.4 & 3.9  & 4.5/4.7   & 2.8/3.1   & 4.9/5.2   & 2.1/2.1 & 11.1/27.6 & 13.5/28.6 & 1.1/0.5 \\
        0.4--0.5 & 4.0  & 5.8/6.1   & 3.5/3.8   & 6.1/6.5   & 3.3/3.3 & 10.6/31.3 & 14.6/32.9 & 1.3/0.6 \\
        0.5--0.6 & 4.4  & 6.5/6.9   & 3.6/4.0   & 6.5/6.9   & 3.9/3.9 & 11.0/32.2 & 15.6/34.0 & 1.5/0.7 \\
        0.6--0.7 & 6.0  & 7.5/8.1   & 4.6/5.2   & 8.3/9.0   & 4.9/4.8 & 12.2/33.6 & 18.3/36.2 & 2.2/1.1 \\
        0.7--0.8 & 7.6  & 7.9/8.5   & 5.1/5.8   & 9.2/10.1  & 5.0/5.0 & 12.9/36.9 & 19.7/39.6 & 3.0/1.5 \\
        0.8--0.9 & 11.4 & 9.3/10.2  & 7.0/8.4   & 12.4/14.1 & 6.2/6.3 & 13.4/39.1 & 24.0/43.3 & 4.9/2.7 \\
        0.9--1.0 & 13.6 & 11.3/12.6 & 8.9/11.1  & 15.0/17.4 & 8.4/8.3 & 13.8/40.5 & 28.6/46.7 & 6.4/3.9 \\
        1.0--1.3 & 16.1 & 10.3/11.4 & 8.3/10.0  & 14.1/16.2 & 7.5/7.6 & 13.8/41.8 & 26.8/46.9 & 7.6/4.3 \\
        \hline
        \hline
    \end{tabular}
\end{table*}

Poisson uncertainties on the number of detected events are computed using asymmetric $\chi^2$ confidence intervals appropriate for low-count statistics \citep{Gehrels1986}:
\[
\sigma_\mathrm{Poisson}^{-} = N - \frac{1}{2} \chi^2_{0.16, 2N}, \quad
\sigma_\mathrm{Poisson}^{+} = \frac{1}{2} \chi^2_{0.84, 2N+2} - N.
\]

Detection efficiency, defined as the fraction of simulated CC SNe that are recovered, is assigned 68~per~cent confidence intervals using the exact binomial method \citep{Clopper1934}: 
\begin{equation}
    \sigma^-_{\epsilon_\mathrm{det}} = p - p_\mathrm{low}, \qquad
    \sigma^+_{\epsilon_\mathrm{det}} = p_\mathrm{high} - p, \qquad
    p = \frac{k}{n},
\end{equation}
where $k = N_\mathrm{detected}$ and $n = N_\mathrm{intrinsic}$. The lower and upper bounds are determined by:
\begin{equation}
    p_\mathrm{low} = \mathrm{B}^{-1}\left(\frac{\alpha}{2}; k, n-k+1\right), \quad
    p_\mathrm{high} = \mathrm{B}^{-1}\left(1-\frac{\alpha}{2}; k+1, n-k\right),
\end{equation}
where $\mathrm{B}^{-1}(P; a, b)$ is the quantile function (inverse cumulative distribution) of the Beta distribution with shape parameters $a$ and $b$. To correspond to a 68~per~cent ($1\sigma$) confidence level, we set $\alpha = 0.32$.

To account for misclassification in our photometric sample, we utilize the classification correction factor $\epsilon_{\mathrm{class}}$ defined in Section~\ref{sec:efficiency}. This factor, as defined in Equation \ref{eq:eps_class}, is calculated as the ratio $(TP + FP) / (TP + FN)$, incorporating the $TP$, $FP$, and $FN$ counts determined by comparing classifier predictions against ground-truth labels in our simulated samples. For each redshift bin, the uncertainties $\sigma^{\pm}_{\epsilon_{\mathrm{class}}}$ are derived by propagating exact Poisson confidence limits \citep{Garwood1936Biometrika, Gehrels1986} for the $TP$ and $FP$ counts in the numerator, normalized by the total simulated population ($TP+FN$) in each bin. This statistical treatment accounts for the discrete nature of the simulated samples, particularly in high-redshift bins where event counts are low, resulting in characteristic asymmetric uncertainties.

MC redshift uncertainties are computed for each bin following the procedure described in Section~\ref{sec:redshift_uncertainties}. For each bin, the 16th, 50th, and 84th percentiles of the simulated counts are used to define the asymmetric uncertainties:
\[
\sigma_{N_\mathrm{MC}}^- = P_{50} - P_{16}, \quad 
\sigma_{N_\mathrm{MC}}^+ = P_{84} - P_{50},
\]
thereby capturing the statistical spread of the counts due to redshift uncertainties \citep{Neill2006}.

The missing fraction uncertainty reflects the uncertainty in the fraction of CC~SNe missed due to host-galaxy and Milky Way extinction \citep{Mattila2012}. The fractional uncertainty associated with the extinction correction in each redshift bin is defined as:
\[
\mathrm{Fraction}_{\mathrm{Low}} =
\frac{N_\mathrm{CC} - N_{\mathrm{CC,low}}}{N_\mathrm{CC}}, \quad
\mathrm{Fraction}_{\mathrm{High}} =
\frac{N_{\mathrm{CC,high}} - N_\mathrm{CC}}{N_\mathrm{CC}},
\]
where $N_\mathrm{CC}$ is the extinction-corrected number of CC~SNe, and $N_{\mathrm{CC,low}}$ and $N_{\mathrm{CC,high}}$ are the corresponding lower and upper bounds, as reported by \citet{Mattila2012}.

The corresponding absolute uncertainties in the extinction-corrected counts are:
\[
\sigma^{-}_\mathrm{ext} = N_\mathrm{CC} - N_{\mathrm{CC,low}}, \qquad
\sigma^{+}_\mathrm{ext} = N_{\mathrm{CC,high}} - N_\mathrm{CC}.
\]
These are propagated into the rate uncertainties as fractional terms $\sigma^{-}_\mathrm{ext}/N_\mathrm{CC}$ and $\sigma^{+}_\mathrm{ext}/N_\mathrm{CC}$.

Together, these five uncertainties provide a comprehensive estimate of the statistical and systematic uncertainties on $R_\mathrm{CC}$, enabling robust propagation into the volumetric CC~SN rate measurement. The upper and lower uncertainties in $R_\mathrm{CC}$ are then propagated as:

{\small
\begin{equation}
\sigma^{+}_{R_\mathrm{CC}} = R_\mathrm{CC}\,
\sqrt{
\left(\!\frac{\sigma^{+}_{N_\mathrm{MC}}}{N_\mathrm{MC}}\!\right)^2+
\left(\!\frac{\sigma^{-}_{\epsilon_\mathrm{class}}}{\epsilon_\mathrm{class}}\!\right)^2+
\left(\!\frac{\sigma^{-}_{\epsilon_\mathrm{det}}}{\epsilon_\mathrm{det}}\!\right)^2+
\left(\!\frac{\sigma^{+}_\mathrm{Poiss}}{N_\mathrm{MC}}\!\right)^2+
\left(\!\frac{\sigma^{+}_\mathrm{ext}}{N_\mathrm{CC}}\!\right)^2
},
\end{equation}
}

{\small
\begin{equation}
\sigma^{-}_{R_\mathrm{CC}} = R_\mathrm{CC}\,
\sqrt{
\left(\!\frac{\sigma^{-}_{N_\mathrm{MC}}}{N_\mathrm{MC}}\!\right)^2+
\left(\!\frac{\sigma^{+}_{\epsilon_\mathrm{class}}}{\epsilon_\mathrm{class}}\!\right)^2+
\left(\!\frac{\sigma^{+}_{\epsilon_\mathrm{det}}}{\epsilon_\mathrm{det}}\!\right)^2+
\left(\!\frac{\sigma^{-}_\mathrm{Poiss}}{N_\mathrm{MC}}\!\right)^2+
\left(\!\frac{\sigma^{-}_\mathrm{ext}}{N_\mathrm{CC}}\!\right)^2
}.
\end{equation}
}

Due to relatively low counts and high uncertainties in the highest redshift bins ($z > 1.0$), these bins were combined into a single bin to reduce uncertainty. A full breakdown of the rate uncertainties per redshift bin, with individual contributions propagated into the uncorrected rate uncertainties and combined with the extinction fraction uncertainties, is shown in Table~\ref{tab:rate_uncertainties}.

\begin{figure}
\centering
\includegraphics[width=\linewidth]{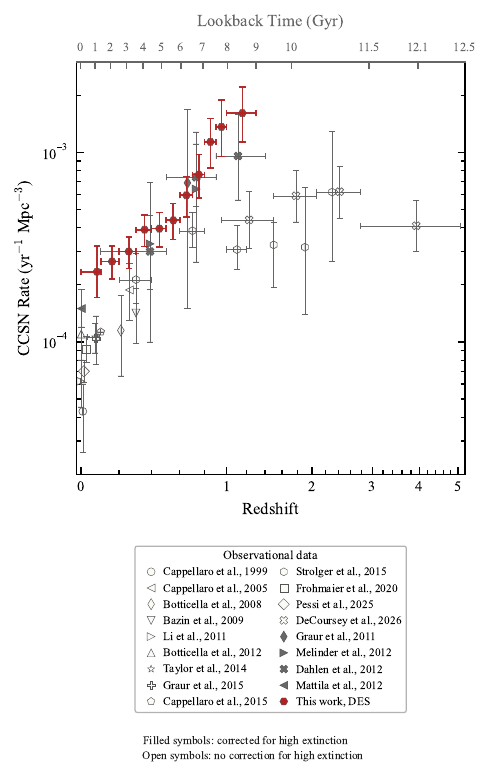}
\vspace{-14pt}
\caption{Comparison of DES CC~SN rates with previous observations (horizontal bars denote the redshift interval). The $x$-axis is scaled by $\log(1+z)$, providing a linear mapping to the logarithmic cosmic scale factor, $\log(a^{-1})$. There are two main reasons for the use of this scale: (i) it linearizes the power-law evolution of the rate, making the trend in our data easier to interpret as a straight line, and (ii) it stretches the low-redshift regime to prevent data crowding, while allowing the x-axis to extend to $z = 5.2$, to show the high-redshift results from \citet{Strolger2015} and \citet{2026arXiv260712018D} }.
\label{fig:salpeter2}
\end{figure}

The final DES volumetric CC SN rates are presented in the top panel of Table~\ref{tab:CC_SN_rates_redshift}, and are shown in Figure \ref{fig:salpeter2}.
A comparison of the rates measured in this work with the rates from other surveys in the literature is presented in Table~\ref{tab:CC_SN_rates} and is shown in Figure \ref{fig:salpeter2}. It can be seen that our rates (shown in dark red) align closely with the other extinction-corrected rates in the literature  \citep{Graur2011, Mattila2012, Melinder2012, Dahlen2012} which are shown as filled grey symbols.

\subsection{MC Propagation of Uncertainties}
\label{sec:mc_error_propagation}

To robustly propagate the combination of discrete counting statistics and continuous, asymmetric systematic uncertainties through our analysis, we implement a hybrid MC framework utilizing $5 \times 10^4$ stochastic realizations per redshift bin. While analytical error propagation via traditional split-summing in quadrature provides a baseline approximation, it inherently struggles to track the transformation of skewed probability density functions (PDFs) through non-linear pipeline elements. In contrast, our numerical MC framework directly maps the joint multidimensional probability space through the volumetric rate inversion, naturally accounting for non-linearities and variable interdependencies without relying on linear approximations.

For each redshift bin $i$, the numerical pipeline executes a dual-stage sampling procedure:
\begin{enumerate}
    \item Statistical Base: Intrinsic poissonian counting fluctuations are established by sampling a discrete Poisson distribution centered on the raw expected CC~SN counts, $N_{\text{poisson}} \sim \text{Poisson}(\lambda = N_{\text{MC}, i})$.
    \item Systematic Perturbations: Non-Gaussian systematic variations are simultaneously superimposed onto this statistical base. We apply a split-Gaussian ($\mathcal{SG}$) noise framework, generating an asymmetric shift vector centered at zero, $\Delta N_{\text{sys}} \sim \mathcal{SG}(0, \sigma_{\text{lo}}, \sigma_{\text{hi}})$, where the lower and upper standard deviations correspond to the uncalibrated count uncertainty bounds ($N_{\text{MC, lo}, i}$ and $N_{\text{MC, hi}, i}$). The combined count distribution is defined as $N_{\text{MC}} = N_{\text{poisson}} + \Delta N_{\text{sys}}$, truncated at $10^{-1}$ to avoid unphysical non-positive values.
\end{enumerate}

Multi-variable systematic convolution is carried out concurrently by independently sampling asymmetric split-Gaussian error profiles for the detection efficiency ($\epsilon_{\text{det}}$) and classification efficiency ($\epsilon_{\text{cls}}$). The systematic uncertainty inherent to the host-galaxy dust extinction correction—derived using the framework of \citet{Mattila2012} as detailed in Sections~\ref{sec:extinction} and~\ref{sec:uncertainties}—is propagated simultaneously. This is achieved by sampling a dimensionless scaling multiplier, $f_{\text{ext}}$, from an asymmetric split-Gaussian distribution centered on unity, $f_{\text{ext}} \sim \mathcal{SG}(1.0, \sigma_{\text{ext, lo}}, \sigma_{\text{ext, hi}})$, where the limits correspond to the fractional extinction uncertainty envelopes ($\text{ext}_{\text{lo}, i}$ and $\text{ext}_{\text{hi}, i}$) for each bin. For each simulation iteration, these joint realizations are propagated through the core volumetric rate inversion equation:
\begin{equation}
    R_{\text{CC}, i, \text{sample}} = R_{\text{CC}, i} \times \left( \frac{N_{\text{MC}}}{N_{\text{MC}, i}} \right) \times \left( \frac{\epsilon_{\text{det}, i}}{\epsilon_{\text{det, sample}}} \right) \times \left( \frac{\epsilon_{\text{cls}, i}}{\epsilon_{\text{cls, sample}}} \right) \times f_{\text{ext, sample}},
\end{equation}
where variables with an $i$ subscript denote the baseline deterministic values (which already include the central host-extinction corrections), and variables with a $\text{sample}$ subscript represent the stochastically drawn values.

\subsubsection{Comparison of Uncertainty Frameworks}
\label{sub:mc_results_comparison}

\begin{figure}
    \centering
    \includegraphics[width=\columnwidth]{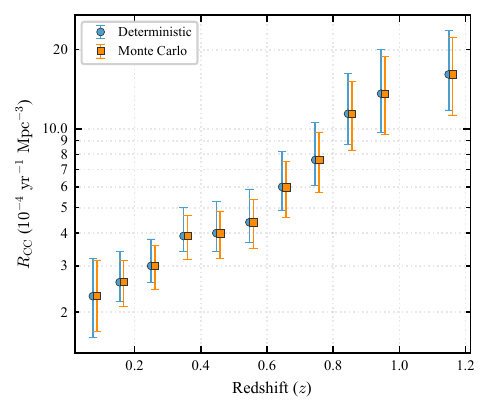}
    \caption{Comparison of CC~SN volumetric rates ($R_{\mathrm{CC}}$) as a function of redshift ($z$). The blue circles show the original deterministic rates derived via standard quadrature error propagation. The orange squares present the median rates from the hybrid MC simulation, which self-consistently convolves Poisson counting statistics with asymmetric systematic uncertainty profiles (classification, efficiency, and dust extinction). For visual clarity, markers are horizontally offset by $\Delta z = \pm 0.006$.}
    \label{fig:ccsn_rate_mc_vs_det}
\end{figure}

A comparison between the volumetric rates derived via deterministic quadrature summing ($R_{\mathrm{CC}}$ (Det)) and the hybrid MC framework ($R_{\mathrm{CC}}$ (MC)) across all redshift bins and subtypes is presented in Table~\ref{tab:CC_SN_rates_redshift}, and visualized in Fig.~\ref{fig:ccsn_rate_mc_vs_det}. The median volumetric rates remain consistent between the two methodologies (varying by $<0.3$ per~cent), confirming the stability of the central rates against numerical sampling effects.

However, the uncertainty envelopes exhibit a clear divergence; the overall MC uncertainty intervals are notably less asymmetric than those generated via deterministic propagation, while maintaining an approximately equivalent total error budget (high-minus-low span). Specifically, the MC propagation yields larger lower-bound uncertainties and smaller upper-bound uncertainties, effectively shifting the weight of the probability density toward the lower bound when compared to the deterministic approach.
Given its superior capacity to map skewed uncertainties through non-linear space and naturally minimize statistical fitting biases, we adopt the MC framework as our definitive analytical baseline throughout the remainder of this work. The comparative deterministic result is shown in Appendix~\ref{app:deterministic_rates}.Appendix~\ref{app:deterministic_rates}.

\subsection{CC~SN subtype rates}
\label{sec:subtype_rates}

Because the H-rich and SE~SNe samples are smaller than the full CC~SN sample, we adopt broader redshift bins for the subtype rate measurements. We use six bins in total: five bins of width $\Delta z = 0.2$ spanning $0 < z < 1.0$, and a final bin of width $\Delta z = 0.3$ covering $1.0 < z < 1.3$, which encompasses the full DES redshift range. We tested alternative binning schemes---including 13 bins of width $\Delta z = 0.1$ and a five-bin scheme with $\Delta z = 0.25$ over $0 < z < 1.0$ along with a final $1.0 < z < 1.3$ bin---and found that they yield consistent results, with increased scatter in the $\Delta z = 0.1$ finer binning case, as might be expected.

The estimation of subtype rates involves corrections at two distinct stages of the analysis. First, a constant calibration factor of 1.33 was applied directly to the SE~SNe number counts to correct for the systematic underclassification identified via comparison with the \citet{Grayling2023} reference sample (see Section~\ref{sec:subtypes}). To ensure the total number of CC~SNe is conserved per bin within our error framework, the H-rich counts were adjusted such that the integrated combination aligns with the total core-collapse budget ($N_{\text{H-rich, cal}} = N_{\text{CC, total}} - N_{\text{SE, cal}}$), accounting for the non-linear shifts introduced during asymmetric error propagation. 

Volumetric rate uncertainties for both H-rich and SE~SNe are evaluated independently within each of the broader redshift bins using the identical five-component analytical framework established for the full CC~SN sample (detailed in Section~\ref{sec:uncertainties}). For each individual subtype rate, we propagate Gehrels Poisson counting statistics, Clopper-Pearson exact binomial detection efficiencies, redshift-specific classification corrections ($\epsilon_{\mathrm{class}}$), MC redshift variations, and host-galaxy/Milky Way dust extinction fractions in quadrature using Equations~7 and 8. By applying this uniform analytical treatment directly to the calibrated subtype bins, we ensure that the discrete statistical limitations and systematic variances inherent to the smaller H-rich and SE~SNe samples are captured identically and rigorously without introducing further independent simulation loops.

\begin{figure}
    \centering
    \includegraphics[width=\linewidth]{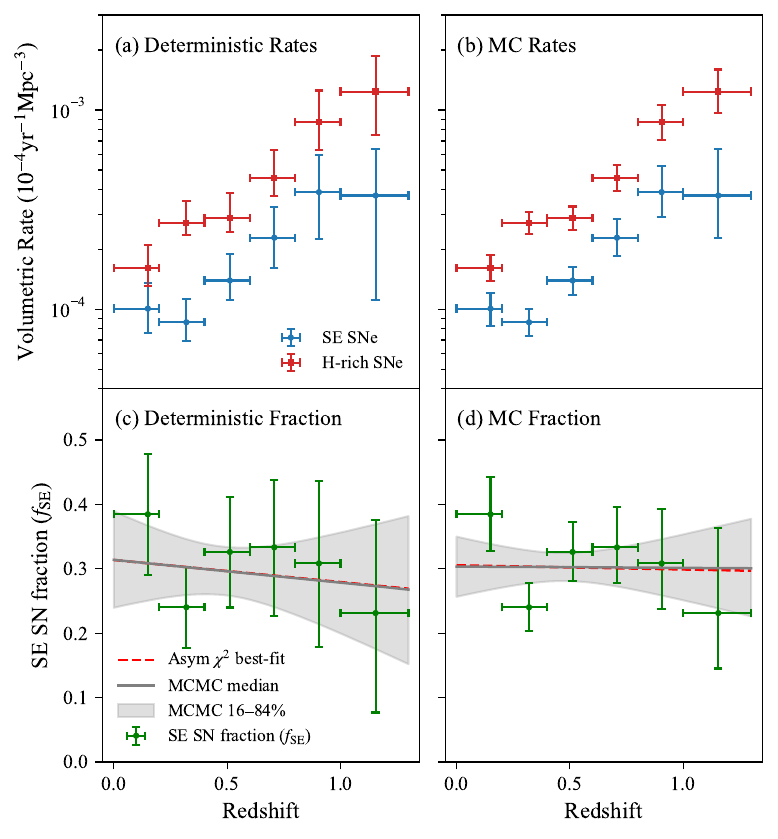}
    \caption{DES volumetric rates of SE and H-rich CC SNe, together with the SE fraction as a proportion of all CC~SN rates, in six redshift bins spanning $z = 0$ to $1.3$. The widths of the horizontal bars indicate the bin sizes.}
    \label{fig:SE_rates}
\end{figure}

\begin{table*}
\centering
\caption{DES CC~SN and subtype number counts, detection efficiencies, classification correction factors, and volumetric rates by redshift bin.}
\label{tab:CC_SN_rates_redshift}

\small 
\setlength{\tabcolsep}{6pt} 
\renewcommand{\arraystretch}{1.25} 

\begingroup
\setlength{\tabcolsep}{8pt}
\begin{tabular}{l cccc cc cc}
\hline
Redshift Bin & $\langle z \rangle_i$ & $N_{\rm raw}$ & $N_{\rm MC}$ & $N_{\rm CC}$ & $\epsilon_{\rm detect}$ & $\epsilon_{\rm class}$ & $R_{\rm CC}$ (Det) & $R_{\rm CC}$ (MC) \\
\hline
\multicolumn{9}{c}{\textbf{All CC SNe}} \\
\hline
0.0--0.1 & 0.080 & 45  & $57_{-4}^{+4}$   & $71_{-17}^{+27}$   & $0.4444_{-0.0751}^{+0.0773}$ & $0.96_{-0.12}^{+0.16}$ & $2.3_{-0.7}^{+0.9}$ & $2.4_{-0.6}^{+0.9}$ \\
0.1--0.2 & 0.161 & 285 & $292_{-7}^{+7}$  & $368_{-50}^{+107}$ & $0.3871_{-0.0201}^{+0.0205}$ & $0.85_{-0.03}^{+0.04}$ & $2.6_{-0.4}^{+0.8}$ & $2.7_{-0.5}^{+0.6}$ \\
0.2--0.3 & 0.256 & 396 & $430_{-10}^{+10}$ & $557_{-71}^{+155}$ & $0.1888_{-0.0096}^{+0.0100}$ & $1.02_{-0.03}^{+0.03}$ & $3.0_{-0.4}^{+0.8}$ & $3.0_{-0.6}^{+0.6}$ \\
0.3--0.4 & 0.354 & 550 & $500_{-11}^{+10}$ & $670_{-85}^{+190}$ & $0.1151_{-0.0056}^{+0.0060}$ & $0.95_{-0.03}^{+0.03}$ & $3.9_{-0.5}^{+1.1}$ & $3.9_{-0.7}^{+0.8}$ \\
0.4--0.5 & 0.453 & 320 & $301_{-10}^{+10}$ & $421_{-55}^{+137}$ & $0.0479_{-0.0029}^{+0.0031}$ & $1.03_{-0.04}^{+0.04}$ & $4.0_{-0.6}^{+1.3}$ & $4.0_{-0.8}^{+0.9}$ \\
0.5--0.6 & 0.553 & 231 & $239_{-9}^{+9}$   & $352_{-49}^{+118}$ & $0.0287_{-0.0019}^{+0.0020}$ & $1.04_{-0.04}^{+0.04}$ & $4.4_{-0.7}^{+1.5}$ & $4.5_{-0.7}^{+1.3}$ \\
0.6--0.7 & 0.652 & 143 & $178_{-9}^{+8}$   & $272_{-44}^{+96}$  & $0.0133_{-0.0011}^{+0.0012}$ & $1.08_{-0.05}^{+0.06}$ & $6.0_{-1.1}^{+2.2}$ & $5.9_{-1.4}^{+1.5}$ \\
0.7--0.8 & 0.752 & 171 & $162_{-8}^{+8}$   & $253_{-44}^{+98}$  & $0.0083_{-0.0008}^{+0.0008}$ & $1.13_{-0.06}^{+0.07}$ & $7.6_{-1.5}^{+3.0}$ & $7.6_{-1.9}^{+2.1}$ \\
0.8--0.9 & 0.851 & 89  & $115_{-7}^{+7}$   & $184_{-37}^{+78}$  & $0.0037_{-0.0005}^{+0.0005}$ & $1.12_{-0.08}^{+0.09}$ & $11.4_{-2.7}^{+4.9}$ & $11.4_{-3.1}^{+3.7}$ \\
0.9--1.0 & 0.951 & 132 & $78_{-7}^{+7}$    & $127_{-30}^{+58}$  & $0.0022_{-0.0003}^{+0.0004}$ & $1.04_{-0.09}^{+0.12}$ & $13.6_{-3.9}^{+6.4}$ & $13.6_{-4.1}^{+5.2}$ \\
1.0--1.3 & 1.155 & 83  & $94_{-7}^{+7}$    & $155_{-35}^{+71}$  & $0.0006_{-0.0001}^{+0.0001}$ & $1.16_{-0.10}^{+0.12}$ & $16.1_{-4.3}^{+7.6}$ & $16.1_{-4.8}^{+6.0}$ \\
\hline
\multicolumn{9}{c}{\textbf{H-rich SNe (Type II)}} \\
\hline
0.0--0.2 & 0.150 & 230 & $208_{-7}^{+7}$   & $261_{-38}^{+78}$  & $0.3987_{-0.0235}^{+0.0239}$ & $0.97_{-0.04}^{+0.04}$ & $1.6_{-0.3}^{+0.5}$ & $1.6_{-0.2}^{+0.3}$ \\
0.2--0.4 & 0.320 & 787 & $716_{-13}^{+13}$ & $946_{-119}^{+266}$ & $0.1500_{-0.0062}^{+0.0063}$ & $0.96_{-0.02}^{+0.02}$ & $2.7_{-0.4}^{+0.8}$ & $2.7_{-0.3}^{+0.4}$ \\
0.4--0.6 & 0.512 & 418 & $361_{-13}^{+12}$ & $518_{-70}^{+170}$ & $0.0361_{-0.0019}^{+0.0020}$ & $1.01_{-0.03}^{+0.03}$ & $2.9_{-0.4}^{+1.0}$ & $2.9_{-0.4}^{+0.4}$ \\
0.6--0.8 & 0.708 & 226 & $225_{-11}^{+11}$ & $347_{-58}^{+129}$ & $0.0114_{-0.0008}^{+0.0009}$ & $0.94_{-0.04}^{+0.04}$ & $4.6_{-0.9}^{+1.7}$ & $4.6_{-0.6}^{+0.7}$ \\
0.8--1.0 & 0.905 & 156 & $110_{-9}^{+9}$   & $178_{-40}^{+80}$  & $0.0030_{-0.0003}^{+0.0004}$ & $0.87_{-0.06}^{+0.07}$ & $8.7_{-2.4}^{+3.8}$ & $8.7_{-1.6}^{+1.9}$ \\
1.0--1.3 & 1.155 & 60  & $67_{-7}^{+7}$    & $111_{-27}^{+52}$  & $0.0007_{-0.0001}^{+0.0001}$ & $0.86_{-0.09}^{+0.12}$ & $12.4_{-4.9}^{+6.3}$ & $12.4_{-2.8}^{+3.6}$ \\
\hline
\multicolumn{9}{c}{\textbf{SE SNe (Type Ibc)}} \\
\hline
0.0--0.2 & 0.150 & 100 & $141_{-5}^{+5}$   & $178_{-26}^{+53}$  & $0.3762_{-0.0350}^{+0.0365}$ & $1.08_{-0.07}^{+0.08}$ & $1.0_{-0.2}^{+0.4}$ & $1.0_{-0.2}^{+0.2}$ \\
0.2--0.4 & 0.320 & 161 & $214_{-8}^{+8}$   & $282_{-37}^{+80}$  & $0.1189_{-0.0087}^{+0.0093}$ & $1.11_{-0.05}^{+0.05}$ & $1.2_{-0.2}^{+0.4}$ & $1.2_{-0.2}^{+0.2}$ \\
0.4--0.6 & 0.512 & 132 & $179_{-9}^{+9}$   & $256_{-36}^{+84}$  & $0.0373_{-0.0030}^{+0.0033}$ & $0.99_{-0.05}^{+0.05}$ & $1.6_{-0.3}^{+0.6}$ & $1.6_{-0.2}^{+0.3}$ \\
0.6--0.8 & 0.708 & 88  & $115_{-7}^{+7}$   & $177_{-31}^{+66}$  & $0.0087_{-0.0011}^{+0.0012}$ & $1.17_{-0.08}^{+0.09}$ & $2.3_{-0.7}^{+1.0}$ & $2.3_{-0.4}^{+0.6}$ \\
0.8--1.0 & 0.851 & 56  & $83_{-6}^{+6}$    & $133_{-30}^{+60}$  & $0.0028_{-0.0005}^{+0.0006}$ & $1.32_{-0.12}^{+0.15}$ & $5.3_{-2.2}^{+2.8}$ & $5.3_{-1.3}^{+1.8}$ \\
1.0--1.3 & 1.155 & 23  & $26_{-5}^{+4}$    & $44_{-12}^{+21}$   & $0.0006_{-0.0002}^{+0.0002}$ & $1.38_{-0.20}^{+0.28}$ & $3.8_{-2.7}^{+2.7}$ & $3.8_{-1.5}^{+2.7}$ \\
\hline
\end{tabular}
\endgroup

\smallskip
\begin{flushleft}
\small \textbf{Note:} Columns list effective redshift ($\langle z \rangle_i$), raw count ($N_{\rm raw}$), MC count ($N_{\rm MC}$), corrected count ($N_{\rm CC}$), detection efficiency ($\epsilon_{\rm detect}$), classification correction factor ($\epsilon_{\rm class}$), deterministic volumetric rates $R_{\rm CC}$ (Det), and MC integrated volumetric rates $R_{\rm CC}$ (MC) in units of $10^{-4}$~yr$^{-1}$~Mpc$^{-3}$. In the lower two panels, for H-rich and SE SNe, $N_{\rm MC}$ and $N_{\rm CC}$ have been calibration corrected using \citet{Grayling2023}, and all subsequent efficiency corrections and asymmetric error bars are derived and propagated using the five-component quadrature framework detailed in Section~\ref{sec:uncertainties} Subtype rates ($R_{\mathrm{H\text{-}rich}}$ and $R_{\mathrm{SE}}$) are evaluated across broader redshift intervals ($\Delta z = 0.2$) by combining fine-bin MC realizations using subtype-specific re-evaluations of $\epsilon_{\mathrm{detect}}$ and $\epsilon_{\mathrm{class}}$. We perform a rate addition consistency check comparing the sum of the individual subtype rates ($R_{\mathrm{H\text{-}rich}} + R_{\mathrm{SE}}$) to the total CC SN rate ($R_{\mathrm{CC}}$); minor residual differences are expected due to volume-weighted averaging and non-linear propagation of asymmetric uncertainties across combined bins.
\end{flushleft}
\end{table*}

Subsequently, at the rate-calculation stage, we apply a redshift-specific classification correction factor, $\epsilon_{\mathrm{class}}(z_i)$, following Equation~\ref{eq:Rcc_integral}. This factor is defined as the ratio of simulated SNe classified as CC~SNe to the total number of true CC~SNe simulated. While the number-count calibration (constant at 1.33) increases the numerator of the rate equation, the application of $\epsilon_{\mathrm{class}}(z_i)$ in the denominator modulates the final rate. Depending on the simulated classifier performance in each bin, this factor can either amplify or offset the initial calibration. The interaction between the physical count calibration and this simulation-derived correction results in a net rate multiplier of $N_{\rm SE}$ ranging from 0.99 to 1.38. This range represents the cumulative effect of our calibration and the classifier's efficiency as characterized through our simulated samples across the survey's redshift range.

Table~\ref{tab:CC_SN_rates_redshift} (bottom two panels) and Figure~\ref{fig:SE_rates} show the resulting subtype rates in the six-bin configuration for both the deterministic (left panels) and MC (right panels) pipelines. In Figure~\ref{fig:SE_rates}, the H-rich (red) and SE (blue) volumetric CC~SN rates are plotted alongside the SE fraction (green points) as a function of redshift. The horizontal error bars indicate the redshift bin widths. 

A linear model of the form $f_{\mathrm{SE}}(z) = m z + c$ is fit to the SE fraction using asymmetric $\chi^2$ minimization. For the deterministic pipeline, this yields $m = -0.03 \pm 0.13$ and $c = 0.31 \pm 0.08$, with a reduced $\chi^2$ of $\chi^2_\nu = 0.52$ ($\chi^2 = 2.07$ for 4 degrees of freedom) and a $p$-value of $0.72$. An independent Markov Chain Monte Carlo (MCMC) analysis gives consistent results, with median posterior values $m = -0.04 \pm 0.13$ and $c = 0.31_{-0.07}^{+0.08}$. ~For the fiducial MC pipeline, the asymmetric $\chi^2$ minimization gives $m = -0.01 \pm 0.09$ and $c = 0.31 \pm 0.05$ ($\chi^2_\nu = 1.43$, $p\text{-value} = 0.22$), while the MCMC analysis yields $m = -0.00 \pm 0.09$ and $c = 0.30 \pm 0.05$. In Figure~\ref{fig:SE_rates} (c) - (d) the red dashed line shows the asymmetric $\chi^2$ best-fit, the grey solid line indicates the MCMC median, and the grey shaded region represents the 16th--84th percentile credible interval.The fiducial MC fitting approach implies a local SE fraction of $31 \pm 5$ per cent. The fitted slope is consistent with zero at the $0.1\sigma$ level, indicating no statistically significant evolution of the SE fraction with redshift across either analysis framework. This conclusion is robust to the choice of redshift binning, and is similarly apparent in the deterministic fitting where the slope is consistent with zero at the $0.2\sigma$ level.

The inferred local SE fraction of $31 \pm 5$ per cent is higher than the lowest literature estimates of 22--25 per cent reported by \citet{Cappellaro1999} and \citet{Prieto2008}, neither of which included Type~IIb SNe in their SE samples, and also higher than the recent ASAS-SN value reported by \citet{Pessi2025} of 24--25~per~cent. It is likewise above the 26--29 per cent fractions found by \citet{Frohmaier2020} and \citet{Schulze2021} using PTF data. However, our result is in excellent agreement with two independent sets of intermediate literature values: the 30--35 per cent fractions reported by \citet{Smartt2009} and \citet{Perley2020} and the 30--34 per cent fractions obtained by \citet{Graur2017b} and \citet{Shivvers2017} from re-analyses of the LOSS sample. The higher early LOSS estimates of 36--37 per cent \citep{Li2011b, Smith2011} were previously suggested to be affected by misclassification and incomplete spectroscopic coverage.

Almost all published SE fraction measurements are based on surveys confined to the local Universe ($z < 0.1$), with only \citet{Schulze2021} extending to slightly higher redshifts ($z < 0.3$). In contrast, the DES sample spans $0 < z < 1.3$, allowing the SE fraction to be measured in multiple independent redshift bins and demonstrating no significant evolution across this range (Figure~\ref{fig:SE_rates}).

\setlength{\tabcolsep}{4pt}
\renewcommand{\arraystretch}{1.15}

\begin{table}
\centering
{\small
\caption{Volumetric CC~SN rates from the literature and this work.} 
\label{tab:CC_SN_rates}
\renewcommand{\arraystretch}{1.3}
\begin{tabular}{@{}ccc P{3.2cm}@{}}
\hline
Redshift & $R_{\mathrm{CC}}$(MC) & Extinction & Source \\
 &  & correction & \\
\hline
0.0000 & $0.6^{+0.2}_{-0.2}$ & No  & Li et al. (2011) \\
0.0026 & $1.1^{+0.4}_{-0.3}$ & No  & Botticella et al. (2012) \\
0.0035 & $1.5^{+0.4}_{-0.3}$ & Yes & Mattila et al. (2012) \\
0.0100 & $0.4^{+0.2}_{-0.2}$ & No  & Cappellaro et al. (1999) \\
0.0149 & $0.7^{+0.1}_{-0.1}$ & No  & Pessi et al. (2025) \\
0.0280 & $0.9^{+0.2}_{-0.1}$ & No  & Frohmaier et al. (2020) \\
0.0720 & $1.1^{+0.2}_{-0.2}$ & No  & Taylor et al. (2014) \\
0.0750 & $1.0^{+0.3}_{-0.3}$ & No  & Graur et al. (2015) \\
\textbf{0.0800} & \textbf{2.4}\bm{$^{+0.9}_{-0.6}$} & \textbf{Yes} & \textbf{This work} \\
0.1000 & $1.1^{+0.0}_{-0.0}$ & No  & Cappellaro et al. (2015) \\
\textbf{0.1606} & \textbf{2.7}\bm{$^{+0.6}_{-0.5}$} & \textbf{Yes} & \textbf{This work} \\
0.2100 & $1.2^{+0.6}_{-0.5}$ & No  & Botticella et al. (2008) \\
\textbf{0.2563} & \textbf{3.0}\bm{$^{+0.6}_{-0.6}$} & \textbf{Yes} & \textbf{This work} \\
0.2600 & $1.9^{+0.7}_{-0.6}$ & No  & Cappellaro et al. (2005) \\
0.3000 & $1.4^{+0.5}_{-0.4}$ & No  & Bazin et al. (2009) \\
0.3000 & $2.1^{+0.8}_{-0.5}$ & No  & Strolger et al. (2015) \\
\textbf{0.3544} & \textbf{3.9}\bm{$^{+0.8}_{-0.7}$} & \textbf{Yes} & \textbf{This work} \\
0.3900 & $3.0^{+1.6}_{-1.1}$ & Yes & Dahlen et al. (2012) \\
0.3900 & $3.3^{+3.7}_{-2.3}$ & Yes & Melinder et al. (2012) \\
\textbf{0.4533} & \textbf{4.0}\bm{$^{+0.9}_{-0.8}$} & \textbf{Yes} & \textbf{This work} \\
\textbf{0.5525} & \textbf{4.5}\bm{$^{+1.0}_{-0.9}$} & \textbf{Yes} & \textbf{This work} \\
\textbf{0.6521} & \textbf{5.9}\bm{$^{+1.5}_{-1.4}$} & \textbf{Yes} & \textbf{This work} \\
0.6600 & $6.9^{+9.9}_{-5.4}$ & Yes & Graur et al. (2011) \\
0.7000 & $3.9^{+1.0}_{-0.7}$ & No  & Strolger et al. (2015) \\
0.7300 & $6.4^{+6.4}_{-3.8}$ & Yes & Melinder et al. (2012) \\
0.7300 & $7.4^{+3.7}_{-2.2}$ & Yes & Dahlen et al. (2012) \\
\textbf{0.7517} & \textbf{7.6}\bm{$^{+2.1}_{-1.9}$} & \textbf{Yes} & \textbf{This work} \\
\textbf{0.8515} & \textbf{11.4}\bm{$^{+3.7}_{-3.1}$} & \textbf{Yes} & \textbf{This work} \\
\textbf{0.9512} & \textbf{13.6}\bm{$^{+5.2}_{-4.1}$} & \textbf{Yes} & \textbf{This work} \\
1.1000 & $3.1^{+1.1}_{-0.7}$ & No  & Strolger et al. (2015) \\
1.1100 & $9.6^{+6.2}_{-4.0}$ & Yes & Dahlen et al. (2012) \\
\textbf{1.1550} & \textbf{16.1}\bm{$^{+6.0}_{-4.8}$} & \textbf{Yes} & \textbf{This work} \\
1.2250 & $4.4^{+1.8}_{-1.3}$ & No  & DeCoursey et al. (2026) \\
1.5000 & $3.3^{+1.0}_{-1.3}$ & No  & Strolger et al. (2015) \\
1.7800 & $5.9^{+2.1}_{-1.6}$ & No  & DeCoursey et al. (2026) \\
1.9000 & $3.2^{+3.4}_{-1.8}$ & No  & Strolger et al. (2015) \\
2.3000 & $6.2^{+6.8}_{-3.5}$ & No  & Strolger et al. (2015) \\
2.4200 & $6.2^{+2.2}_{-1.7}$ & No  & DeCoursey et al. (2026) \\
3.9200 & $4.1^{+1.5}_{-1.1}$ & No  & DeCoursey et al. (2026) \\
\hline
\end{tabular}
}
\smallskip
\begin{flushleft}
\footnotesize \textbf{Note:}
 Columns list the effective redshift $\langle z \rangle_i$, the volumetric rate $R_{\mathrm{CC}}$ (in units of $10^{-4}\,\mathrm{yr}^{-1}\,\mathrm{Mpc}^{-3}$), whether or not the rate was corrected for dust extinction, and the source of the measurement.
\end{flushleft}
\end{table}

\section{Discussion}
\label{sec:discussion}
\subsection{CC SN rates as tracers of the cosmic star formation history}

\label{sec:ccsn_csfh_tracers}

CC SNe provide a direct observational probe of the CSFH. Their progenitor stars, with initial masses typically in the range $8$--$40~M_\odot$, have lifetimes shorter than $10^8$~yr, which is negligible on cosmological timescales. As a result, the volumetric CC~SN rate, $R_{\mathrm{CC}}(z)$, closely traces the instantaneous star formation rate density, $\psi(z)$. The two quantities are related through a proportionality constant, $k$, commonly referred to as the yield:

\begin{equation}
R_{\mathrm{CC}}(z) = k \, \psi(z),
\label{eq:rcc}
\end{equation}

where $k$ (in units of $M_\odot^{-1}$) represents the number of CC~SNe produced per unit stellar mass formed. The yield depends directly on the fraction of stars formed within the mass range capable of undergoing core collapse relative to the total stellar mass formed, governed by the initial mass function (IMF; $\phi(M)$). The detailed mathematical formulation of $k$ across different IMF choices and progenitor mass windows is provided in Appendix~\ref{app:csfh_math}.

To interpret our fiducial DES CC~SN rates (derived using the hybrid MC framework described in Section~\ref{sec:mc_error_propagation}), four widely used empirical or semi-empirical CSFH models were adopted for comparison: \citet{Madau2014}, \citet{Madau2017}, \citet{Harikane2022}, and \citet{Kim2024}. These models span a range of observational tracers, redshift coverage, and IMF assumptions (see Appendix~\ref{app:csfh_math} for their explicit analytical forms and theoretical yields).

\begin{figure*}
    \centering
    \includegraphics[width=0.98\linewidth]{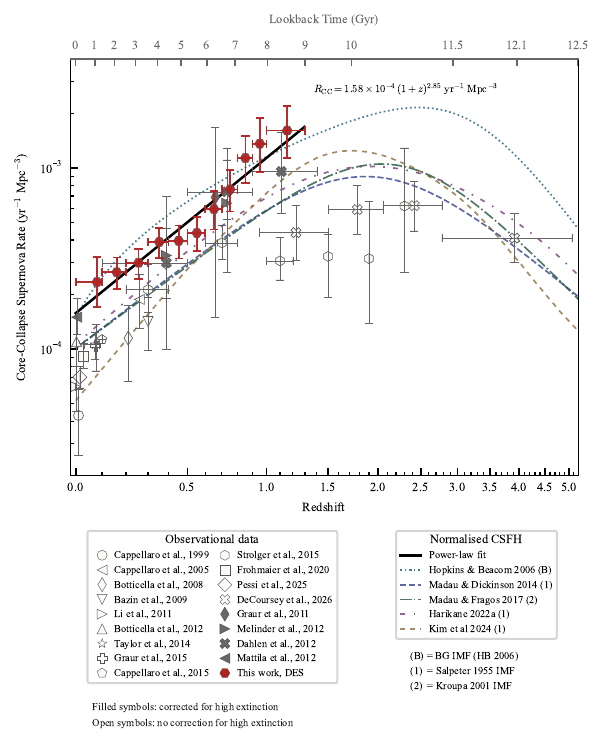}
    \vspace{-12pt}
    \caption{Comparison of DES CC~SN rates with previous observations and CSFH models using each model's IMF assumptions and assuming a progenitor range of $8$--$40~M_\odot$ and an IMF stellar mass range of $0.1$--$100~M_\odot$ (horizontal bars denote the redshift interval). The $x$-axis is scaled by $\log(1+z)$, providing a linear mapping to the logarithmic cosmic scale factor, $\log(a^{-1})$. There are two main reasons for the use of this scale: (i) it linearizes the power-law evolution of the rate, making the trend in our data easier to interpret as a straight line, and (ii) it stretches the low-redshift regime to prevent data crowding, while allowing the x-axis to extend to $z = 5.2$. This extended range is necessary so that the high-redshift results from \citet{Strolger2015} and \citet{2026arXiv260712018D} can be shown, and to visualize the expected turnover in the normalized CSFH models as they turn over beyond cosmic dawn.}
    \label{fig:salpeter}
\end{figure*}

Figure~\ref{fig:salpeter} shows these CSFH models scaled to the DES CC~SN rates using Equation~\ref{eq:rcc}, with the CC~SN yield, $k$, derived under the native IMF assumptions of each model. It is evident from Figure~\ref{fig:salpeter} that all four of the recent scaled CSFH models lie systematically below the empirical DES CC~SN rates and other literature measurements that have been corrected for high-extinction losses. The older \citet{Hopkins2006} CSFH, parameterized according to \citet{Cole2001} and scaled to a \citet{Baldry2003} IMF (with the shallower $-2.15$ high-mass slope, shown as the dotted blue curve in Figure~\ref{fig:salpeter}), lies slightly above most of the extinction-corrected rates. Of the four recent models evaluated, only the \textit{JWST}-constrained framework of \citet{Harikane2022} is consistent with the baseline DES CC~SN rates at approximately the $3\sigma$ level.

In their review, \citet{Madau2014} presented in their figure 10 both extinction-corrected \citep{Dahlen2012, Mattila2012} and uncorrected \citep{Botticella2008, Bazin2009, Li2011a} CC~SN rates. The extinction-corrected rates exceed by up to $\sim 70$~per~cent predictions based on their CSFH scaled using Equation~\ref{eq:rcc} with a Salpeter IMF derived yield. This is consistent with the offsets observed in this work.

Conversely, in \citet[figure~3]{Melinder2012} the \citet{Hopkins2006} CSFH (\citealt{Baldry2003} IMF with $\alpha = 2.15$), needed to be slightly scaled down in order to provide a good fit to extinction-corrected CC~SN rates. This is consistent with our Figure~\ref{fig:salpeter}, where the same \citet{Hopkins2006} CSFH lies slightly above both our DES measurements and the other extinction-corrected rates in the literature \citep{Graur2011, Dahlen2012, Mattila2012, Melinder2012}.

The systematic shortfall in rates predicted by recent CSFH models and our measurements is most naturally explained by structural limitations in the independent CSFH estimates themselves, which are derived from integrations of galaxy luminosity functions traced by UV, IR, and emission-line indicators rather than calibrated directly via stellar counting metrics. In this context, the discrepancy may reflect an incomplete census of star formation in the underlying galaxy models, particularly if a substantial population of underluminous, low-mass galaxies is missed by conventional imaging pipelines \citep[e.g.,][]{Boogaard2018A&A619, Kim2023, tacchella2024starformationcosmicdawngalaxies, Mosleh2024, Lapasia2026}. Such systems could contribute significantly to the total CC~SN rate while remaining faint or undetected in luminosity-based CSFH measurements, especially at higher redshift.

Alternatively, the discrepancy could be driven by physical variations in the underlying stellar mass distribution, or by an incomplete census of star formation resulting from extreme dust attenuation. Top-heavy IMFs have been reported in local starbursts such as M82 \citep{Rieke1993}, in high-redshift systems including AzTEC-3 \citep{dwek2011} and dusty starbursts at $z \sim 2{-}3$ \citep{zhang2018}, as well as in massive star-forming regions such as 30~Doradus \citep{schneider2018}. These extreme dusty environments preferentially enhance the high-mass end of the mass function.

This $1.7\times$ scaling relative to the standard Salpeter baseline can therefore be understood in terms of observational completeness. If the fraction of obscured massive stars aligns with the missing CC~SN fraction estimated by \citet{Mattila2012}, an empirical dust correction factor rising from $\sim 1.3$ to $\sim 1.7$ over $0 < z < 1.3$ naturally reconciles the canonical yield with this higher value, bringing all four CSFH models into consistency with our definitive MC measurements within $1\sigma$.

This tension between rate-derived yields and recent CSFH models is also evident on local, Galactic scales. Standard star formation models, which rely on conventional scaling rules and a standard \citet{Salpeter1955} IMF, predict a Milky Way CC~SN rate of roughly 1.0 to 1.4 per century \citep{Robitaille2010, Chomiuk2011, Kennicutt2012}. In contrast, three independent observational techniques point to significantly higher Galactic rates: neutron star population synthesis models require $2.0_{-1.0}^{+1.9}$ CC~SNe per century to explain observed pulsar and magnetar birth rates \citep{Araujo2026}, dust-unbiased $^{26}\mathrm{Al}$ gamma-ray measurements estimate $1.9 \pm 0.4$ CC~SNe per century \citep{Diehl2006, Diehl2021}, and targeted surveys scaled by $B$-band luminosity yield $2.2 \pm 0.6$ CC SNe per century \citep{Li2011b}. Together, these higher independent estimates reinforce our finding that traditional SFHs with a standard Salpeter IMF significantly underestimate the CC~SN rate.

The choice of uncertainty framework directly impacts our tension with the four recent CSFH models. Summing errors quadratically in the deterministic framework yields a mean best-fit yield of $k_{\text{CC(Det)}} = 0.0129 \pm 0.0011~\text{M}_\odot^{-1}$ (Appendix~\ref{app:deterministic_rates}). The non-linear MC propagation shifts bounds downwards, producing a lower mean (fiducial) yield of $k_{\text{CC(MC)}} = 0.0114 \pm 0.0009~\text{M}_\odot^{-1}$ (or $0.0112 \pm 0.0009\,\text{M}_\odot^{-1}$ evaluated without empirical error floors; Appendix~\ref{app:no_ef_comparison}).

Instead of adjusting the IMF to fit our rates, an alternative approach could be to adjust the progenitor mass range in the context of a standard \citet{Salpeter1955} IMF. For this IMF, integrated over a total stellar mass range of $0.1\text{--}100\,\mathrm{M}_\odot$, the theoretical CC SN yield for canonical progenitor mass limits of $8\text{--}40\,\mathrm{M}_\odot$ is $0.0067~\mathrm{M}_\odot^{-1}$. To reconcile this baseline with our rate derived yield of  $0.0114~\mathrm{M}_\odot^{-1}$ purely by adjusting progenitor mass boundaries, the lower limit $M_{\mathrm{min}}$ would need to extend down to $\approx 5.6\,\mathrm{M}_\odot$. However, extending $M_{\mathrm{min}}$ below $\sim 8\,\mathrm{M}_\odot$ directly conflicts with empirical white dwarf (WD) initial--final mass relations and asymptotic giant branch (AGB) evolutionary limits \citep[e.g.,][]{Smartt2009, Cummings2018}. Furthermore, due to the steep slope of the Salpeter IMF ($\alpha = 2.35$), extending the upper mass cutoff $M_{\mathrm{max}} \to \infty$ yields an asymptotic maximum of only $0.0077~\mathrm{M}_\odot^{-1}$, demonstrating that high-mass progenitors alone cannot bridge the gap.

Adopting a narrower explosion window (e.g., a lower upper bound driven by direct collapse into black holes; \citealt{Smartt2009, Kochanek2020MNRAS4934945}) requires an extremely flat IMF slope of $\alpha \sim 2.00$, pushing the required top-heaviness well beyond established literature models such as \citet{Baldry2003}. Adjusting progenitor mass boundaries alone cannot solve this without violating low-mass AGB and WD limits or invoking an unphysical slope. 

Consequently, as our fiducial MC yield aligns very well with the theoretical yield of $k_{\mathrm{BG03(\alpha = 2.20)}} = 0.0116 ~\text{M}_\odot^{-1}$ derived using the well established \citet[Table 2]{Baldry2003} IMF parametrization across standard total ($0.1\text{--}100\,\mathrm{M}_\odot$) and progenitor ($8\text{--}40\,\mathrm{M}_\odot$) integration ranges, we adopt this framework as our benchmark reference for subsequent analysis in Section~\ref{sec:progenitors}.

Contextualizing the theoretical $k_{\mathrm{BG03}}$ yield alongside previous literature calculations highlights the impact of these IMF choices. Earlier studies \citep[e.g.,][]{Graur2011, Dahlen2012, Mattila2012, Melinder2012} calculated baseline theoretical yields (integrating over standard stellar mass ranges of $0.1\text{--}100 $ or $ 125\,\mathrm{M}_\odot$, and canonical progenitor mass limits of $8\text{--}40 $ or $  50\,\mathrm{M}_\odot$) of $k \approx 0.0067\text{--} 0.0069\,\mathrm{M}_\odot^{-1}$ under a standard \citet{Salpeter1955} IMF or $0.0088\text{--}0.0103\,\mathrm{M}_\odot^{-1}$ under a "diet-Salpeter" or \citet{Kroupa2001} IMF.  

More recently, fitting to dust-corrected JADES rate measurements, \citet{Vassallo2026} inferred an elevated empirical yield of $k_{\mathrm{CC}} \approx 0.0078\,\mathrm{M}_\odot^{-1}$ (up from an uncorrected $0.0050\,\mathrm{M}_\odot^{-1}$), which would require an extreme $8\text{--}120\,\mathrm{M}_\odot$ integration window under a Salpeter IMF that leaves no room for direct black hole collapse. By adopting the \citet{Baldry2003} high-mass slope ($-2.2$), our empirical rate result alleviates this tension naturally while keeping progenitor boundaries within physically grounded limits.

\begin{figure*}
    \centering
    \begin{subfigure}[t]{0.48\textwidth}
        \centering
        \includegraphics[width=\textwidth]{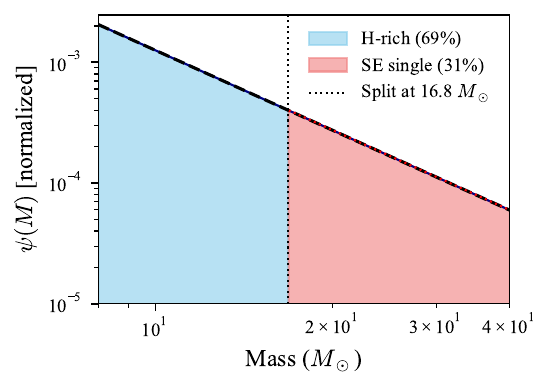}
        \caption{}
        \label{fig:mass_split_single}
    \end{subfigure}
    \hfill
    \begin{subfigure}[t]{0.48\textwidth}
        \centering
        \includegraphics[width=\textwidth]{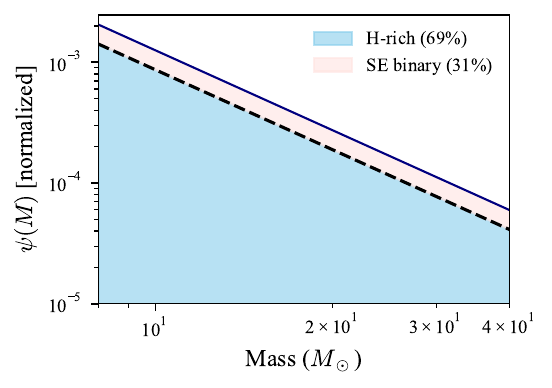}
        \caption{}
        \label{fig:mass_split_binary}
    \end{subfigure}

    \vspace{0.5em} 

    \begin{subfigure}[t]{0.48\textwidth}
        \centering
        \includegraphics[width=\textwidth]{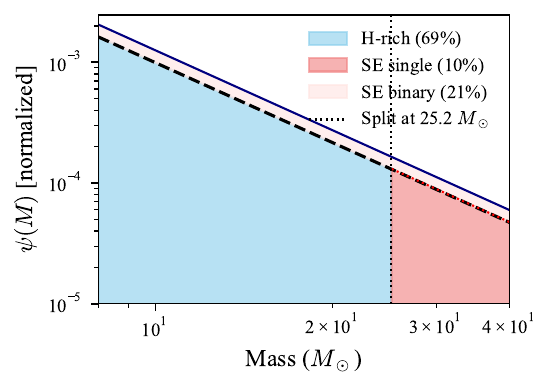}
        \caption{}
        \label{fig:mass_split_mixed}
    \end{subfigure}

    \caption{Subplots show different SE SN progenitor scenarios: (a) single, massive stars; (b) binary systems; (c) a combination of single stars and binaries. The area of the IMF that contain H-rich SN progenitors is coloured light blue; single-star SE SN progenitors in light red; and binary SE SN progenitors in light pink.}
    \label{fig:mass_split_combined}
\end{figure*}

Although the four-model mean yield fits to our rates differ slightly between deterministic and MC approaches due to uncertainty asymmetry, fitting a simple power law of the form $R_{\mathrm{CC}}(z) = R_0 (1 + z)^\alpha$ yields identical best-fit parameters across both framework choices: $R_0 = (1.58 \pm 0.17) \times 10^{-4}\,\mathrm{yr}^{-1}\,\mathrm{Mpc}^{-3}$ and $\alpha = 2.85 \pm 0.25$ for our primary All Fields baseline ($\chi^2_{\mathrm{red}} = 0.41$, $p = 0.932$, $\sigma_{\mathrm{eq}} = 0.08$). Evaluating this fit against literature measurements explicitly corrected for dust extinction \citep[e.g.,][]{Graur2011, Mattila2012, Dahlen2012, Melinder2012} demonstrates strong consistency, yielding a reduced $\chi^2$ of $0.184$ ($p = 0.989$, $0.014\sigma$ discrepancy). As expected, the rate rises continuously out to $z \approx 1.3$, with the expected cosmic noon turnover ($z \approx 2$) lying beyond our redshift baseline. Comprehensive field depth diagnostic tests confirming that selection completeness does not bias these power-law fits are detailed in Appendix~\ref{sec:field_depth} ~(Table~\ref{tab:field_comparison} and Figure~\ref{fig:des_ccsne_rates_app}). We retain the ``All Fields'' MC sample as our primary fiducial baseline because it maximizes survey volume, provides the tightest constraints, and mitigates cosmic variance across independent lines of sight.

\subsection{H-rich and SE~SNe progenitors}
\label{sec:progenitors}

The subtype rates derived in Section~\ref{sec:subtype_rates} provide additional constraints on the progenitor mass ranges of H-rich and SE~SNe. Despite the success in measuring cosmic rates, the specific nature of CC~SN progenitors remains a subject of significant debate. For H-rich Type II~SNe, the `Red Supergiant Problem' persists, where the observed maximum mass of progenitors appears systematically lower than theoretical predictions for stars that should undergo core collapse \citep{Smartt2009, smartt2015}. Furthermore, for stripped-envelope (SE)~SNe, it remains unclear whether the loss of the hydrogen envelope is driven primarily by extreme stellar winds in massive single stars or through Roche-lobe overflow in lower-mass binary systems \citep{Smith2011, Sana2012}. Determining the host galaxy properties and relative rates of these events is therefore critical for providing the environmental constraints needed to break these degeneracies. A summary of progenitor masses for Type~IIP/L~SNe (which constitute $\sim$80--90~per~cent of the H-rich population) based on direct imaging, theoretical modelling and population synthesis is presented in Table~\ref{tab:progenitor_masses}.

Following \citet{Smith2011}, and adopting the \citet{Baldry2003} IMF (with a high-mass slope of $\alpha = 2.20$) integrated over a total stellar mass range of $0.1\text{--}100\,\mathrm{M}_\odot$ and canonical CC~SN progenitor limits of $8\text{--}40\,\mathrm{M}_\odot$ ($k_{\mathrm{BG03(\alpha = 2.20)}}$), three progenitor scenarios were explored. In all three scenarios, we assume that the progenitors of H-rich SNe are exclusively single stars. SNe~IIn are included in the H-rich progenitor sample and may extend the H-rich SN progenitor mass range somewhat.

In the massive single-star progenitor scenario (e.g., \citealt{MaederLequeux1982, Heger2003,Georgy2012, Groh2013A&A558A131, Zapartas2021}), all SE~SNe are assumed to originate from stars initially more massive than \(20{-}25~\mathrm{M}_\odot\). In this framework, the removal of the hydrogen envelope,  and, in the case of SNe Ic, the subsequent stripping of helium occurs via strong stellar winds prior to core collapse. 

Under these assumptions, the resulting progenitor mass range for H-rich SNe lies between $\sim 8$ and $16.8~\mathrm{M}_\odot$, as shown in Figure~\ref{fig:mass_split_single}. These relationships are consistent with earlier population studies such as those by \citet{Smith2011}. The upper limit of the H-rich SN progenitor mass range found here is consistent with the sample of measured progenitor masses compiled by \citet{smartt2015}. However, \citet{Graur2017a} showed that the evolution of the fraction of mass-normalized H-rich to SE~SN rates as a function of metallicity was inconsistent with SE~SNe originating solely in the explosions of massive, single stars. This progenitor scenario is also inconsistent with our results in section~\ref{sec:subtype_rates}, where we show that the fraction of SE SNe out of all CC SNe remains constant from z = 0 to 1.3. The winds of massive, single stars that are needed to strip their outer envelopes are metallicity-driven, so their strength--and hence the SE SN fraction--should increase as the average stellar metallicity increases with decreasing redshift (e.g., \citealt{Vink2001, Mokiem2007A&A473603}).

    \begin{table}
    \centering
    \caption{Summary of IIP/L progenitor mass estimates from different methods.}
    \label{tab:progenitor_masses}
    \setlength{\tabcolsep}{3pt}
    \renewcommand{\arraystretch}{1.2}
    \small
    \begin{tabular}{@{} l l p{3.8cm} @{}}
    \hline
    Method & Mass (M$_\odot$) & Notes \\
    \hline
    Direct imaging & 8--16.5$^a$ to 18.5$^b$ & Optical limits$^{1,2,3}$ \\
    Dust-corr. imaging & Up to $\sim$21 & Incl. extinction$^{4,5}$ \\
    Hydro. modelling & 13--20 & From light curves$^{6,7,8}$ \\
    Theor. (explodability) & Varies & Engine-based models$^{9}$ \\
    Pop. synthesis & 8--20 & IMF/fallback$^{10,11,12}$ \\
    Nebular spectra & 8--18 & Core mass from lines$^{10}$ \\
    \hline
    \end{tabular}
    
    \vspace{0.8em}
    \begin{minipage}{\columnwidth}
    \footnotesize
    \textbf{References:} 
    (1)~\citet{Smartt2009MNRAS3951409}; (2)~\citet{Smartt2009}; (3)~\citet{smartt2015}; 
    (4)~\citet{davies2018}; (5)~\citet{DaviesBeasor2020MNRAS493}; (6)~\citet{morozova2015}; 
    (7)~\citet{Bose2015ApJ806160}; (8)~\citet{Martinez2020A_A642A143}; (9)~\citet{sukhbold2016}; 
    (10)~\citet{zapartas2017}; (11)~\citet{Eldridge2018PASA35e049}; (12)~\citet{Kochanek2020MNRAS4934945}
    
    \vspace{0.4em}
    $^a$ Upper limit derived from stellar luminosities by \citet{smartt2015} using STARS \citep{Eldridge2004} and Geneva \citep{Hirschi2004} stellar evolutionary models. \\
    $^b$ Upper limit derived from stellar luminosities by \citet{smartt2015} using the KEPLER \citep{WoosleyHeger2015} stellar evolutionary model.
    \end{minipage}
    \end{table}
An alternative interpretation posits that all SE~SNe originate in binary systems (e.g., \citealt{Vartanyan2021, Heger2003}), where mass transfer through Roche-lobe overflow or common-envelope evolution removes the outer hydrogen and helium layers. In this model, the stripping process does not require high stellar masses, and hence the progenitors are generally less massive than in the single-star scenario.

Assuming that all SE~SNe arise from binary interactions, the progenitor mass range for H-rich CC~SNe extends from \(8\) to \(40~\mathrm{M}_\odot\) (Figure~\ref{fig:mass_split_binary}). The model further assumes that binary progenitors are drawn proportionally from the IMF-weighted CC~SN progenitor distribution, such that a fixed fraction of events is assigned to the binary channel across the full mass range. Such binary-dominated channels (with the likely exception of SNe Ic) are supported by the population synthesis and observational results presented by \citet{Smith2011} and \citet{Graur2017a}. However, the high H-rich SN mass limit of $40~\mathrm{M}_\odot$
reignites the missing SN II progenitor problem highlighted by \citet{smartt2015}.

 A third scenario combines both progenitor scenarios \citep{Eldridge2013}. In the mixed progenitor model, roughly two-thirds of SE~SNe are attributed to binary interactions, while the remaining one-third originate from massive single stars. This hybrid approach is motivated by both observational and theoretical evidence suggesting that binary interactions dominate, but do not exclusively account for, envelope stripping among massive stars \citep{Sana2012, Smith2011}. 

Under this framework, the inferred progenitor mass range for H-rich SNe spans \(8\)--\(25.2~\mathrm{M}_\odot\) (Figure~\ref{fig:mass_split_mixed}). This composite scenario provides a flexible interpretation that reconciles the observed subtype fractions with expected stellar population demographics. By accounting for a significant binary contribution, this range accommodates the higher-mass red supergiant progenitors suggested by recent dust-correction studies \citep{DaviesBeasor2020MNRAS493} and `explodability' models \citep{sukhbold2016}, whilst avoiding the unrealistically high \(40~\mathrm{M}_\odot\) limit of the fully binary scenario. It is also consistent with the SN~II progenitor mass range found by \citet{smartt2015} (allowing for an SNe~IIn contribution and uncertainties) and with the SE~SN rates measured by \citet{Graur2017b}.

\section{Conclusions}
\label{sec:conclusions}

The five-year DES-SN survey has yielded one of the largest homogeneous samples of CC~SN candidates to date. Over the course of the survey, 31,636 difference-imaged light curves were recorded. Using a two-step, binary random-forest classification algorithm, we identified a sample of 3,169 CC~SNe. Following a human quality control and a precise check to verify these SNe were within the strict survey footprint used for rate calculation, a sample of 2,445 were retained. Redshift uncertainties were propagated using an MC approach, while detection efficiency and classification biases were rigorously quantified for each redshift bin. The resulting volumetric CC~SN rates, corrected for the estimated fraction of SNe missed due to high dust extinction, were measured in $\Delta z = 0.1$ 
intervals spanning $0 < z < 1$. Above $z = 1$, the data were grouped into a single interval spanning $1.0 < z < 1.3$.
This broader high-$z$ bin smooths out localized Poisson fluctuations 
in both the observed and simulated samples, thereby mitigating small-number 
statistics and providing a significantly more robust estimation of the 
detection efficiency. Within this range, the DES CC~SN rates are well fit by a power law of the form $R_{\mathrm{CC}}(z) = R_0 (1 + z)^{\alpha}$, with $R_0 = (1.58 \pm 0.17) \times 10^{-4}\,\mathrm{yr^{-1}\,Mpc^{-3}}$ and $\alpha = 2.85 \pm 0.25$. The expected turnover of the CC SN rates around cosmic noon ($z \approx 2$) is not yet apparent in our measurements, consistent with our redshift boundaries. 

The observed DES CC SN rates were then compared to predictions from five recent CSFH models \citep{Hopkins2006, Madau2014, Madau2017, Harikane2022, Kim2024}. The comparison assumed the same IMFs as used in each model, adopting a progenitor mass range of $8\text{--}40\,\text{M}_\odot$ and a stellar IMF range of $0.1\text{--}100\,\text{M}_\odot$.

The DES rates are systematically $\sim 70$\,per\,cent higher than the CC~SN rates predicted by the four most recent CSFH models. However, the DES measurements align with earlier extinction-corrected rate estimates and older CSFHs such as \citet{Hopkins2006}, which adopt more top-heavy IMF formulations. Adopting a \citet{Baldry2003} IMF rather than a \citet{Salpeter1955} IMF fully reconciles this discrepancy. Alternatively, the higher observed rates may imply that current CSFH models miss a population of low-luminosity galaxies.

We note that in the past some CSFH works compared their models against a combination of extinction-corrected and uncorrected CC~SN rates, which gave an impression of agreement between the models and observations that does not hold under uniform treatment (e.g., \citealt{Madau2014}). Going forward, we suggest that CSFH models should only be compared against fully extinction-corrected CC~SN rates, such as those presented here.

A further subclassification of the DES CC SN sample into distinct H-rich and SE~SNe populations revealed a local SE~SN fraction of $\sim 31 \pm 5$~per~cent, consistent with recent literature values (e.g., \citealt{Graur2017b, Shivvers2017, Perley2020}). Within the uncertainties, this fraction appears to remain constant across the redshift range. An analysis of the single-star and binary SE SN progenitor scenarios, in light of our CC SN rates, favours a combination of both models. The constant fraction of SE SNe across $z = 0\text{--}1.3$ provides further evidence against a single, massive progenitor scenario for SE SNe, as the strength of the metallicity-dependent stellar winds required to strip the progenitors of hydrogen and helium should rise with decreasing redshift.

\section*{Acknowledgments}

We thank Rick Kessler for his comments and suggestions.
Funding for the DES Projects has been provided by the U.S. Department of Energy, the U.S. National Science Foundation, the Ministry of Science and Education of Spain, the Science and Technology Facilities Council of the United Kingdom, the Higher Education Funding Council for England, the National Center for Supercomputing Applications at the University of Illinois at Urbana-Champaign, the Kavli Institute of Cosmological Physics at the University of Chicago, the Center for Cosmology and Astro-Particle Physics at the Ohio State University, the Mitchell Institute for Fundamental Physics and Astronomy at Texas A\&M University, Financiadora de Estudos e Projetos, Funda{\c c}{\~a}o Carlos Chagas Filho de Amparo {\`a} Pesquisa do Estado do Rio de Janeiro, Conselho Nacional de Desenvolvimento Cient{\'i}fico e Tecnol{\'o}gico and the Minist{\'e}rio da Ci{\^e}ncia, Tecnologia e Inova{\c c}{\~a}o, the Deutsche Forschungsgemeinschaft and the Collaborating Institutions in the Dark Energy Survey.

The Collaborating Institutions are Argonne National Laboratory, the University of California at Santa Cruz, the University of Cambridge, Centro de Investigaciones Energ{\'e}ticas, Medioambientales y Tecnol{\'o}gicas-Madrid, the University of Chicago, University College London, the DES-Brazil Consortium, the University of Edinburgh, the Eidgen{\"o}ssische Technische Hochschule (ETH) Z{\"u}rich, Fermi National Accelerator Laboratory, the University of Illinois at Urbana-Champaign, the Institut de Ci{\`e}ncies de l'Espai (IEEC/CSIC), the Institut de F{\'i}sica d'Altes Energies, Lawrence Berkeley National Laboratory, the Ludwig-Maximilians Universit{\"a}t M{\"u}nchen and the associated Excellence Cluster Universe, the University of Michigan, the National Optical Astronomy Observatory, the University of Nottingham, The Ohio State University, the University of Pennsylvania, the University of Portsmouth, SLAC National Accelerator Laboratory, Stanford University, the University of Sussex, Texas A\&M University, and the OzDES Membership Consortium.

Based in part on observations at Cerro Tololo Inter-American Observatory, National Optical Astronomy Observatory, which is operated by the Association of Universities for Research in Astronomy (AURA) under a cooperative agreement with the National Science Foundation. 

The DES data management system is supported by the National Science Foundation under Grant Numbers AST-1138766 and AST-1536171. The DES participants from Spanish institutions are partially supported by MICINN under grants ESP2017-89838, PGC2018-094773, PGC2018-102021, SEV-2016-0588, SEV-2016-0597, and MDM-2015-0509, some of which include ERDF funds from the European Union. IFAE is partially funded by the CERCA program of the Generalitat de Catalunya. Research leading to these results has received funding from the European Research Council under the European Union’s Seventh Framework Program (FP7/2007-2013) including ERC grant agreements 240672, 291329, and 306478. We acknowledge support from the Brazilian Instituto Nacional de Ci{\^e}ncia e Tecnologia (INCT) do e-Universo (CNPq grant 465376/2014-2).

This work is based in part on data acquired at the Anglo-Australian Telescope. We acknowledge the traditional owners of the land on which the AAT stands, the Gamilaraay people, and pay our respects to elders past and present.

\section*{Data Availability}

The data underlying this article are publicly available as part of the DES-SN Data Release v0.1, accessible at
\url{https://des-sn-dr.readthedocs.io/en/latest/}. 
This release includes transient light curves from the DES-SN survey, and analyses of SN Ia light curves for cosmology.
Simulations used were generated using SNANA. 
All data products and associated documentation are provided by the DES Collaboration and hosted by the National Center for Supercomputing Applications (NCSA) and the University of Chicago Research Computing Center (RCC).

\clearpage 
\appendix

\section*{\centering APPENDICES}


\section{Confusion Matrices}
\label{sec:appendix_A}

A confusion matrix summarizes classification performance by comparing true astrophysical classes against model predictions. The vertical axis represents the true labels while the horizontal axis indicates the predicted labels. Values along the primary (dark blue) diagonal denote correctly classified instances (true positives and true negatives), whereas off-diagonal (light blue) values represent misclassifications (false positives and false negatives). Overall classification accuracy is reported with each matrix, calculated as the fraction of total correct predictions relative to the total sample size in that redshift bin:
\begin{equation}
    \text{Accuracy} = \frac{\text{True Positives} + \text{True Negatives}}{\text{Total Sample Size}}.
\end{equation}

\begin{figure*}
    \centering
    \includegraphics[width=0.65\textwidth]{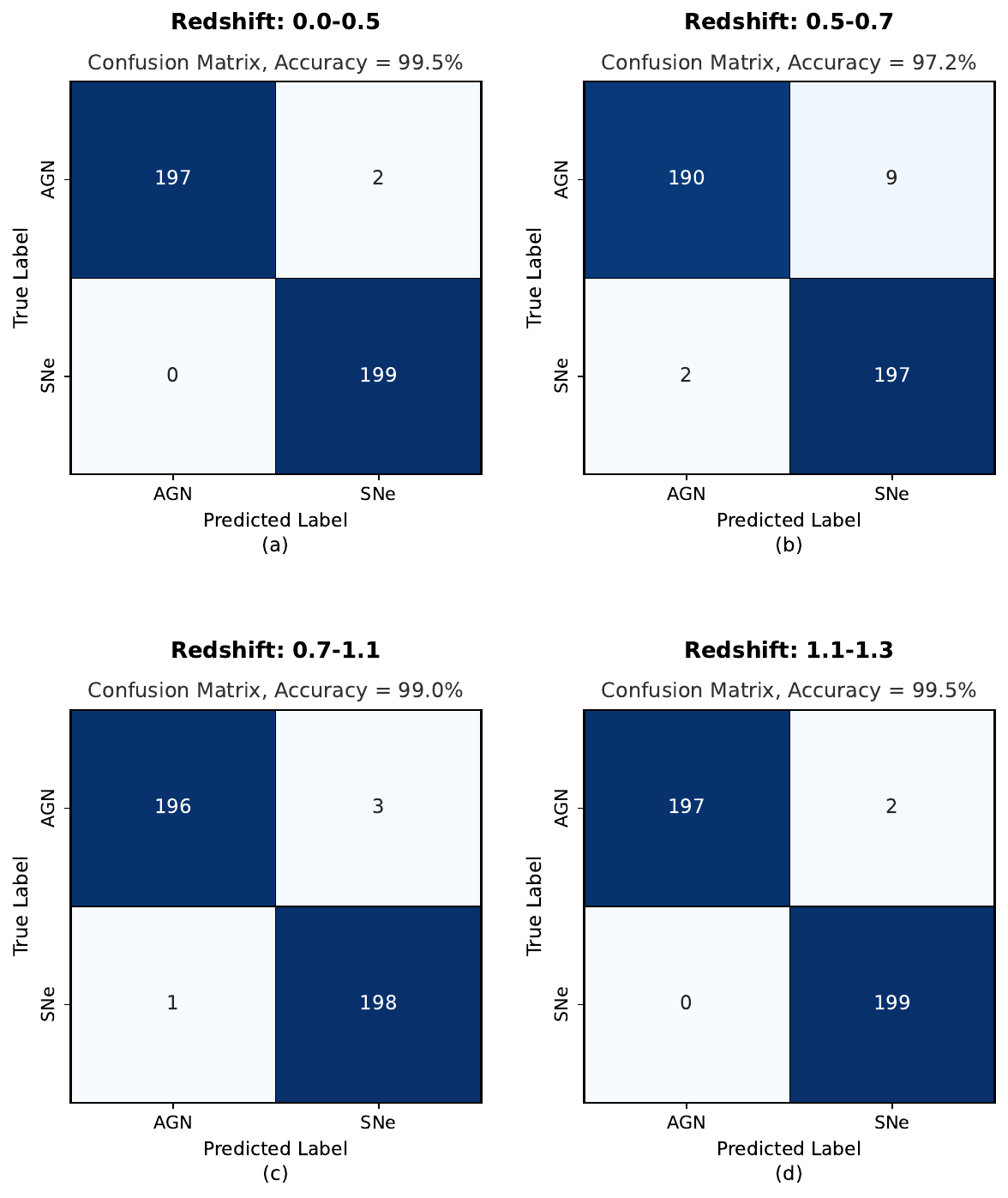}
    \caption{Four AGN redshift-specific confusion matrices and corresponding accuracy scores for our Stage~1 SN vs.\ AGN classification pipeline across intervals: (a) $0.0 \le z < 0.5$; (b) $0.5 \le z < 0.7$; (c) $0.7 \le z < 1.1$; and (d) $1.1 \le z < 1.3$.}
    \label{fig:stage1_cms}
\end{figure*}

Figure~\ref{fig:stage1_cms} presents the four AGN redshift-specific confusion matrices and corresponding accuracy scores for our Stage~1 SN vs.\ AGN classification pipeline.

\begin{figure*}
    \centering
    \includegraphics[width=0.98\textwidth]{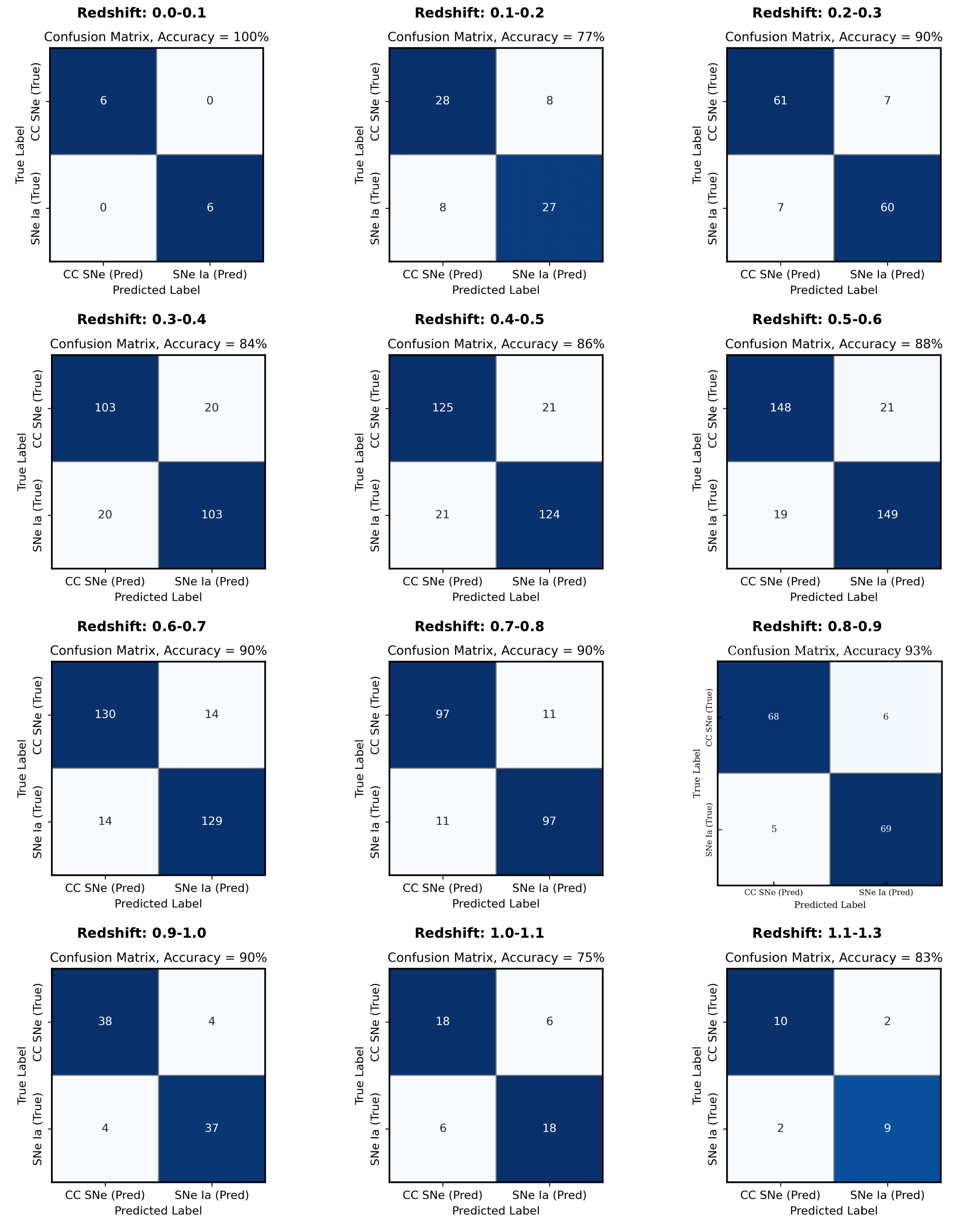}
    \caption{Twelve confusion matrices and corresponding accuracies for our Stage~2 CC~SNe vs.\ SNe~Ia classification pipeline, evaluated across individual $\Delta z = 0.1$ redshift intervals from $z = 0.0$ to $z = 1.3$ (except the last $z = 1.1 - 1.3$ interval which has a width of $z = 0.2$).}
    \label{fig:figA5}
\end{figure*}

Figure~\ref{fig:figA5} presents the 12 confusion matrices and corresponding accuracies for our Stage~2 CC~SN vs.\ SN~Ia classification pipeline, evaluated across individual $\Delta z = 0.1$ redshift intervals (except for the last $z = 1.1 - 1.3$ interval, which has a width of $z = 0.2$).

%
\section{Anderson-Darling Tests for Peak Average Absolute Magnitude}
\label{sec:appendix_ad_tests}

To evaluate the classification power of the single most influential feature of our stage 2 classifier—the four-band average peak absolute magnitude—we performed two-sample Anderson-Darling (AD) tests across 12 redshift bins. These tests assess how closely the peak magnitude distributions of simulated CC~SNe and simulated SNe~Ia match the distribution of photometrically classified, observed CC~SNe.

\begin{figure*}
    \centering
    \includegraphics[width=\textwidth]{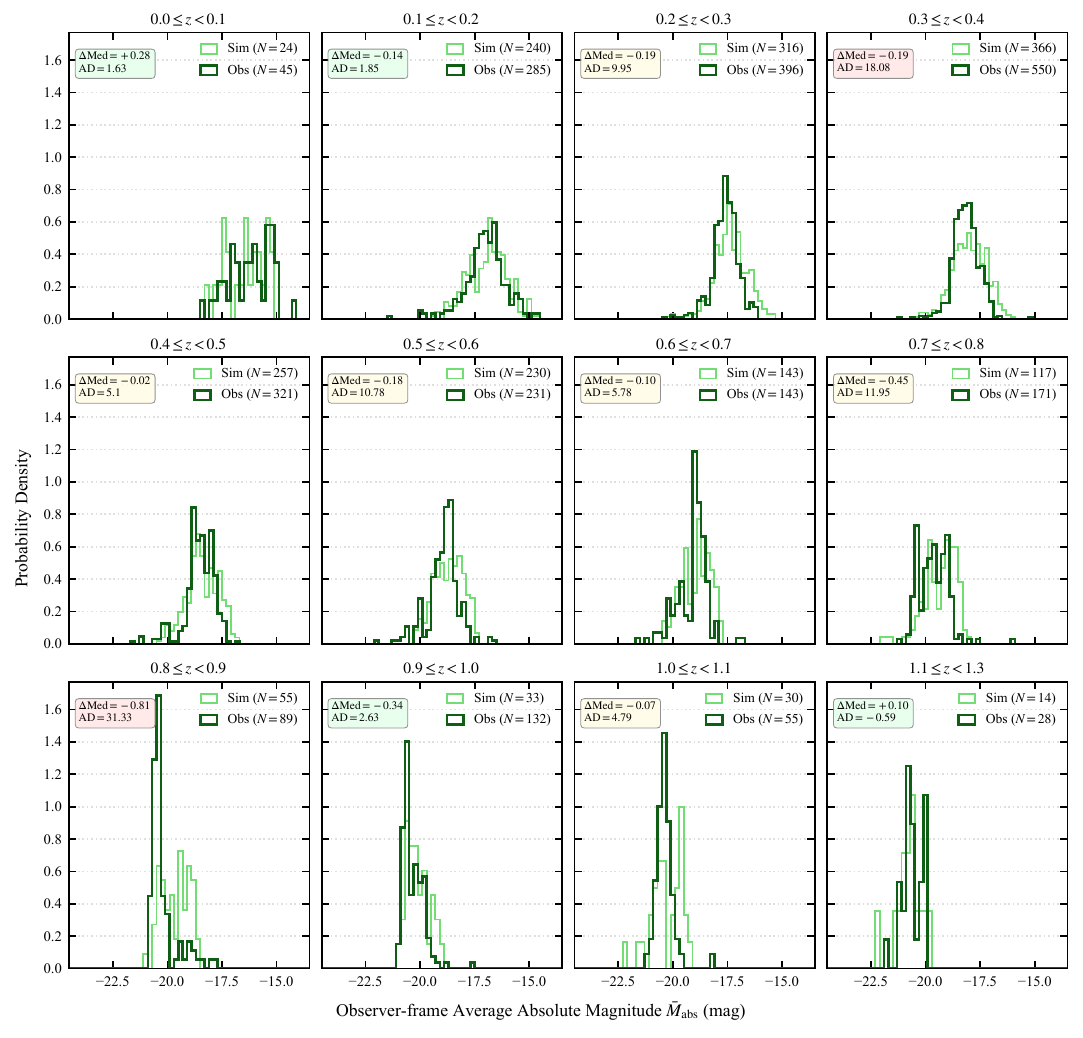}
    \caption{Anderson-Darling test results comparing the peak average absolute magnitude distributions of simulated CC~SNe against observed, photometrically classified CC~SNe across 12 redshift bins. In the top left box, Good agreement is shown in green, moderate agreement in yellow and poor agreement in pink. The $\Delta$Med statistic shows the observed minus simulated median magnitude, so a negative $\Delta$Med means that the observed value is brighter than the simulated one.}
    \label{fig:figC1}
\end{figure*}

Figure~\ref{fig:figC1} shows the comparison between simulated CC~SNe and observed, photometrically classified CC~SNe. The two distributions demonstrate strong statistical agreement, yielding a reasonably good fit ($\text{AD statistic} < 12$) in 10 out of 12 redshift bins, with 4 bins achieving exceptionally good agreement ($\text{AD statistic} < 3$). Only two redshift intervals ($z = 0.3\text{--}0.4$ and $z = 0.8\text{--}0.9$) exhibit elevated test statistics ($\text{AD} = 18$ and $31$, respectively), reflecting localized boundary noise and completeness variations rather than systematic classification failure.

\begin{figure*}
    \centering
    \includegraphics[width=\textwidth]{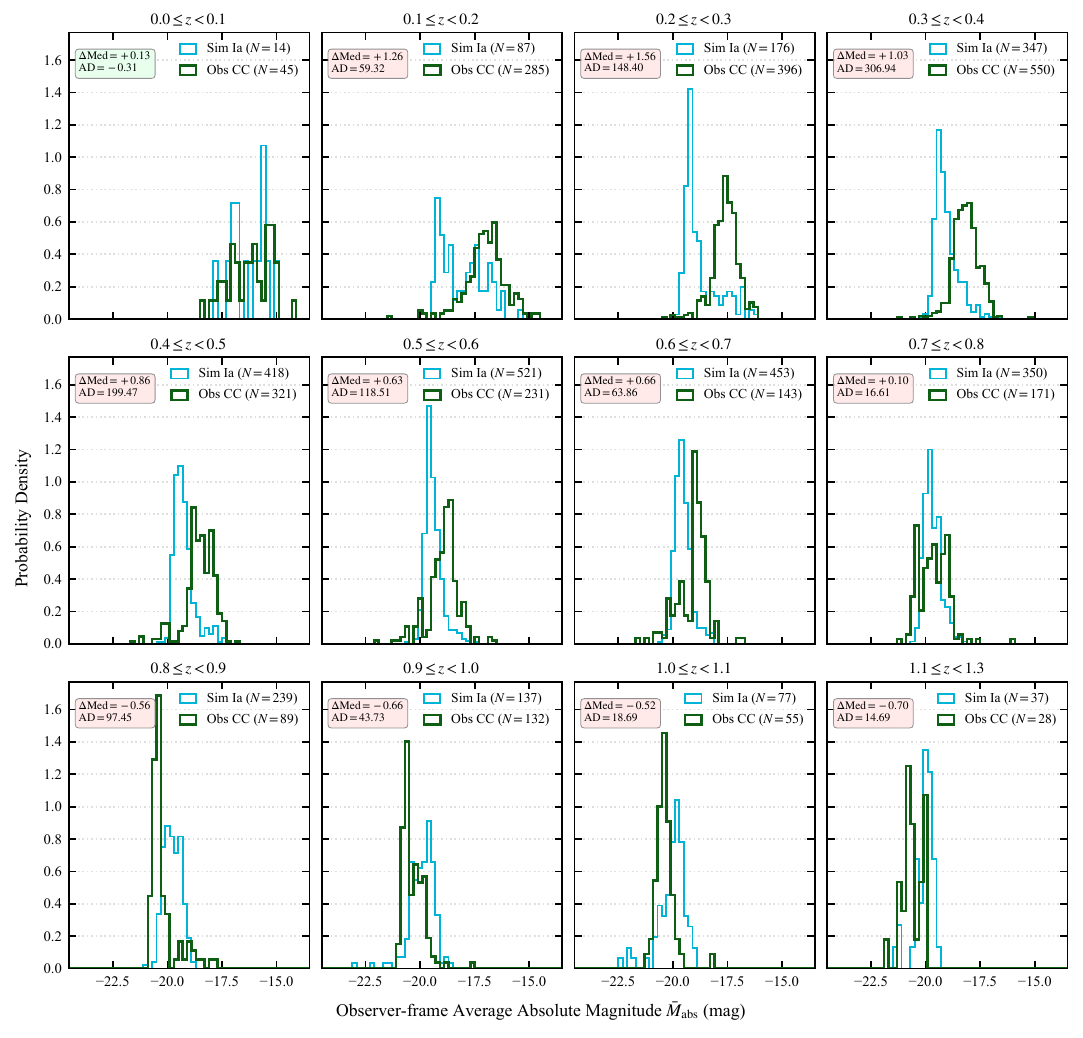}
    \caption{Anderson-Darling test results comparing the peak average absolute magnitude distributions of simulated SNe~Ia against observed, photometrically classified CC~SNe across 12 redshift bins. In the top left box, Good agreement is shown in green, moderate agreement in yellow and poor agreement in pink. The $\Delta$Med statistic shows the observed minus simulated median magnitude, so a negative $\Delta$Med means that the observed value is brighter than the simulated one.}
    \label{fig:figC2}
\end{figure*}

Conversely, Figure~\ref{fig:figC2} displays the corresponding AD test comparing simulated SNe~Ia to the observed CC~SN sample. In stark contrast to the CC~SN comparison, the SN~Ia distributions show clear statistical incompatibility: all bins except the lowest ($z = 0.0\text{--}0.1$) yield poor fits ($\text{AD statistic} > 12$), with 7 bins exhibiting severe mismatch ($\text{AD statistic} > 50$). Across all 12 redshift bins, the AD test statistic for simulated SNe~Ia is universally worse than that of simulated CC~SNe when compared against the observed CC~SN population. These test results confirm that the peak absolute magnitude feature provides robust, unambiguous separation between SNe~Ia and CC~SNe across the entire survey redshift range.

\section{Validation of PSNID-\texorpdfstring{$z$}{z} Reliability and Sample Unbiasedness}
\label{app:PSNID_independence}

Because a subset of CC~SNe in our sample relies solely on light-curve photometric redshifts (\texttt{PSNID-}z), we perform two diagnostic checks to ensure their inclusion introduces no systematic bias: (1) assessing \texttt{PSNID-}z accuracy under an unconstrained flat prior (\texttt{ZPRIOR=0}) against spectroscopic host redshifts (\texttt{spec-}z), and (2) verifying that including \texttt{PSNID-}z-only objects preserves the measured rate evolution.

CC~SNe that rely on \texttt{PSNID-}z as their redshift source necessarily adopt the mean of their \texttt{ZPRIOR=0} measurements as no other priors are available. To prevent external priors from biasing the \texttt{PSNID-}z uncertainty evaluation, the spectroscopic sample used must be explicitly filtered to retain strictly unconstrained \texttt{ZPRIOR=0} measurements (excluding \texttt{spec-}z and \texttt{photo-}z priors). Because our primary redshift hierarchy ($\text{spec-}z > \text{photo-}z > \text{PSNID-}z$) automatically prioritizes higher-quality constraints whenever present, this explicit filtering is only required for the uncertainty evaluation subset.

For CC~SNe with both spectroscopic and alternative redshift estimates, \texttt{photo-}z shows a strong correlation with \texttt{spec-}z: 337 out of 461 events ($73.1$ per~cent) lie within $\Delta z < 0.05$, and 407 out of 461 ($88.3$ per~cent) within $\Delta z < 0.1$. This corresponds to a Root Mean Square Error (RMSE, representing the standard deviation of the residuals) of $0.139$ and a Normalised Median Absolute Deviation ($\sigma_{\rm NMAD}$) of $0.026$. By contrast, \texttt{PSNID-}z exhibits a broader core scatter across the unselected sample, with 337 out of 1199 events ($31.2$ per~cent) within $\Delta z < 0.05$ and 613 out of 1199 ($51.1$ per~cent) within $\Delta z < 0.1$, yielding an RMSE of $0.254$ and $\sigma_{\mathrm{NMAD}} = 0.108$.

\begin{table}
    \centering
    \caption{\texttt{PSNID-}z (\texttt{ZPRIOR=0}) accuracy vs.\ host \texttt{spec-}z across SN template sub-types and probability-weighted combination ($N=1199$).}
    \label{tab:tabC1}
    \setlength{\tabcolsep}{5pt}
    \renewcommand{\arraystretch}{1.15}
    \small
    \begin{tabular}{cccccc}
        \hline
        \hline
        \textbf{Template / Metric} & \textbf{$R$} & \textbf{$\mu_{\Delta z}$} & \textbf{$\sigma_z$} & \textbf{$|\Delta z| \le 0.05$} & \textbf{$|\Delta z| \le 0.10$} \\
        \hline
        $z_{\mathrm{Ia}}$   & 0.34 & $+0.149$ & 0.31 & 23.4\% & 37.0\% \\
        $z_{\mathrm{Ib/c}}$ & 0.47 & $-0.144$ & 0.20 & 19.3\% & 38.8\% \\
        $z_{\mathrm{II}}$   & 0.37 & $+0.163$ & 0.38 & 21.1\% & 37.7\% \\
        \hline
        \textbf{Weighted}   & \textbf{0.49} & \textbf{$+0.041$} & \textbf{0.26} & \textbf{31.0\%} & \textbf{49.8\%} \\
        \hline
    \end{tabular}
\end{table}

\begin{figure}
    \centering
    \includegraphics[width=\columnwidth]{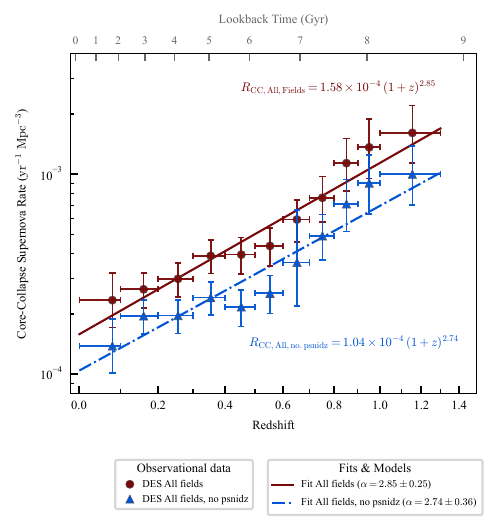}
    \caption{Volumetric CC~SN rate evolution for the full sample ($\alpha = 2.85 \pm 0.25$) versus the subset excluding \texttt{PSNID-}z-only events ($\alpha = 2.74 \pm 0.35$). Omitting \texttt{PSNID-}z events yields a fully consistent slope ($\Delta \alpha = 0.11$, $0.42\sigma$).}
    \label{fig:figC3}
\end{figure}

As summarized in Table~\ref{tab:tabC1}, \texttt{PSNID-}z performance across individual templates shows moderate correlations ($R = 0.34\text{--}0.47$) and template-dependent systematic offsets ($\mu_{\Delta z} = -0.144 \text{ to } +0.163$, $\sigma_z = 0.20\text{--}0.38$). Type~Ia and Type~II templates yield positive biases because lower-luminosity CC~SNe are shifted to higher redshifts to match model fluxes, whereas Type~Ib/c templates exhibit a negative bias as the fitter overestimates absolute luminosity and places events at lower redshifts.

The probability-weighted \texttt{PSNID-}z combines model likelihoods across all templates, substantially improving overall accuracy: $R$ increases to $0.49$, mean bias drops near zero ($\mu_{\Delta z} = +0.041$), and dispersion decreases to $\sigma_z = 0.26$, with $31.0$ per~cent and $49.8$ per~cent of fits falling within $|\Delta z| \le 0.05$ and $\le 0.10$, respectively.

To check for rate systematic bias, we compare the baseline rate against a filtered sample excluding all \texttt{PSNID-}z-only SNe ($37$ per~cent sample reduction; Figure~\ref{fig:figC3}). While removing these fainter, hostless, or high-redshift events reduces the overall rate normalization ($R_0$) by $34$ per~cent (from $1.58$ to $1.04 \times 10^{-4}\,\mathrm{yr}^{-1}\,\mathrm{Mpc}^{-3}$), the power-law evolutionary slope remains fully consistent ($\alpha_{\mathrm{baseline}} = 2.85 \pm 0.25$ vs.\ $\alpha_{\mathrm{no.\,\texttt{PSNID-}z}} = 2.74 \pm 0.35$). The slight slope difference ($\Delta \alpha = 0.11$) represents a negligible $0.42\sigma$ shift. This significance is evaluated using variance subtraction to properly account for covariance in nested samples ($\sigma_{\mathrm{diff}} = \sqrt{\sigma_{\mathrm{sub}}^2 - \sigma_{\mathrm{full}}^2} = \sqrt{0.35^2 - 0.25^2} = 0.259$). We conclude that including \texttt{PSNID-}z-only events introduces no systematic bias into the measured redshift evolution while expanding effective survey volume and statistical power.

\section{Impact of Redshift MC Uncertainty Floors}
\label{app:no_ef_comparison}

To evaluate the sensitivity of our results to empirical error floors in sparse boundary regimes, we perform a control MC resampling analysis without applying error floors. For transients with non-zero redshift uncertainty (\texttt{photo-}z or \texttt{PSNID-}z), MC redshift realizations are drawn from a skew-normal distribution:
\begin{equation}
z^{(i)} \sim \mathrm{SkewNormal}(\xi = z, \omega = \bar{\sigma}_z, \alpha),
\end{equation}
where $\bar{\sigma}_z = (\sigma_z^- + \sigma_z^+)/2$ is the mean uncertainty and the skewness parameter is defined as $\alpha = (\sigma_z^+ - \sigma_z^-)/\bar{\sigma}_z$. Spectroscopic redshifts are treated as fixed values with no uncertainties.

\begin{figure}
    \centering
    \includegraphics[width=\columnwidth]{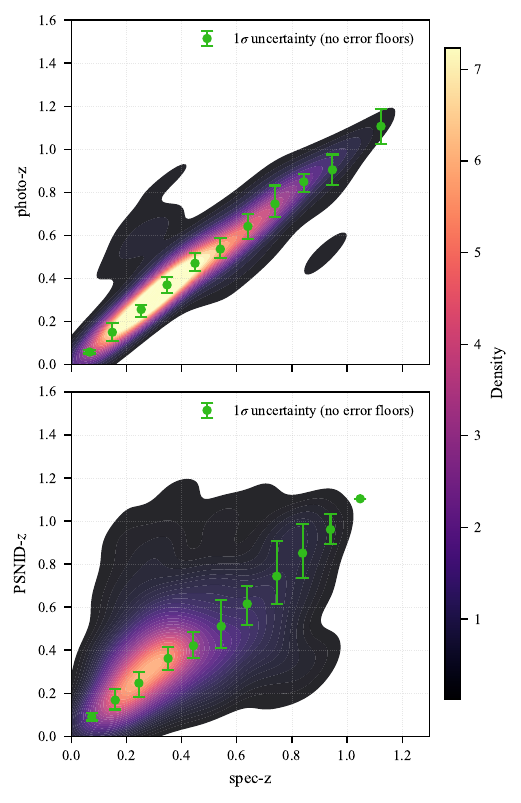}
    \caption{Estimated photometric redshifts (\texttt{photo-}z; top panel) and PSNID redshifts (\texttt{PSNID-}z; bottom panel) plotted against spectroscopic host redshifts (\texttt{spec-}z) for the control MC analysis without error floors. Green uncertainty bands illustrate the unfloored, non-monotonic uncertainty bounds at the high- and low-$z$ boundaries ($z < 0.1$ and $z > 0.8$), which stem from sparse data pairing \texttt{spec-}z with \texttt{photo-}z and unconstrained \texttt{PSNID-}z (\texttt{ZPRIOR=0}) during bin-level error estimation.}
    \label{fig:figC5}
\end{figure}

In contrast to our baseline analysis—where enforced redshift-bin error floors ($\sigma_{\text{photo-}z, \text{floor}} = 0.06\text{--}0.20$ and $\sigma_{\text{PSNID-}z, \text{floor}} = 0.09\text{--}0.25$ across boundary bins $z < 0.1$ and $z > 0.8$) produce smooth, monotonically expanding uncertainty contours as shown in Figure~\ref{fig:contour final}—the unfloored control case exhibits severe boundary distortions.

Both \texttt{photo-}z and \texttt{PSNID-}z uncertainty ranges are distorted at the redshift extremes due to:
\begin{itemize} 
    \item data sparsities: the low availability of \texttt{photo-}z and \texttt{PSNID-}z measurements at low redshift ($z < 0.1$) and limited \texttt{spec-}z reference objects at high redshift ($z > 0.8$), and
    \item inherently high intrinsic scatter in \texttt{PSNID-}z when evaluated against spectroscopic redshifts without external host priors. 
\end{itemize}
As shown in Figure~\ref{fig:figC5}, this combination of intrinsic scatter and small-sample variance in sparse boundary bins distorts error propagation if unconstrained. In the absence of enforced error floors, small-sample variance and localized bias uncertainties generate erratic, unphysically narrow uncertainty ranges in the lowest ($z \le 0.1$) and highest ($z > 0.8$) redshift regimes.

\begin{figure}
    \centering
    \includegraphics[width=\columnwidth]{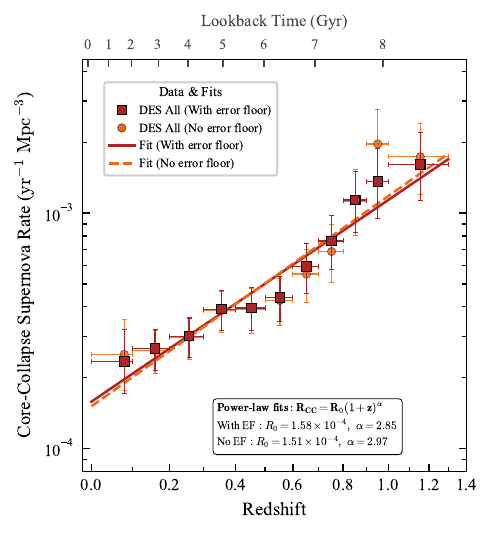}
    \caption{Volumetric core-collapse supernova rate ($R_{\mathrm{CC}}$) as a function of redshift ($z$) and lookback time, comparing the baseline DES sample with enforced error floors (dark red squares, solid line) to the control sample without error floors (orange circles, dashed line). Power-law fits of the form $R_{\mathrm{CC}}(z) = R_0 (1 + z)^\alpha$ yield consistent parameters between the two treatments: $R_0 = 1.58 \times 10^{-4}\,\mathrm{yr}^{-1}\,\mathrm{Mpc}^{-3}$ with $\alpha = 2.85$ (with error floor) versus $R_0 = 1.51 \times 10^{-4}\,\mathrm{yr}^{-1}\,\mathrm{Mpc}^{-3}$ with $\alpha = 2.97$ (no error floor). While omitting error floors introduces slight bin-level shifts in central values and uncertainties—particularly at boundary redshifts—the overall rate evolution remains robust.}
    \label{fig:ef_comparison}
\end{figure}


The omission of empirical error floors directly propagates these boundary distortions into the volumetric rate fits and subsequent yield determinations. Power-law fits of the form $R_{\mathrm{CC}}(z) = R_0 (1 + z)^\alpha$ shift slightly from the baseline values ($R_0 = 1.58 \times 10^{-4}\,\mathrm{Mpc}^{-3}\,\mathrm{yr}^{-1}$, $\alpha = 2.85$) to $R_0 = 1.51 \times 10^{-4}\,\mathrm{Mpc}^{-3}\,\mathrm{yr}^{-1}$ and $\alpha = 2.97$ in the unfloored case (Figure~\ref{fig:ef_comparison}). 

Removing the error floors causes only a marginal downward shift in the mean best-fit yield across the four CSFH models, from our fiducial MC baseline of $k_{\text{CC(MC)}} = 0.0114 \pm 0.0009~\text{M}_\odot^{-1}$ to $k_{\text{CC(MC, no\ EF)}} = 0.0112 \pm 0.0009~\text{M}_\odot^{-1}$. The choice to apply error floors therefore plays a minor role compared to the uncertainty framework itself; with the noticeable shift from the deterministic estimate ($k_{\text{CC(Det)}} = 0.0129 \pm 0.0011~\text{M}_\odot^{-1}$; Appendix~\ref{app:deterministic_rates}) being driven by how non-linear error propagation is naturally handled in the MC framework.

\section{Cross-Band \texorpdfstring{$K$}{K}-Correction Formalism and Linear Selection Approximation}
\label{sec:appendix_b5}

\begin{figure}
    \centering
    \includegraphics[width=\columnwidth]{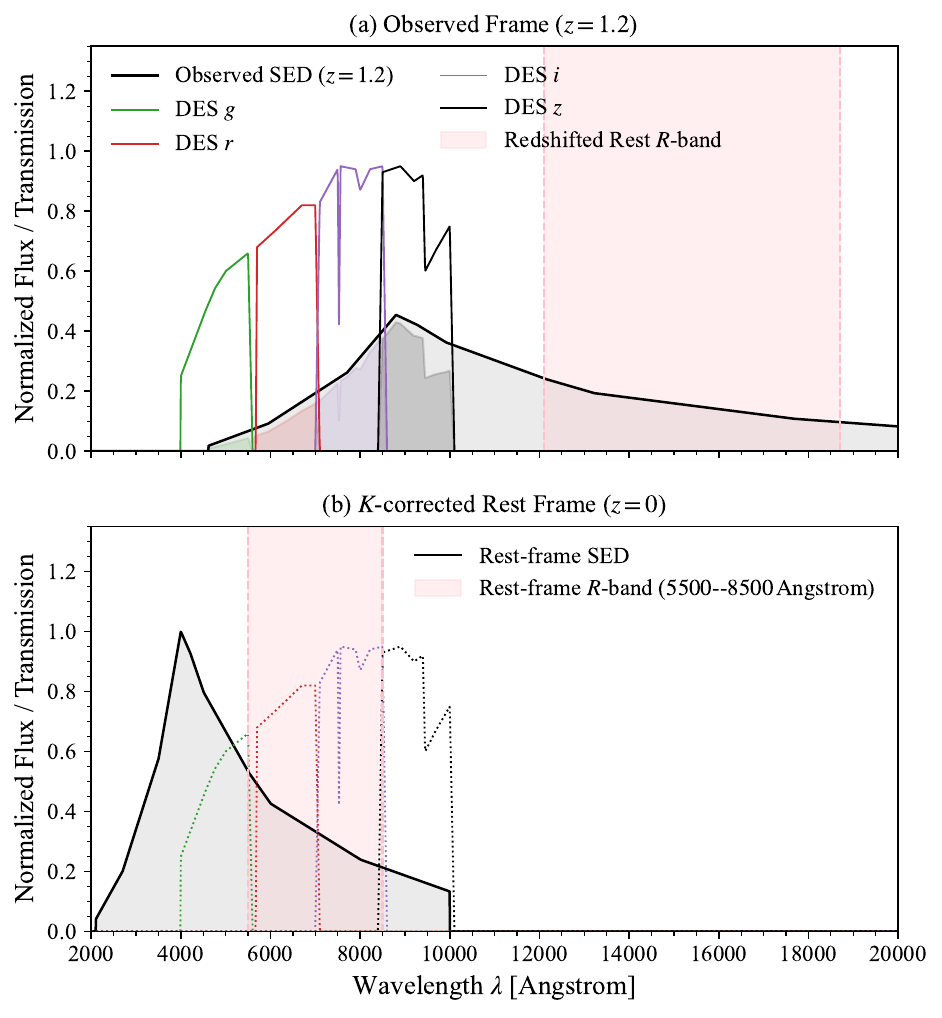}
    \caption{Cross-band $K$-correction schematic for DES optical observations at $z = 1.2$. (a) Redshifted Window: Rest-frame $R$-band ($5500\text{--}8500\,\text{\AA}$, pink) redshifts into near-infrared ($12,100\text{--}18,700\,\text{\AA}$), beyond DES optical coverage ($griz$). Synthetic DES $i$- and $z$-band fluxes sample rest-frame near-UV ($\sim 3200\text{--}4500\,\text{\AA}$). (b) Rest Frame: $R$-band baseline at $z=0$ defining luminosity parameters ($M_0, \sigma$).}
    \label{fig:figC4_schematic}
\end{figure}

Figure~\ref{fig:figC4_schematic} illustrates the cross-band $K$-correction mapping at $z = 1.2$, showing how observer-frame optical filters ($griz$) sample the rest-frame near-UV. 

\begin{figure}
    \centering
    \includegraphics[width=\columnwidth]{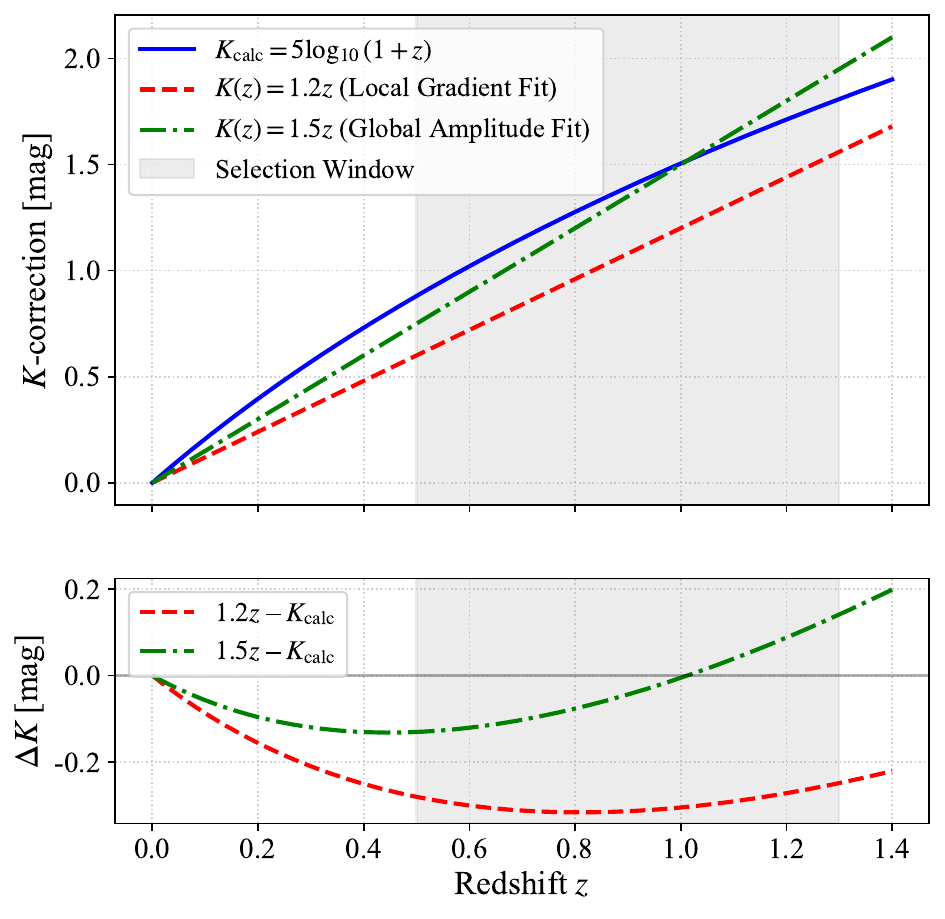}
    \caption{Comparison of cross-band $K$-correction parametrizations as a function of redshift. Top panel: Analytic cross-band $K$-correction $K_{\text{calc}} = 5\log_{10}(1+z)$ (solid blue curve) alongside the linear local gradient fit $K(z) = 1.2z$ (dashed red line) and the global amplitude fit $K(z) = 1.5z$ (dash-dotted green line). The shaded region indicates the primary survey selection window ($z \in [0.5, 1.3]$). Bottom panel: Residual differences $\Delta K = K_{\text{fit}} - K_{\text{calc}}$ for both linear models.}
    \label{fig:figC4_models}
\end{figure}

We adopt $\mathrm{d}K/\mathrm{d}z \approx 1.2$ (the local gradient fit) to prevent distorting selection boundary slopes across redshift. A systematic check using a global amplitude fit ($1.5z$) yields negligible shifts ($\Delta M_0 = -0.03\,\mathrm{mag}$, $\Delta \sigma = +0.10\,\mathrm{mag}$), confirming rate inference stability. Figure~\ref{fig:figC4_models} details these models and residual differences $\Delta K$ across $z \in [0.0, 1.3]$.

\subsection*{Analytical Derivation of the Linear Selection Boundary}
At high redshift, where the survey probes only the extreme bright tail of the intrinsic luminosity function ($m_c \ll M_0 + \mu(z)$), the integrated efficiency $\bar{\epsilon}(z)$ simplifies to:
\begin{equation}
    \bar{\epsilon}(z) \propto \epsilon_0 \exp\left[ -\frac{(M_c(z)-M_0)^2}{2\sigma^2} \right],
\end{equation}
where $M_c(z) \equiv m_c - \mu(z) - K(z) - A$. Taking the negative logarithm provides the mathematical justification for the behavior of the volumetric correction factor ($1/\bar{\epsilon}$):
\begin{equation}
    \log_{10} \left( \frac{1}{\bar{\epsilon}(z)} \right) \approx \frac{\log_{10}(e)}{2\sigma^2} \left[ \mu(z)+K(z)+A+M_0-m_c \right]^2 + C,
\end{equation}
where $C$ is a constant offset. 

While this expression is formally quadratic, governing the characteristic non-linear curvature observed at lower redshifts, it transitions toward a stable, quasi-linear slope at higher redshifts ($z \gtrsim 0.6$). The slope of the correction factor, $a$, is given by the total derivative:
\begin{equation}
    a = \frac{\mathrm{d}}{\mathrm{d}z} \left[ \log_{10} \left( \frac{1}{\bar{\epsilon}(z)} \right) \right] \approx \frac{\log_{10}(e)}{\sigma^2} \left( M_c(z) - M_0 \right) \left( \frac{\mathrm{d}\mu}{\mathrm{d}z} + \frac{\mathrm{d}K}{\mathrm{d}z} \right).
\end{equation}
As the survey probes deeper into the bright tail of the core-collapse luminosity function, the widening magnitude gap $(M_c(z) - M_0)$ is counterbalanced by the combined flattening of the cosmological dimming rate, $\mathrm{d}\mu/\mathrm{d}z$, and the constant linear gradient of the observer-frame $K$-correction ($\mathrm{d}K/\mathrm{d}z \approx 1.2$). When this linear selection boundary intersects the underlying population, the recovered rate correction emerges from the interaction of the boundary slope ($\mathrm{d}\mu/\mathrm{d}z + \mathrm{d}K/\mathrm{d}z$) and the population-driven effect of that cut slicing through the Gaussian tail ($\sim \log_{10}e / \sigma^2$).

\section{Selection Function Fitting Variants} 
\label{sec:app_fitting_details} 

To evaluate the stability of our baseline selection function optimization ($M_0 = -16.40 \pm 0.06$\,mag, $\sigma = 1.53 \pm 0.04$\,mag, $\chi^2_\nu = 1.78$ with fixed $\epsilon_0 = 0.50$), we evaluated alternative efficiency ceiling configurations and field-depth subsets:

Efficiency Ceiling Variations ($\epsilon_0 = 0.55$): Evaluating an alternative configuration with $\epsilon_0 = 0.55$ yields a slightly improved fit with $\chi^2_\nu = 1.60$ and recovers a population mean of $M_0 = -16.22$\,mag (matching the simulation templates of \citealt{Vincenzi2019, Vincenzi2021}). However, this configuration recovers a broader population width ($\sigma = 1.58 \pm 0.04$\,mag), increasing tension with the local \citet{Li2011a} value of $\sigma = 1.38$\,mag. Furthermore, the empirical low-redshift asymptote in our data visually favors $\epsilon_0 = 0.50$. Overall, baseline consistency across statistical fits and empirical asymptotes reliably converges within $\epsilon_0 \in [0.50, 0.60]$.

Field Depth Subsets (Deep-Fields Only): Restricting the forward-modeling analysis to deep-field simulated CC~SNe ($m_c = 24.50$\,mag) with $\epsilon_0 = 0.51$ yields $M_0 = -16.97 \pm 0.08$\,mag, $\sigma = 1.32 \pm 0.06$\,mag, and $\chi^2_\nu = 1.09$. Allowing the efficiency ceiling to reach its empirical asymptote of $\epsilon_0 = 0.75$ recovers $M_0 = -16.25 \pm 0.09$\,mag, but increases the dispersion to $\sigma = 1.59 \pm 0.06$\,mag with a poorer fit ($\chi^2_\nu = 2.10$). Reconciling empirical efficiency asymptotes with robust statistical fits proves difficult in deep-fields-only subsets, whereas both constraints naturally align in our All Fields baseline.

\section{Sensitivity to Luminosity Function Evolution}
\label{sec:app_lf_sensitivity}

Our baseline analysis assumes a redshift-invariant intrinsic CC~SN luminosity function based on local templates \citep{Li2011a, Vincenzi2019}. Under this baseline framework, extinction corrections are not explicitly deconvolved from the intrinsic luminosity function and are assumed to be encompassed within the observed dispersion (setting $A_{\mathrm{ext}} = 0$). While there is currently neither a firm theoretical consensus nor direct observational evidence for luminosity function evolution with redshift, the population mean absolute magnitude $M_0$ and intrinsic dispersion $\sigma$ could hypothetically shift at higher redshifts due to progenitor metallicity trends or evolving subtype fractions \citep[e.g.,][]{Vincenzi2021}. To quantify the sensitivity of our results to such potential population drift, we construct an analytic framework based on our physical forward model, adopting $\Delta M_0 = \pm 0.1\,\mathrm{mag}$ at $z = 1$ as a standard benchmark value for potential mean evolution.

Defining the redshift-dependent limiting absolute magnitude threshold as $M_c(z) \equiv m_c - \mu(z) - K(z)$, our physical forward model parameterizes the efficiency in the high-redshift, flux-limited tail as:
\begin{equation}
    \bar{\epsilon}(z) \propto \exp\left[ -\frac{(m_c - \mu(z) - K(z) - M_0)^2}{2\sigma^2} \right] = \exp\left[ -\frac{(M_c(z) - M_0)^2}{2\sigma^2} \right],
    \label{eq:app_tail_efficiency}
\end{equation}
where $m_c$ is the survey limiting magnitude, $\mu(z)$ is the distance modulus, and $K(z)$ is the $K$-correction. We can analytically evaluate the response of the high-redshift volumetric rate corrections ($1/\bar{\epsilon}$) to hypothetical variations in $M_0$ and $\sigma$ without re-running full image-level simulations.

Taking the base-10 logarithm of the inverse detection efficiency, $y \equiv \log_{10}(1/\bar{\epsilon}(z))$, and applying the logarithmic identity $\log_{10}(x) = \log_{10}(e) \cdot \ln(x)$, we express $y$ as:
\begin{align}
    \log_{10}\left(\frac{1}{\bar{\epsilon}(z)}\right) &= \log_{10}\left( \exp\left[ \frac{(m_c - \mu(z) - K(z) - M_0)^2}{2\sigma^2} \right] \right) + C \nonumber \\
    &= \frac{\log_{10}(e)}{2\sigma^2} \left[ m_c - \mu(z) - K(z) - M_0 \right]^2 + C,
    \label{eq:app_log_inv_eff}
\end{align}
where $C$ incorporates constant normalization factors and offsets.

To evaluate the response of the logarithmic correction factor to potential shifts in the population mean $M_0$, we compute the partial derivative $\frac{\partial \log_{10}(1/\bar{\epsilon})}{\partial M_0}$ from Equation~\ref{eq:app_log_inv_eff}. Applying the chain rule yields:
\begin{equation}
    \left| \frac{\partial \log_{10}(1/\bar{\epsilon})}{\partial M_0} \right| = \frac{\log_{10}(e)}{\sigma^2} \left| m_c - \mu(z) - K(z) - M_0 \right|.
    \label{eq:app_dM0}
\end{equation}

Similarly, computing the partial derivative with respect to the intrinsic population dispersion $\sigma$ yields:
\begin{equation}
    \left| \frac{\partial \log_{10}(1/\bar{\epsilon})}{\partial \sigma} \right| = \frac{\log_{10}(e)}{\sigma^3} \left[ m_c - \mu(z) - K(z) - M_0 \right]^2.
    \label{eq:app_dsigma}
\end{equation}

To evaluate these sensitivity responses at $z = 1.0$, we evaluate the detection depth gap. Accounting for the survey apparent depth ($m_c = 23.96\,\mathrm{mag}$), distance modulus ($\mu(1.0) = 44.10\,\mathrm{mag}$), and $K$-correction ($K(1.0) = 1.20\,\mathrm{mag}$), the absolute magnitude detection boundary reaches $M_c = -21.34\,\mathrm{mag}$. Relative to the population mean ($M_0 = -16.40\,\mathrm{mag}$), this yields a depth gap of $|M_c - M_0| = 4.94\,\mathrm{mag}$ (using baseline fit parameters $\log_{10}(e) \approx 0.4343$ and $\sigma = 1.53\,\mathrm{mag}$).

Under our benchmark shift of $\Delta M_0 = \pm 0.1\,\mathrm{mag}$ in average intrinsic brightness, the logarithmic correction responds by:
\begin{align}
    \Delta \log_{10}(1/\bar{\epsilon}) &\approx \frac{\log_{10}(e)}{\sigma^2} |M_c - M_0| \cdot \Delta M_0 \nonumber \\
    &= \frac{0.4343}{1.53^2} \times 4.94 \times 0.1 \approx 0.0917\,\mathrm{dex},
    \label{eq:dlog_inv_eff_dM0}
\end{align}
which corresponds to a $\sim 23.5$~per~cent fractional shift in the linear correction factor $1/\bar{\epsilon}$ ($10^{0.0917} \approx 1.235$). 

Likewise, a hypothetical shift of $\Delta \sigma = \pm 0.03\,\mathrm{mag}$ in the population width alters the logarithmic correction by:
\begin{align}
    \Delta \log_{10}(1/\bar{\epsilon}) &\approx \frac{\log_{10}(e)}{\sigma^3} |M_c - M_0|^2 \cdot \Delta \sigma \nonumber \\
    &= \frac{0.4343}{1.53^3} \times (4.94)^2 \times 0.03 \approx 0.0887\,\mathrm{dex},
\end{align}
which translates to a $\sim 22.6$~per~cent variation in linear space ($10^{0.0887} \approx 1.226$).

If low-metallicity progenitor physics at high redshift (e.g., reduced wind-driven mass loss or IMF variations) were to induce a progressive population brightening scaling to $\Delta M_0 \approx -0.1\,\mathrm{mag}$ at $z \sim 1$, the linear selection correction ($1/\bar{\epsilon}$) at $z \sim 1$ would decrease by $\sim 23.5$~per~cent. Modeling this hypothetical effect as a progressive scaling factor of $(1 - 0.235z)$ on the efficiency corrections shifts the inferred evolutionary index to $\alpha = 2.44$, with an adjusted local normalization of $R_0 = 1.63 \times 10^{-4}\,\mathrm{Mpc}^{-3}\,\mathrm{yr}^{-1}$.

This sensitivity test confirms that our core physical conclusion—strong volumetric rate growth with redshift—is robust to potential luminosity function evolution, as the evolutionary index remains steep ($\alpha = 2.44$) even under progressive population brightening.

Ultimately, a redshift-invariant luminosity function remains our baseline model due to the lack of empirical or theoretical constraints requiring evolution. Future ultra-deep infrared surveys e.g with the James Webb or Nancy Grace Roman Telescopes will be required to directly probe the faint tail of the population at high redshift.

\section{Deterministic Uncertainty Results}
\label{app:deterministic_rates}

The deterministic uncertainty analysis exhibits a more pronounced positive asymmetry compared to the fiducial MC results. Consequently, the four-model mean CC~SN yield shifts upward to $k = 0.0129 \pm 0.0011\,\mathrm{M}_\odot^{-1}$, compared to the MC baseline of $k = 0.0114 \pm 0.0009\,\mathrm{M}_\odot^{-1}$. 

For the power-law fits, as summarized in Column 1 of Table~\ref{tab:field_comparison}, the deterministic framework yields a lower reduced $\chi^2$ ($\chi^2_{\mathrm{red}} = 0.32$, $\sigma_{\mathrm{eq}} = 0.04$) than the MC baseline ($\chi^2_{\mathrm{red}} = 0.41$, $\sigma_{\mathrm{eq}} = 0.08$). This lower $\chi^2_{\mathrm{red}}$ and smaller equivalent scatter ($\sigma_{\mathrm{eq}}$) stem from standard linear quadrature propagation, which produces broader asymmetric uncertainty bounds that increase the error denominators in the fit evaluations. 

\begin{figure}
    \centering
    \includegraphics[width=\columnwidth]{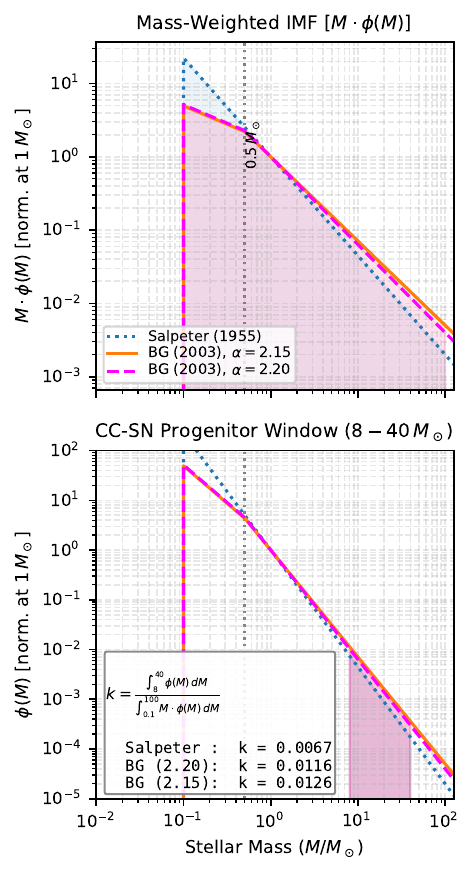}
    \caption{Comparison of a single power-law \citet{Salpeter1955} IMF baseline against two continuity-enforced variants of the \citet{Baldry2003} (BG03) formulation. Vertical dotted lines show the break point mass ($0.5\,\mathrm{M}_\odot$) and the CC~SN progenitor integration window ($8\text{--}40\,\mathrm{M}_\odot$). Inset annotations show the calculation of the yield ($k$) in each case.}
    \label{fig:figB1}
\end{figure}

This higher deterministic yield ($k = 0.0129\,\mathrm{M}_\odot^{-1}$) aligns with the shallower of the two \citet{Baldry2003} IMF variants featuring a high-mass slope of $-2.15$ \citep[as discussed in][]{Hopkins2006}. For a standard progenitor mass range of $8\text{--}40\,\mathrm{M}_\odot$, this configuration results in a theoretical yield of $k_{\mathrm{BG03(\alpha = 2.15)}} = 0.0126$, as illustrated by the solid orange profile in Figure~\ref{fig:figB1}.

\FloatBarrier 

%

\section{CSFH Models and IMF Yield Calculations}
\label{app:csfh_math}

The CC SN yield $k$ (in units of $\mathrm{M}_\odot^{-1}$) represents the number of CC~SNe formed per unit stellar mass. For an initial mass function $\phi(M)$, lower/upper progenitor limits $[M_{\mathrm{P,low}}, M_{\mathrm{P,high}}]$, and total stellar integration limits $[M_{\mathrm{low}}, M_{\mathrm{high}}]$, $k$ is given by:

\begin{equation}
k = \frac{\int_{M_{\mathrm{P,low}}}^{M_{\mathrm{P,high}}} \phi(M)\, dM}{\int_{M_{\mathrm{low}}}^{M_{\mathrm{high}}} M \, \phi(M)\, dM}.
\label{eq:general_yield}
\end{equation}

Assuming a canonical progenitor window of $8\text{--}40\,\mathrm{M}_\odot$ and total stellar range of $0.1\text{--}100\,\mathrm{M}_\odot$, standard functional forms yield the following baseline values:
\begin{align}
k_{\mathrm{Kroupa(\alpha = 2.30)}} &= \frac{\int_{8}^{40} \phi_{\mathrm{Kroupa(\alpha = 2.30)}}(M)\,dM}{\int_{0.1}^{100} M\,\phi_{\mathrm{Kroupa(\alpha = 2.30)}}(M)\,dM} = 0.0101, \label{eq:ccsn_yields1} \\
k_{\mathrm{Salpeter(\alpha = 2.35)}} &= \frac{\int_{8}^{40} \phi_{\mathrm{Salpeter(\alpha = 2.35)}}(M)\,dM}{\int_{0.1}^{100} M\,\phi_{\mathrm{Salpeter(\alpha = 2.35)}}(M)\,dM} = 0.0067. \label{eq:ccsn_yields2}
\end{align}

\begin{figure}
    \centering
    \includegraphics[width=\columnwidth]{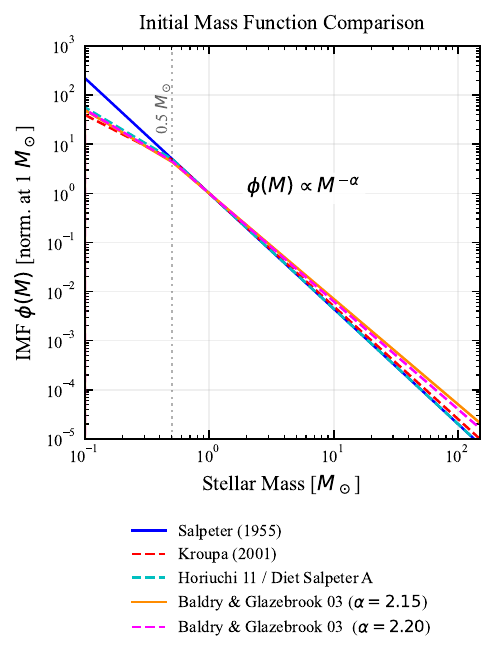}
    \caption{Comparison of alternative IMF ($\phi(M)$) parameter models from $0.1$ to $125\,\mathrm{M}_\odot$ stellar mass, normalized to the standard single power-law \citet{Salpeter1955} IMF at $1\,\mathrm{M}_\odot$. Alongside are shown the broken power-law configurations from \citet{Kroupa2001}, \citet{Horiuchi2011}, and the two \citet{Baldry2003} options (slope ($-\alpha) = -2.15, -2.20$). The dotted vertical line identifies the common break point at $0.5\,\mathrm{M}_\odot$.}
    \label{fig:imf_comparison}
\end{figure}

For the \citet{Baldry2003} IMF—which features a broken power law at $0.5\,\mathrm{M}_\odot$ and high-mass slopes of $-2.15$ or $-2.20$ (Figure~\ref{fig:imf_comparison})—the theoretical yields increase significantly:
\begin{align}
k_{\mathrm{BG03(\alpha = 2.20)}} &= \frac{\int_{8}^{40} \phi_{\mathrm{BG03(\alpha = 2.20)}}(M)\,dM}{\int_{0.1}^{100} M\,\phi_{\mathrm{BG03(\alpha = 2.20)}}(M)\,dM} = 0.0116, \label{eq:ccsn_yields3} \\
k_{\mathrm{BG03(\alpha = 2.15)}} &= \frac{\int_{8}^{40} \phi_{\mathrm{BG03(\alpha = 2.15)}}(M)\,dM}{\int_{0.1}^{100} M\,\phi_{\mathrm{BG03(\alpha = 2.15)}}(M)\,dM} = 0.0126. \label{eq:ccsn_yields4}
\end{align}

Mass allocations under single-star and binary core-collapse evolutionary pathways corresponding to these yield derivations are illustrated in Figure~\ref{fig:mass_split_combined} of Section~\ref{sec:ccsn_csfh_tracers}.

The four recent reference CSFH models ($\psi(z)$ in $\mathrm{M}_\odot\,\mathrm{yr}^{-1}\,\mathrm{Mpc}^{-3}$) evaluated in Section~\ref{sec:ccsn_csfh_tracers} are parametrized as follows:

\begin{description}
    \item[\citet{Madau2014}:] UV and IR fit ($0 < z < 8$) assuming a \citet{Salpeter1955} IMF:
    \begin{equation}
    \psi(z) = 0.015 \, \frac{(1+z)^{2.7}}{1 + \left[(1+z)/2.9\right]^{5.6}}.
    \end{equation}

    \item[\citet{Madau2017}:] X-ray binary reconstruction up to $z \sim 6$ assuming a \citet{Kroupa2001} IMF:
    \begin{equation}
    \psi(z) = 0.010 \, \frac{(1+z)^{2.6}}{1 + \left[(1+z)/3.2\right]^{6.2}}.
    \end{equation}

    \item[\citet{Harikane2022}:] Combined \textit{JWST} UV, IR, and H$\alpha$ data assuming a \citet{Salpeter1955} IMF:
    \begin{equation}
    \psi(z) = \left[\,61.7(1+z)^{-3.13} + 10^{0.22(1+z)} + 2.4 \times 10^{0.5(1+z) - 3}\,\right]^{-1}.
    \end{equation}

    \item[\citet{Kim2024}:] Lensed \textit{JWST} stacked luminosity functions ($z \gtrsim 8$) assuming a \citet{Salpeter1955} IMF.
\end{description}

\section{Field Depth and Completeness}
\label{sec:field_depth}

\begin{table*}
    \centering
    \caption{Power-law fit parameters ($R_{\mathrm{CC}}(z) = R_0(1+z)^\alpha$) and average best-fit CSFH CC~SN yields ($k_{\mathrm{CC}}$) across survey field depth subsets.}
    \label{tab:field_comparison}
    \setlength{\tabcolsep}{8pt}
    \renewcommand{\arraystretch}{1.15}
    \begin{tabular}{l c c c c}
        \hline
        \hline
        Parameter / Metric & 
        \makecell{All Fields\\(Det)} & 
        \makecell{All Fields\\(MC baseline)} & 
        \makecell{Deep Fields\\(MC)} & 
        \makecell{Shallow Fields\\(MC)} \\
        \hline
        Bins & 11 & 11 & 11 & 10 ($z \le 1.0$) \\
        $\langle k_{\mathrm{CC}} \rangle$ ($\mathrm{M}_\odot^{-1}$) & $0.0129 \pm 0.0009$ & $0.0114 \pm 0.0009$ & $0.0108 \pm 0.0009$ & $0.0115 \pm 0.0011$ \\
        $R_0$ ($10^{-4}\,\mathrm{yr}^{-1}\,\mathrm{Mpc}^{-3}$) & $1.58 \pm 0.17$ & $1.58 \pm 0.17$ & $1.4 \pm 0.2$ & $1.7 \pm 0.5$ \\
        $\alpha$ & $2.85 \pm 0.25$ & $2.85 \pm 0.25$ & $3.0 \pm 0.4$ & $3.0 \pm 0.7$ \\
        $\chi^2_{\mathrm{red}}$ & 0.32 & 0.41 & 0.62 & 1.90 \\
        $p$-value & 0.97 & 0.93 & 0.82 & 0.06 \\
        $\sigma_{\mathrm{eq}}$ & 0.04 & 0.08 & 0.22 & 1.91 \\
        \hline
    \end{tabular}
\end{table*}

We select the 11-bin ``All Fields'' MC sample as our primary fiducial baseline for several key reasons. First, the MC framework avoids the simplifying assumptions of standard linear error propagation (see Appendix~\ref{app:deterministic_rates} for a full comparison with deterministic error propagation), directly sampling the underlying parameter probability density functions to yield robust, non-linear percentile-based confidence intervals ($\text{16th--84th}$ percentiles). Second, it properly accounts for asymmetric boundary constraints on efficiency and extinction corrections without artificially inflating the error budget at high redshift. Finally, the ``All Fields'' configuration maximizes the effective statistical volume of the survey and mitigates cosmic variance across independent lines of sight ($\sigma_{\mathrm{eq}} = 0.08$ compared to $\sigma_{\mathrm{eq}} = 0.22$ for deep fields alone).

To verify whether field depth or selection completeness introduces systematic bias into our rate calculations, we evaluated power-law fits across different field subsets (with fit metrics summarized in Table~\ref{tab:field_comparison}). Across all field subsets, the fitted power-law parameters ($R_0$ and $\alpha$) and best-fit CSFH yields ($\langle k_{\mathrm{CC}} \rangle$) remain fully consistent within uncertainties, confirming that our detection efficiency corrections reliably account for variations in survey depth.

\begin{figure}
    \centering
    \includegraphics[width=\columnwidth]{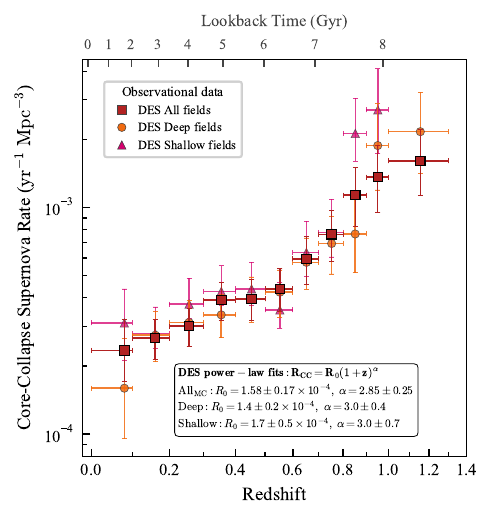}
    \caption{Volumetric CC~SN rates derived from DES, split by field depth subset: All Fields (red squares), Deep Fields (orange circles), and Shallow Fields (magenta triangles). Power-law fits ($R(z) = R_0(1+z)^\alpha$) for each subset are summarized in Table~\ref{tab:field_comparison} and the inset box.}
    \label{fig:des_ccsne_rates_app}
\end{figure}

Restricting the analysis to CC~SNe in the deep fields yields a power-law slope of $\alpha = 3.0 \pm 0.4$ and a local normalization of $R_0 = (1.4 \pm 0.2) \times 10^{-4}\,\mathrm{yr}^{-1}\,\mathrm{Mpc}^{-3}$, with a four-model mean yield value $\langle k_{\mathrm{CC}} \rangle = 0.0108 \pm 0.0009\,\mathrm{M}_\odot^{-1}$. The shallow fields result in $R_0 = (1.7 \pm 0.5) \times 10^{-4}\,\mathrm{yr}^{-1}\,\mathrm{Mpc}^{-3}$, $\alpha = 3.0 \pm 0.7$, and $\langle k_{\mathrm{CC}} \rangle = 0.0115 \pm 0.0011\,\mathrm{M}_\odot^{-1}$ across 10 bins ($z \le 1.0$). Both sub-field fits exhibit slightly steeper slopes ($\alpha \approx 3.0$) than the combined baseline ($R_0 = 1.58 \times 10^{-4}\,\mathrm{yr}^{-1}\,\mathrm{Mpc}^{-3}$, $\alpha = 2.85$, $\langle k_{\mathrm{CC}} \rangle = 0.0114 \pm 0.0009\,\mathrm{M}_\odot^{-1}$), though they remain fully consistent within uncertainties, as illustrated in Figure~\ref{fig:des_ccsne_rates_app}. All reported four-model mean yields and uncertainties in this field depth comparison are derived from MC sampling across the four CSFH models.

\clearpage

\enlargethispage{1\baselineskip}

\let\oldthebibliography\thebibliography
\let\endoldthebibliography\endthebibliography
\renewenvironment{thebibliography}[1]{%
  \begin{oldthebibliography}{#1}%
    \raggedbottom
    \setlength{\itemsep}{0pt plus 0.2pt}
    \setlength{\parsep}{0pt}
    \setlength{\parskip}{0pt}
}{%
  \end{oldthebibliography}%
}

\bibliographystyle{mnras}
\bibliography{example} 




\bsp	
\label{lastpage}
\end{document}